\documentclass{aa} 

\RequirePackage{aas_macros}
\usepackage{graphicx}
\usepackage{natbib}
\bibpunct{(}{)}{;}{a}{}{,} 
\usepackage{txfonts}
\usepackage{aalongtable}
\usepackage{multirow}
\usepackage{placeins}
\usepackage{latexsym}
\usepackage{wasysym}

\begin{document}
\def\r500{$r_{\mathrm{500}}$}
\def\P3P0{P3/P0}
\def\B_P3{B$_{\mathrm{P3}}$}
\def\b_w{B$_{\mathrm{w}}$}
\def\id_P3{P3/P0$_{\mathrm{ideal}}$}
\def\i_w{$w_{\mathrm{ideal}}$}
\def\C_B_P3{B$_{\mathrm{P3,c}}$}
\def\c_b_w{B$_{\mathrm{w,c}}$}
\def\p3_c{P3/P0$_{\mathrm{c}}$}
\def\cp3{P3/P0_c}
\def\w_c{$w_{\mathrm{c}}$}

   \title{Studying the properties of galaxy cluster morphology estimators}

   \author{A. Wei\ss mann\inst{1,2}, H. B\"ohringer\inst{1}, R. \v{S}uhada\inst{3} and S. Ameglio\inst{4}}

   \institute{$^1$ Max-Planck-Institut f\"{u}r extraterrestrische Physik, Postfach 1312, Giessenbachstr., 85741 Garching, Germany,\\ {\tt weissman@mpe.mpg.de}\\
	      $^2$ Institut f\"{u}r Astronomie, Universit\"{a}t Wien, T\"{u}rkenschanzstr. 17, 1180 Wien, Austria\\
	      $^3$ Universit\"{a}ts-Sternwarte, Fakult\"{a}t f\"{u}r Physik der Ludwig-Maximilians-Universit\"{a}t, Scheinerstr. 1, 81679 M\"{u}nchen, Germany \\
	      $^4$ Dipartimento di Astronomia dell'Universit\`{a} di Trieste, Via G.B. Tiepolo 11, 34131 Trieste, Italy\\}
              
   \date{Received xxxx; Accepted yyyy}

  \abstract
{X-ray observations of galaxy clusters reveal a large range of morphologies with various degrees of disturbance, showing that the assumptions of hydrostatic equilibrium and spherical shape which are used to determine the cluster mass from X-ray data are not always satisfied. It is therefore important for the understanding of cluster properties as well as for cosmological applications to detect and quantify substructure in X-ray images of galaxy clusters. Two promising methods to do so are power ratios and center shifts. Since these estimators can be heavily affected by Poisson noise and X-ray background, we performed an extensive analysis of their statistical properties using a large sample of simulated X-ray observations of clusters from hydrodynamical simulations. We quantify the measurement bias and error in detail and give ranges where morphological analysis is feasible. A new, computationally fast method to correct for the Poisson bias and the X-ray background contribution in power ratio and center shift measurements is presented and tested for typical XMM-\emph{Newton} observational data sets. We studied the morphology of 121 simulated cluster images and establish structure boundaries to divide samples into relaxed, mildly disturbed and disturbed clusters. In addition, we present a new morphology estimator - the peak of the 0.3-1 \r500 \P3P0 profile to better identify merging clusters. The analysis methods were applied to a sample of 80 galaxy clusters observed with XMM-\emph{Newton}. We give structure parameters (\P3P0 in \r500, $w$ and \P3P0$_{\mathrm{max}}$) for all 80 observed clusters. Using our definition of the \P3P0 ($w$) substructure boundary, we find 41\% (47\%) of our observed clusters to be disturbed.}


\authorrunning{Wei\ss mann et al.}
\titlerunning{Studying the properties of galaxy cluster morphology estimators}

   \maketitle

\titlerunning{short title}
\authorrunning{Names}
%

\section{Introduction}

Clusters of galaxies form from positive density fluctuations and grow hierarchically through the extremely energetic process of merging and mass accretion. With due time they are thought to reach dynamical equilibrium and form the largest virialized structures in the Universe. This makes them very interesting tools to study cosmology and the evolution of large scale structure in which they appear as nodes at the intersection of filaments. In the soft X-ray band the hot intracluster medium (ICM) which resides in the intergalactic space and makes up about 15\% of the total cluster mass is observed. Already early X-ray observations of galaxy clusters revealed that the ICM distribution is not smooth and azimuthally symmetric for all objects. In the beginning of the 1990s it became more clear from ROSAT observations that galaxy clusters are not relaxed objects but that they contain substructure \citep[e.g.][]{Briel1991,Briel1992}. Since then, a lot of effort was put into the identification and characterization of substructure in the ICM in order to determine the dynamical state of the cluster. \citet{Jones1992} showed that around 30\% of their $\sim$200 clusters observed with the EINSTEIN satellite contain substructure. This was an important step in the understanding of structure formation, because it showed that cluster formation and evolution has not finished yet. In previous studies different parameter boundaries for the distinction of substructured and regular clusters have been used. The fraction of clusters with substructure was estimated to be about 40-70\% for X-ray observations \citep{Mohr1995, Jones1999, Schuecker2001, Kolokotronis2001}. This indicates that the merging and accretion activity, which is reflected by the presence of multiple surface brightness peaks or disturbed morphologies, has not yet ceased in clusters. Substructure as a tracer of merging activity indicates a deviation from the relaxed and virialized state and can make a precise cluster mass determination very difficult. Since hydrostatic equilibrium is one of the main assumptions for cluster mass estimates, large errors can occur, which influence the constraints of cosmological parameters which are derived using cluster masses. Recent studies of simulations \citep[e.g.][]{Nagai2007,Piffaretti2008,Jeltema2008,Lau2009,Meneghetti2010,Rasia2012} and observations \citep[e.g.][]{Zhang2008,Okabe2010} show that the hydrostatic X-ray mass can be biased low between 10\% and 30\%. The largest deviations are expected to occur for galaxy clusters with substructure and it is therefore very important to accurately characterize substructure and the dynamical state of a cluster.\newline

Over the years many methods to characterize and quantify substructure in galaxy clusters were proposed (see \citet{Buote2002} for a review). A simple and descriptive method to reveal substructure in a galaxy cluster is to subtract a smooth elliptical $\beta$ model from the X-ray cluster image and to examine the residuals \citep[e.g.][]{Davis1993,Neumann1997}. Wavelet analysis and decomposition have been applied to many clusters in X-rays \citep[e.g.][]{Slezak1994,Arnaud2000,Maurogordato2011}. This technique enables substructure analysis on different scales and the separation of different components. Another approach is the classification of cluster morphologies by visual inspection for X-ray images \citep[e.g.][]{Jones1992}. Several other methods classify the morphology of galaxy clusters. Measuring e.g. a clusters ellipticity is very common \citep[e.g.][]{McMillan1989,Pinkney1996,Schuecker2001,Plionis2002}, but this property is not a good indicator for a clusters dynamical state because both relaxed and disturbed clusters can have significant ellipticities. Better indicators of the dynamical state of a cluster are power ratios \citep{Buote1995,Buote1996} and center shifts \citep{Mohr1993}, which will be both addressed in this paper. \newline

Most substructure studies were performed on low-redshift clusters \citep[e.g.][]{Mohr1995,Buote1996,Jones1999}. With the recent increase in the detection of high-redshift clusters, also the number of substructure studies of fairly large high-z samples using power ratios and other substructure parameters became important \citep[e.g.][]{Bauer2005,Jeltema2005,Hashimoto2007c}. However, studies of the uncertainties and bias using these methods especially for low-quality (low net counts and/or high background) observations are sparse \citep[e.g.][]{Buote1996,Jeltema2005,Boehringer2009}. \newline

This is the main issue we want to address in this paper. We use a large sample of simulated X-ray cluster images to study the influence of shot noise on the power ratio and center shift calculation and present a method based on \citet{Boehringer2009} (B10 hereafter) to correct for it. We give parameter ranges in which a cluster can be expected to be relaxed or significantly disturbed. In addition, we give updated substructure parameters for a sample of 80 galaxy clusters based on XMM-observations which are part of several well-known samples. We discuss power ratios, center shifts and a new parameter in detail and present possible applications.\newline
The paper is structured as follows. In Sect. \ref{structureparameters} we introduce structure parameters used in this study. We briefly present the set of simulated X-ray cluster images in Sect. \ref{Simulations} which were used to calibrate and test our method. The investigation of the influence of Poisson noise and net counts on the reliability of power ratios and center shifts is given in Sect. \ref{Section4}. We also introduce our method to correct for the noise and background contribution and test its accuracy. In Sect. \ref{threshold} we define different morphological boundaries for power ratios and center shifts. We apply our analysis to a sample of 80 galaxy clusters observed with XMM-\emph{Newton}, which is characterized in Sect. \ref{Section5}. A short overview of the data reduction is given in Sect. \ref{Section6}. In Sect. \ref{secMorph} we show results of the morphological analysis of the observed cluster sample and introduce an improved morphological estimator. We discuss the results in Sect. \ref{Discussion} and conclude with Sect. \ref{Conclusions}. Throughout the paper, the standard $\Lambda$CDM cosmology was assumed: $H_{0}$=70 km s$^{-1}$ Mpc$^{-1}$, $\Omega_{\Lambda}$=0.7, $\Omega_{\mathrm{M}}$=0.3.

\section{Substructure parameters}
\label{structureparameters}
\subsection*{Power ratios}
\label{powerratios}

The power ratio method was introduced by \citet{Buote1995} with the aim to parametrize the amount of substructure in the intracluster medium and to relate it to the dynamical state of a cluster. Only the distribution of structure on cluster scales which dominates the global dynamical state is of interest. Power ratios are based on a 2D multipole expansion of the clusters gravitational potential using the surface mass density distribution. Power ratios are thus giving an account of the azimuthal structure where moments of increasing order describe finer and finer structures. The powers are calculated within a certain aperture radius (e.g. \r500) with the aperture centered on the mass centroid. \newline

The 2D multipole expansion of the two-dimensional gravitational potential $\psi(R,\phi)$ can be written as

\begin{equation}
 \psi(R,\phi)=-2G \left[ a_{0} ln\frac{1}{R} + \sum^{\infty}_{m=1}\frac{1}{mR^{m}}(a_{m}cos(m\phi)+b_{m}sin(m\phi)) \right]
\end{equation} 

\noindent where $a_{m}$ and $b_{m}$ are

\begin{equation}
 a_{m}(R)=\int_{R' \le R} \Sigma(\vec{x}')(R')^{m} cos(m\phi')d^{2}x'
\label{Eq2}
\end{equation}
\begin{equation}
 b_{m}(R)=\int_{R' \le R} \Sigma(\vec{x}')(R')^{m} sin(m\phi')d^{2}x'
\label{Eq3}
\end{equation}

\noindent where $\vec{x}'$=$(R',\phi')$ are the coordinates, $G$ is the gravitational constant and $\Sigma$ represents the surface mass density \citep{Buote1995}. The powers are defined by the integral of the magnitude of $\psi_{m}$, the m-th term in the multipole expansion of the potential, and evaluated in a circular aperture with radius R

\begin{equation}
 P_{m}(R)=\frac{1}{2\pi} \int_{0}^{2\pi} \psi_{m}(R,\phi) \psi_{m}(R,\phi) d\phi .
\label{Equation4}
\end{equation}

Ignoring factors of 2G, this relates to the following relations which are used to calculate the powers, where $a_{m}$ and $b_{m}$ are taken from Eq. \ref{Eq2} and \ref{Eq3}
\begin{equation}
 P_{0}=[a_{0}ln(R)]^{2}
\end{equation}
and
\begin{equation}
 P_{m}=\frac{1}{2m^{2}R^{2m}}(a_{m}^{2}+b_{m}^{2}) .
\end{equation}

In X-rays the surface brightness is used instead of the projected surface mass density, assuming that the X-ray surface brightness distribution traces the gravitational potential \citep{Buote1995}. In order to obtain powers which are independent of the X-ray luminosity, they are normalized by the zeroth-order moment and thus called power ratios. This allows a direct comparison of clusters with different X-ray brightness. P0, the monopole, gives the flux. P1 and P2 represent dipole and quadrupole, P3 and P4 can be associated with hexapole and octopole moments. Higher order moments become more sensitive to disturbances on smaller scales which do not significantly contribute to the characterization of the global dynamical state of a cluster. The power ratios P2/P0 and P4/P0 are strongly correlated, however P4 is more sensitive to smaller scales than P2. While relaxed but elliptical clusters rather yield low P2/P0 and merging systems show higher P2/P0, this power ratio is not a clear indicator of the dynamical state because it is sensitive to both ellipticity or bimodality. Odd moments are sensitive to unequal-sized bimodal structures and asymmetries, while they vanish for relaxed, single-component clusters. \P3P0 is thus the smallest moment which unambiguously indicates substructure in the ICM and provides a clear measure for the dynamical state of a cluster \citep[e.g.][B10]{Buote1995,Jeltema2005}. It is therefore the primary substructure measure in our analysis.
\subsection*{Center shifts}
\label{w}

The center shift parameter $w$ measures the centroid variations in different aperture sizes. The centroid is defined as the "center of mass" of the X-ray surface brightness and obtained for each aperture size separately. The X-ray peak is determined from an image smoothed with a Gaussian with $\sigma$ of 8 arcseconds. We calculate the offset of the X-ray peak from the centroid for 10 aperture sizes (0.1-1 \r500) and obtain the final parameter $w$ as the standard deviation of the different center shifts in units of \r500 \citep[e.g.][B10]{Mohr1993,Ohara2006}:
\begin{equation}
w=\left[\frac{1}{N-1} \sum_{i} (\Delta_{i}- <\Delta>)^{2}  \right]^{1/2} \times \frac{1}{r_{\mathrm{500}}}
\end{equation}

\noindent where $\Delta_{i}$ is the offset between the centroid and the X-ray peak in aperture $i$.

\section{Sample of simulated clusters}
\label{Simulations}

We use a set of 121 simulated cluster X-ray images to test the power ratio and center shift method, their bias due to shot noise and their uncertainties. This set includes 117 simulations from \citet{Borgani2004} and 4 from \citet{Dolag2009} to populate the desired mass range. All clusters were simulated using the TreePM/SPH code GADGET-2 \citep{Springel2005}. The clusters were extracted from the simulation at z=0 and the X-ray images were created by \citet{Ameglio2007,Ameglio2009}. The simulated cluster images do not include any observational artifacts (noise, bad pixels etc.) or background and were already used by B10. Due to the so-called overcooling problem in galaxy cluster simulations \citep[e.g.][]{Borgani}, the images may contain clumps of cold gas which appear as point-like sources. \citet{Ameglio2007} detected and removed these gas clumps. All remaining structures are therefore infalling groups or clusters. Keeping cold gas clumps in the simulated X-ray images may lead to a larger fraction of disturbed clusters and a different distribution of substructure parameters than is observed \citep[][B10 - Sect. 5.2.]{Nagai2007, Piffaretti2008}. The distribution of the parameters, however, is only critical for a direct comparison of simulations and observations, which is not the scope of this paper.\newline
Although the clusters are drawn from two sets of simulations they cover the full range of morphologies of clusters in the Universe and include a wide mass range ($0.8\times 10^{14} - 2.2 \times 10^{15} h^{-1}$ M$_{\astrosun}$). This sample is used exclusively to test the bias correction method and to calibrate the structure boundaries, thus to relate the visual impression of the image to a P3/P0 and $w$ range. For these purposes it is not crucial to use a representative sample of the full mass range, especially since the simulated cluster morphology distribution is only weakly mass dependent. We only required the sample to cover the full range of morphological parameters and do not take into account any global cluster properties.\newline

\begin{figure}[h]
\centering
    \includegraphics[width=\columnwidth]{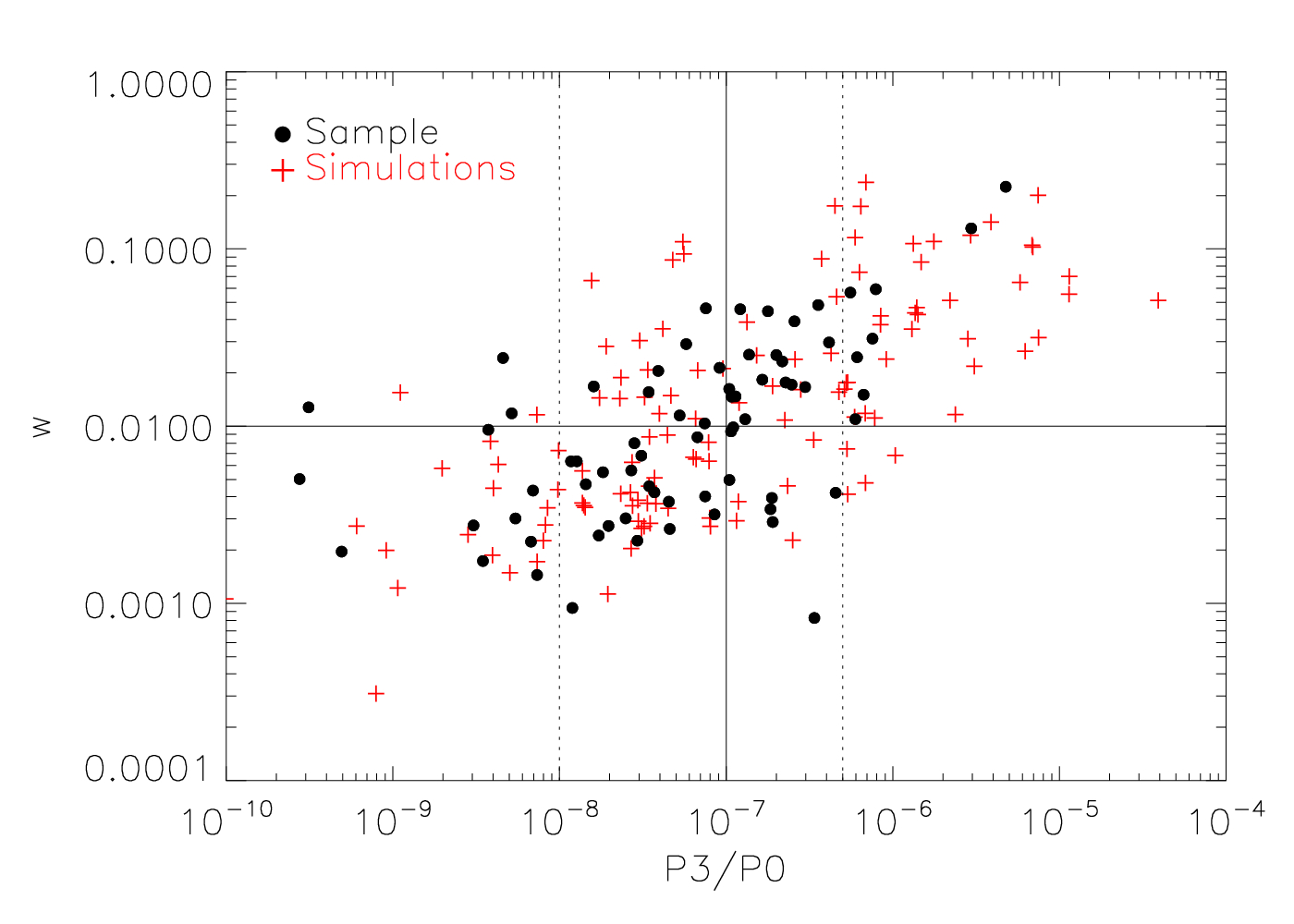}
\caption{Comparison of the sample of 80 clusters observed with XMM-\emph{Newton} (black circles) and 121 simulated X-ray cluster images (red crosses) in the \P3P0-$w$ plane. The solid and dotted lines show the different morphological ranges as discussed below in Sect. \ref{threshold}.}
\label{SIMREXP3w}
\end{figure}

A comparison between the substructure parameters $w$ and \P3P0 of the sample of 80 observed clusters and the simulations without noise is given in Fig. \ref{SIMREXP3w}. This figure also gives a first impression of the parameter range clusters occupy in this diagram - namely $10^{-10}< \mathrm{P3/P0} <10^{-4}$ and $10^{-4}<w<1$. Clusters sometimes yield negative \P3P0 values after the bias correction (\p3_c, see Sect. \ref{method}) with an uncertainty indicating that the result is consistent with zero. Such clusters are not displayed in the figures. 

\section{Study of the systematics of substructure measures}
\label{Section4}

Observations, in particular those with low photon statistics, suffer from shot noise which will produce artificial structure and lead to inaccurate results in the substructure analysis. It is therefore important to characterize this bias (difference between real and spuriously detected amount of structure). Power ratios are applied to clusters since 1995 and several studies regarding the influence of photon noise on the measured power ratios and center shifts were performed \citep[e.g.][B10]{Jeltema2005,Hart2008}. In this paper we extend the work of B10 who introduced two methods (azimuthal redistribution and repoissonization) to estimate the bias and the uncertainties. However it was left open which approach yields better results in which signal-to-noise range. Using the repoissonization algorithm of B10, we make a comprehensive investigation of the performance of the bias and uncertainty estimates for a wide range of observational parameters and derive recipes on how to best correct the bias. 

\subsection{Study of shot noise bias and uncertainties}
\label{characterizationofbias}

Let's consider an idealized, radially symmetric cluster. Such an object should yield substructure parameters (power ratios and $w$) equal to zero. Once noise is added, the parameters of the same cluster increase significantly. We therefore denote the difference between the power ratio signal of the ideal image of a cluster (P$_{\mathrm{ideal}}$) and the signal of the same cluster with noise as true bias. For the simulations and if not stated otherwise, we give the bias as the true bias in \% of the ideal value:\newline
\begin{equation}
 \frac{\mathrm{P}-\mathrm{P}_{\mathrm{ideal}}}{\mathrm{P}_{\mathrm{ideal}}}\times 100 = \mathrm{B_{P}}
\label{EqbiasP3}
\end{equation}

For center shifts, the bias (\b_w) is defined analogously. In this and all following sections, we focus our analysis on the power ratio \P3P0, which is more sensitive to shot noise than the center shift parameter $w$.\newline

Shot noise makes very symmetric clusters appear more structured (positive bias). On the other hand, it can smooth out structure and a very structured cluster may actually seem more relaxed (negative bias). How the amount of shot noise and thus the reliability of the identification of substructure depend on the photon statistics of an observation and the measured substructure value is investigated using our set of 121 simulated clusters with different morphologies. In order to perform a realistic study, we create four images with different total count numbers (1\,000, 2\,000, 30\,000 and 170\,000 counts within \r500) for each simulated cluster. These four different count levels were chosen to sample a range of XMM-\emph{Newton} cluster observations, e.g. $1000-2000$ counts are typical for high redshift systems, while the values for the REXCESS sample for example range between 30\,000 and 170\,000 counts. \newline

\begin{figure}[h]
\begin{center}
    \includegraphics[width=\columnwidth]{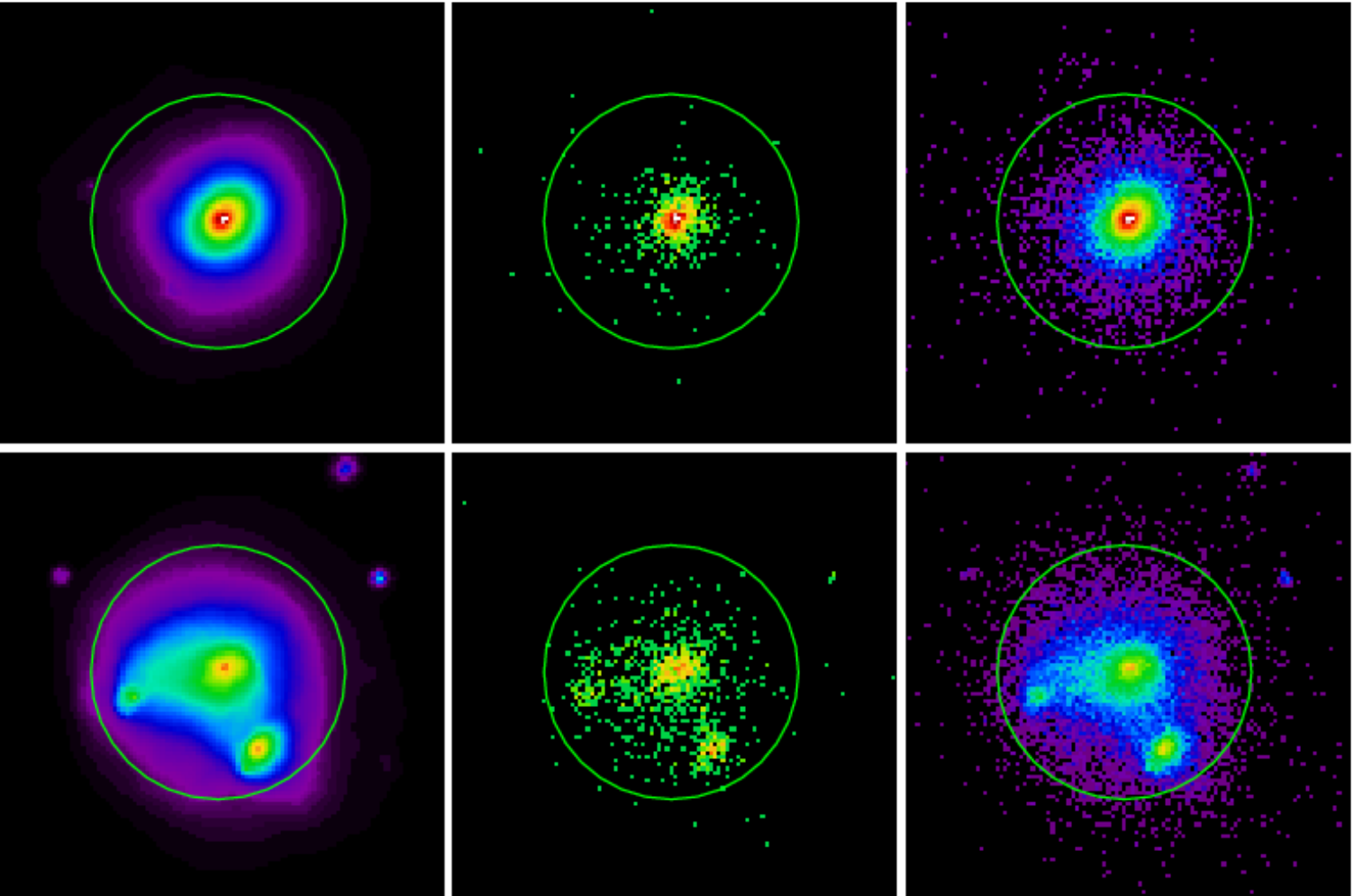}
\caption{Example of a relaxed (upper panels) and a disturbed (lower panels) simulated cluster X-ray image including no noise (left) and poissonized images with 1\,000 (middle) and 30\,000 counts (right) within \r500 (indicated by circle).}
\label{HANSSIMS}
\end{center}
\end{figure}

\begin{figure*}[!h]
\begin{center}
 \includegraphics[width=\columnwidth]{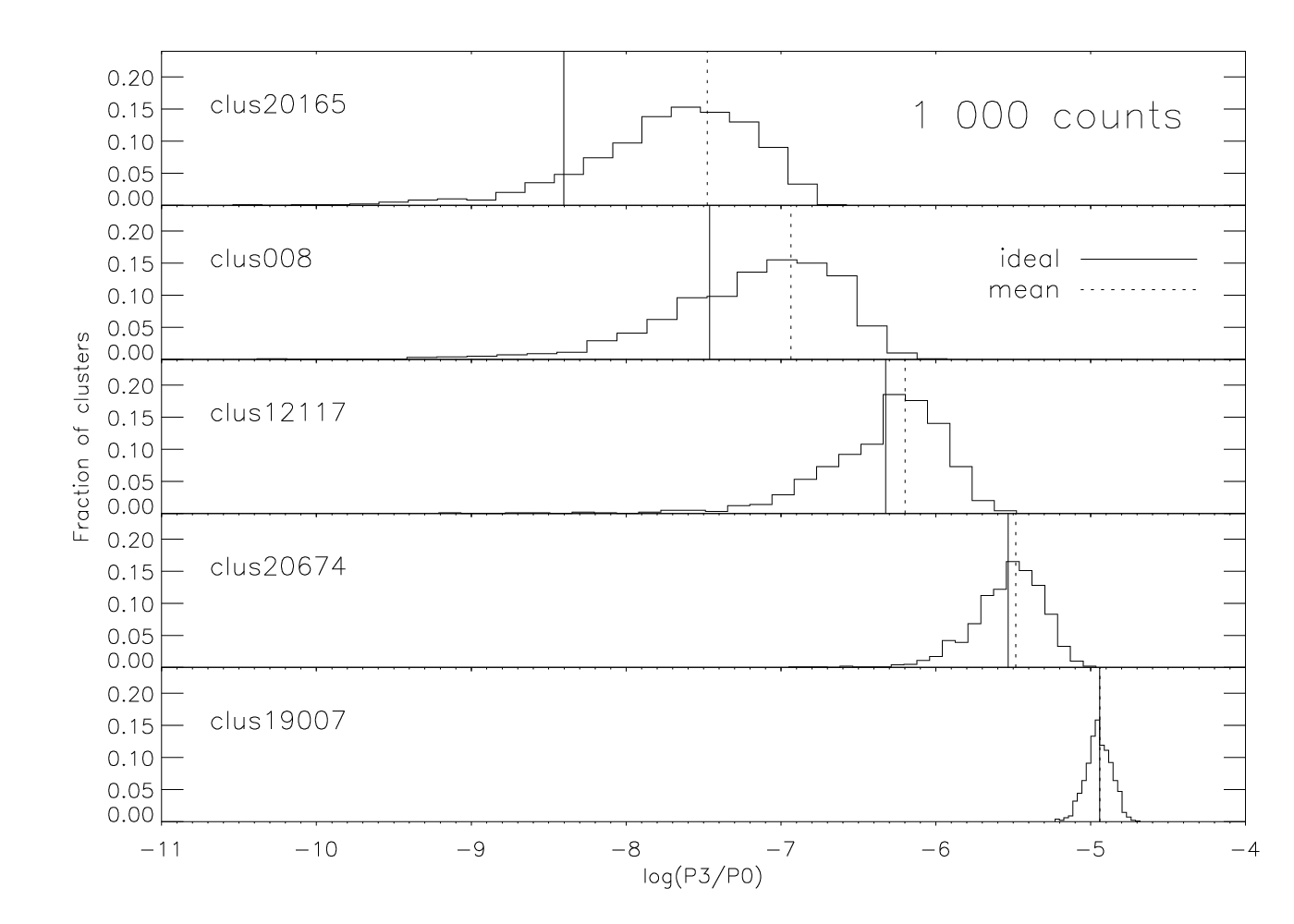}
 \includegraphics[width=\columnwidth]{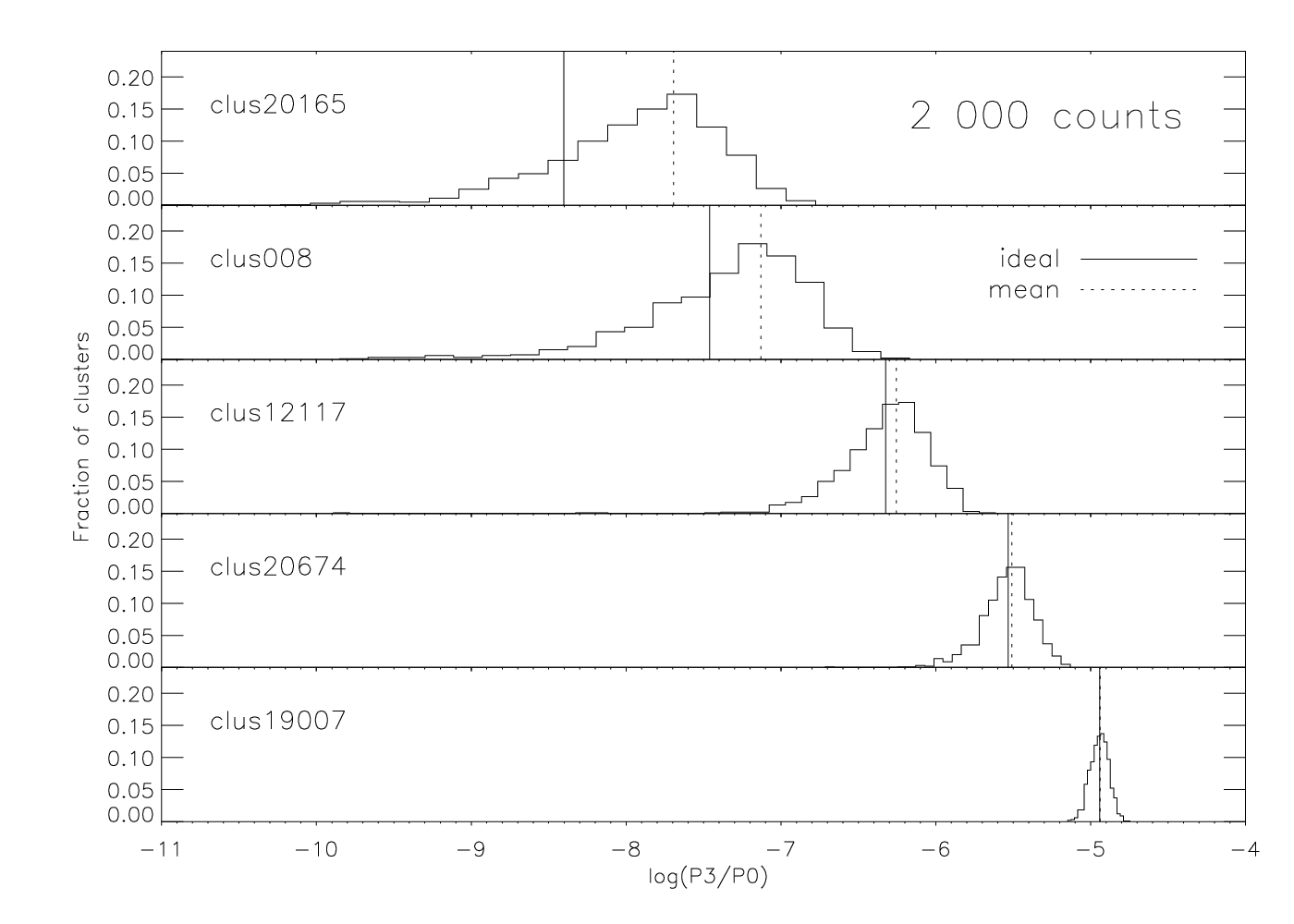}
 \includegraphics[width=\columnwidth]{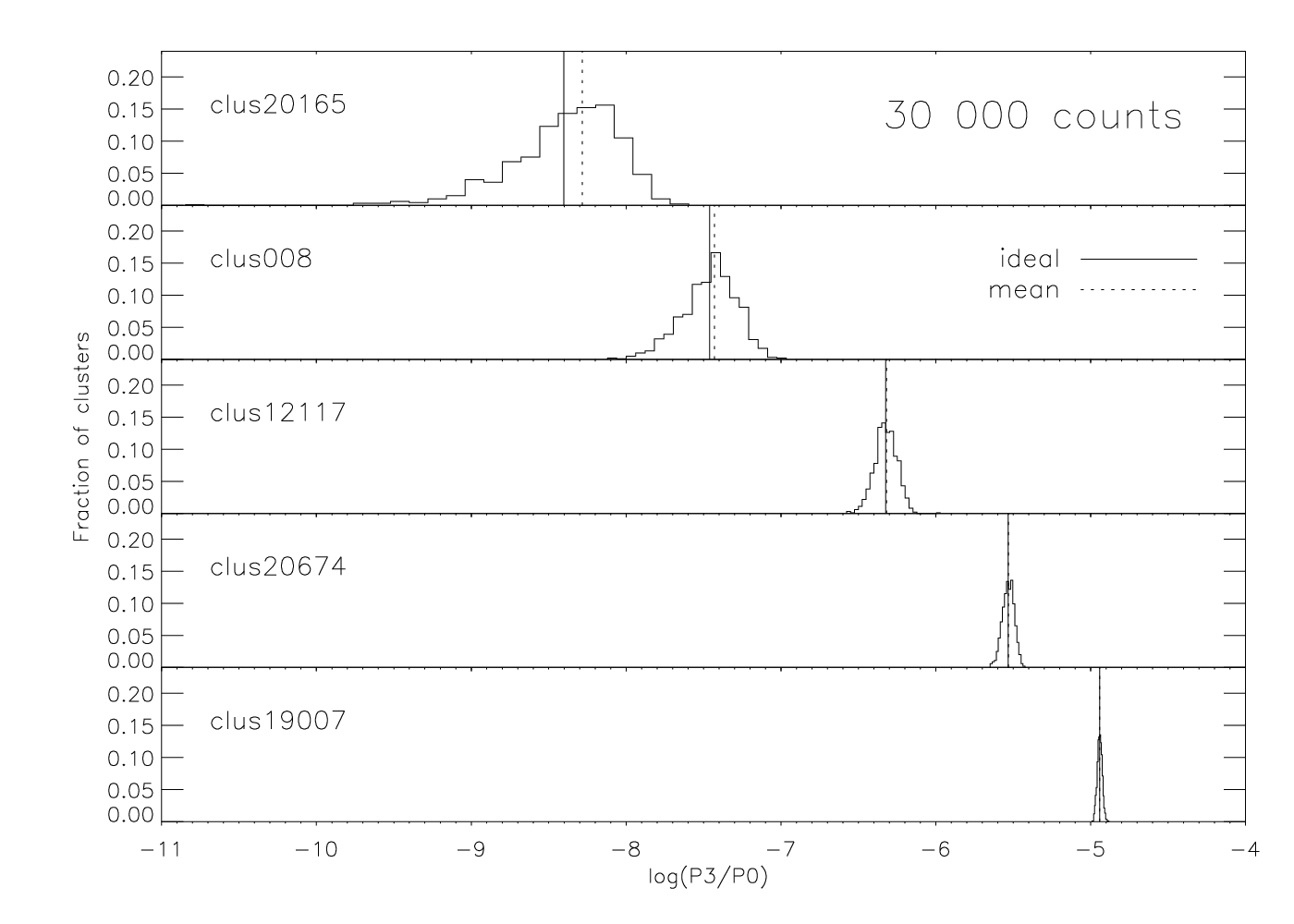}
 \includegraphics[width=\columnwidth]{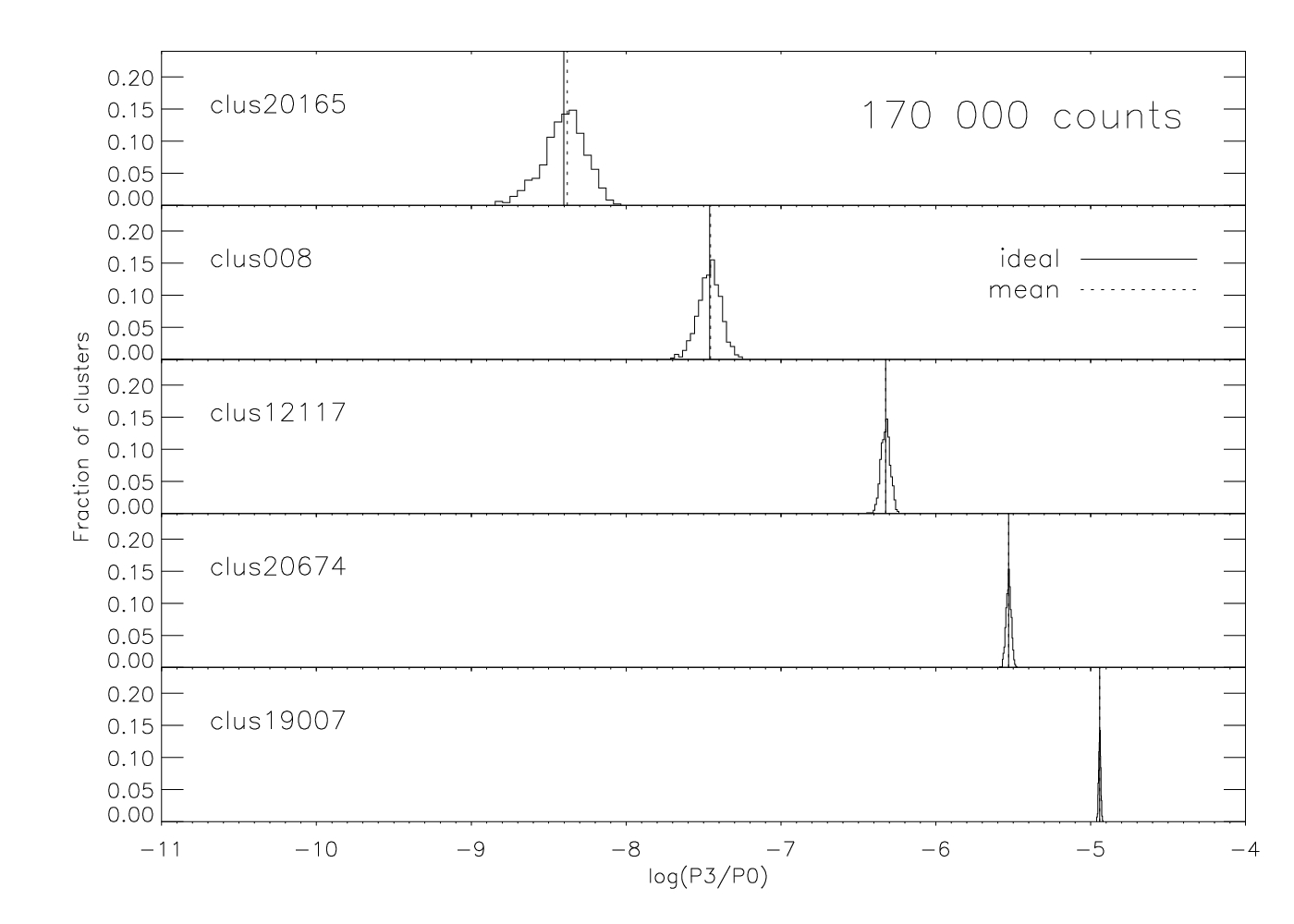}
\caption{\P3P0 distribution (reflecting the bias) for different structured clusters and counts. The solid line marks the ideal \P3P0 value, the dotted line indicates the mean of 1\,000 realizations with noise. Details are given in Table \ref{biastable100}. A comparison between this figure and Table \ref{biastable100} shows that 100 realizations are sufficient to estimate the bias.}
\label{P3depcts}
\end{center}
\end{figure*}

First, we take the simulated cluster image and normalize the surface brightness in such a way that the counts equal the chosen total count number. At this point, the pixel content is still a real number. In a second step, we poissonize the ideal cluster image (introducing shot noise) using the \textit{zhtools}\footnote{hea-www.harvard.edu/RD/zhtools} task \textit{poisson}. We call such images poissonized images or realizations, with integers as pixel content. \newline
As is apparent from the visual inspection of two simulated clusters in Fig. \ref{HANSSIMS}, the effect of photon noise is severe at low counts (middle), but also high counts images (right) are affected. It is therefore important to estimate and correct the bias as accurately as possible. The influence of shot noise and the uncertainties can be explained using Fig. \ref{P3depcts}, which provides a summary of our study. In the 4 subpanels we show the behavior of \P3P0 for different total count numbers (top left: 1\,000, top right: 2\,000, bottom left: 30\,000 and bottom right: 170\,000 counts) and several dynamical states (5 simulated cluster observations). The solid line indicates \id_P3, the power ratio of the ideal image without shot noise. The mean \P3P0 of 1\,000 poissonizations of the ideal cluster image is shown by the dotted line. In addition this figure shows the uncertainty ($\sigma$) of the mean \P3P0 as the width of the \P3P0 distribution.\newline

We find that the bias introduced to the \P3P0 results behaves differently for different morphologies (subpanels in all 4 figure panels). The upper panel shows the case of clus20165 - a cluster with little intrinsic structure. Photon noise boosts the power ratio signal and the whole \P3P0 distribution is shifted to higher substructure values. This is reflected by the obtained mean signal (dotted line), which is significantly larger than the real signal (solid line). This effect is strong, especially below 30\,000 counts. In addition, the uncertainty (width of the distribution) is large. Going step by step to more disturbed clusters (from top to bottom panel) shows the dependence of the bias on the degree of disturbance and the total count number. While clus008 still shows a large bias up to 2\,000 counts, it is already very small for 30\,000 counts. More disturbed clusters therefore are not as affected by photon noise as relaxed objects. This is apparent when looking at the \P3P0 distribution in the bottom panels (clus20674 and clus19007). Even at 1\,000 counts the bias is very small and the \P3P0 distribution narrow, which reflects a mean \P3P0 signal with a relatively small bias and uncertainty. The statistical summary of these results is given in Table \ref{biastable100}, where we list the ideal and mean structure parameter of poissonized images along with the bias in percent and the uncertainties in real (top) and log space (bottom). For all values in the table we have repeated the poissonization process using just 100 instead of 1\,000 realizations and found this lower number to be sufficient to obtain accurate statistical results. We thus work in the following studies with 100 poissonizations per case. In addition, we studied the influence of Poisson noise on the individual powers - P0 and P3. The flux P0 is only marginally sensitive to Poisson noise and does not contribute to the bias of P3/P0. The bias of P3/P0 thus reflects the influence of Poisson noise on P3.\newline

The dependence on the counts is due to the increasing effect of photon noise when dealing with low photon statistics. For relaxed clusters this leads to a very large bias and uncertainties, especially for low counts. In the case of very structured clusters with e.g. two components, the bias is negligible and the uncertainties small. For clusters with only a moderate amount of structure, we find a clear dependence on the counts. Therefore, one should be careful when applying this method to low counts observations (significantly less than 30\,000 counts). \newline

\begin{table*}
\begin{center}
\caption{Statistical results on \P3P0 and $w$ for poissonized simulated cluster images. We give the ideal substructure values and the mean for 100 realizations including their 1-$\sigma$ uncertainties in real (top) and log space (bottom). The bias (\B_P3 and \b_w) is listed in \% of the ideal value, as defined in Eq. \ref{EqbiasP3}. The results are given for 4 different total count numbers. This table corresponds to Fig. \ref{P3depcts}, but with less realizations.}
\scalebox{0.77}{
 \begin{tabular}{lccccccccc}
\hline\hline\\[-1.2ex]
\multirow{2}{*}{\P3P0} &\multirow{2}{*}{\id_P3} & \multicolumn{2}{c}{1\,000 cts} & \multicolumn{2}{c}{2\,000 cts}& \multicolumn{2}{c}{30\,000 cts}  & \multicolumn{2}{c}{170\,000 cts}\\[0.5ex]
& & \multicolumn{1}{c}{mean \P3P0}  &\multicolumn{1}{c}{\B_P3}& \multicolumn{1}{c}{mean \P3P0}   &\multicolumn{1}{c}{\B_P3}& \multicolumn{1}{c}{mean \P3P0} & \multicolumn{1}{c}{\B_P3}& \multicolumn{1}{c}{mean \P3P0}   & \multicolumn{1}{c}{\B_P3} \\[0.5ex]
\hline\\[-1.2ex]
clus20165 & $4.0\times10^{-9}$ & $3.7\times10^{-8} \pm 3.5\times10^{-8}$ & 839 & $1.9\times10^{-8} \pm 2.0\times10^{-8}$ & 376 & $4.9\times10^{-9} \pm 3.1\times10^{-9}$&  24 & $4.1\times10^{-9} \pm 1.4\times10^{-9}$&  3 \\
clus008   & $3.5\times10^{-8}$ & $8.7\times10^{-8} \pm 8.4\times10^{-8}$ & 152 & $7.2\times10^{-8} \pm 6.7\times10^{-8}$ & 109 & $3.6\times10^{-8} \pm 1.3\times10^{-8}$&  5  & $3.4\times10^{-8} \pm 5.7\times10^{-9}$& -1 \\
clus12117 & $4.7\times10^{-7}$ & $6.1\times10^{-7} \pm 5.0\times10^{-7}$ & 28  & $5.4\times10^{-7} \pm 2.7\times10^{-7}$ & 14  & $4.8\times10^{-7} \pm 7.9\times10^{-8}$&  2  & $4.7\times10^{-7} \pm 3.1\times10^{-8}$& -0.3 \\
clus20674 & $3.0\times10^{-6}$ & $3.2\times10^{-6} \pm 1.4\times10^{-6}$ & 8   & $3.2\times10^{-6} \pm 1.1\times10^{-6}$ & 9   & $2.9\times10^{-6} \pm 2.6\times10^{-7}$& -0.4& $3.0\times10^{-6} \pm 1.0\times10^{-7}$& -0.1 \\
clus19007 & $1.1\times10^{-5}$ & $1.1\times10^{-5} \pm 1.9\times10^{-6}$ & -4  & $1.1\times10^{-5} \pm 1.5\times10^{-6}$ & 0.1 & $1.1\times10^{-5} \pm 3.9\times10^{-7}$&  0.4& $1.1\times10^{-5} \pm 1.6\times10^{-7}$& 0.05 \\
\hline\\[-1.2ex]
\multirow{2}{*}{$w$} &\multirow{2}{*}{\i_w} & \multicolumn{2}{c}{1\,000 cts} & \multicolumn{2}{c}{2\,000 cts}& \multicolumn{2}{c}{30\,000 cts}  & \multicolumn{2}{c}{170\,000 cts}\\[0.5ex]
& & \multicolumn{1}{c}{mean $w$}& \multicolumn{1}{c}{\b_w}& \multicolumn{1}{c}{mean $w$} &  \multicolumn{1}{c}{\b_w}& \multicolumn{1}{c}{mean $w$}& \multicolumn{1}{c}{\b_w}& \multicolumn{1}{c}{mean $w$} & \multicolumn{1}{c}{\b_w} \\[0.5ex]
\hline\\[-1.2ex]
clus20165 & 0.0019 & 0.0029 $\pm$ 0.0012 &  55 & 0.0025 $\pm$ 0.0010 &  35 & 0.0020  $\pm$ 0.0003 & 6.2 & 0.0019 $\pm$ 0.0001 & -0.13 \\
clus008   & 0.0087 & 0.0090 $\pm$ 0.0030 &  4  & 0.0094 $\pm$ 0.0024 &  8  & 0.0088  $\pm$ 0.0007 & 1.4 & 0.0087 $\pm$ 0.0003 &  0.3 \\
clus12117 & 0.0156 & 0.0163 $\pm$ 0.0036 &  5  & 0.0153 $\pm$ 0.0025 &  -2 & 0.0157  $\pm$ 0.0008 & 0.6 & 0.0156 $\pm$ 0.0003 &  0.3 \\
clus19007 & 0.0700 & 0.0662 $\pm$ 0.0061 &  -5 & 0.0679 $\pm$ 0.0046 &  -3 & 0.0662  $\pm$ 0.0061 & 0.1 & 0.0700 $\pm$ 0.0004 &  -0.01 \\
clus20674 & 0.1193 & 0.1194 $\pm$ 0.0099 &0.03 & 0.1193 $\pm$ 0.0076 &-0.05& 0.1193  $\pm$ 0.0018 &-0.07& 0.1193 $\pm$ 0.0008 & 0.005 \\[0.5ex]
\hline\hline\\[-1.2ex]
\multirow{2}{*}{\P3P0} &\multirow{2}{*}{log(\id_P3)} & \multicolumn{2}{c}{1\,000 cts} & \multicolumn{2}{c}{2\,000 cts}& \multicolumn{2}{c}{30\,000 cts}  & \multicolumn{2}{c}{170\,000 cts}\\[0.5ex]
& & \multicolumn{1}{c}{log(mean \P3P0)}  &\multicolumn{1}{c}{\B_P3}& \multicolumn{1}{c}{log(mean \P3P0)}   &\multicolumn{1}{c}{\B_P3}& \multicolumn{1}{c}{log(mean \P3P0)} & \multicolumn{1}{c}{\B_P3}& \multicolumn{1}{c}{log(mean \P3P0)}   & \multicolumn{1}{c}{\B_P3} \\[0.5ex]
\hline\\[-1.2ex]
clus20165 & -8.40 & $-7.66 \pm 6.9$ & 839  & $-7.99 \pm 8.5$ & 376 & $-8.42 \pm 5.3$ &  24  & $ -8.41 \pm 2.8$& 3 \\
clus008   & -7.46 & $-7.26 \pm 6.8$ & 152  & $-7.32 \pm 6.6$ & 109 & $-7.47 \pm 2.7$ &  5   & $ -7.47 \pm 1.2$& -1 \\
clus12117 & -6.32 & $-6.38 \pm 5.2$ & 28   & $-6.34 \pm 3.1$ & 14  & $-6.32 \pm 1.0$ &  2   & $ -6.33 \pm 0.4$& -0.3 \\
clus20674 & -5.53 & $-5.54 \pm 2.5$ & 8    & $-5.52 \pm 1.9$ & 9   & $-5.54 \pm 0.5$ & -0.4 & $ -5.53 \pm 0.2$& -0.1 \\
clus19007 & -4.94 & $-4.97 \pm 0.9$ & -4   & $-4.95 \pm 0.6$ & 0.  & $-4.94 \pm 0.17$&  0.4 & $ -4.94 \pm 0.1$& 0.05 \\
\hline\\[-1.2ex]
\multirow{2}{*}{$w$} &\multirow{2}{*}{log(\i_w)} & \multicolumn{2}{c}{1\,000 cts} & \multicolumn{2}{c}{2\,000 cts}& \multicolumn{2}{c}{30\,000 cts}  & \multicolumn{2}{c}{170\,000 cts}\\[0.5ex]
& & \multicolumn{1}{c}{log(mean $w$)}& \multicolumn{1}{c}{\b_w}& \multicolumn{1}{c}{log(mean $w$)} &  \multicolumn{1}{c}{\b_w}& \multicolumn{1}{c}{log(mean $w$)}& \multicolumn{1}{c}{\b_w}& \multicolumn{1}{c}{log(mean $w$)} & \multicolumn{1}{c}{\b_w} \\[0.5ex]
\hline\\[-1.2ex]
clus20165 & -2.721 & -2.539 $\pm$ 1.012 &  55 & -2.598 $\pm$ 1.081 &  35 & -2.702 $\pm$ 0.463 & 6.2 & -2.729 $\pm$ 0.211 & -0.13 \\
clus008   & -2.061 & -2.045 $\pm$ 0.681 &  4  & -2.028 $\pm$ 0.520 &  8  & -2.054 $\pm$ 0.162 & 1.4 & -2.059 $\pm$ 0.067 &  0.3 \\
clus12117 & -1.807 & -1.787 $\pm$ 0.391 &  5  & -1.815 $\pm$ 0.297 &  -2 & -1.805 $\pm$ 0.090 & 0.6 & -1.806 $\pm$ 0.038 &  0.3 \\
clus19007 & -1.155 & -1.179 $\pm$ 0.108 & -5  & -1.168 $\pm$ 0.079 &  -3 & -1.179 $\pm$ 0.108 & 0.1 & -1.155 $\pm$ 0.006 &  -0.01 \\
clus20674 & -0.923 & -0.923 $\pm$ 0.076 & 0.03& -0.923 $\pm$ 0.059 &-0.05& -0.924 $\pm$ 0.014 &-0.07& -0.923 $\pm$ 0.006 & 0.005 \\[0.5ex]
\hline
\label{biastable100}
\end{tabular}
}
\end{center}
\end{table*}

\begin{figure*}[!h]
\begin{center}
\begin{minipage}[h]{\columnwidth}
 \includegraphics[width=\columnwidth]{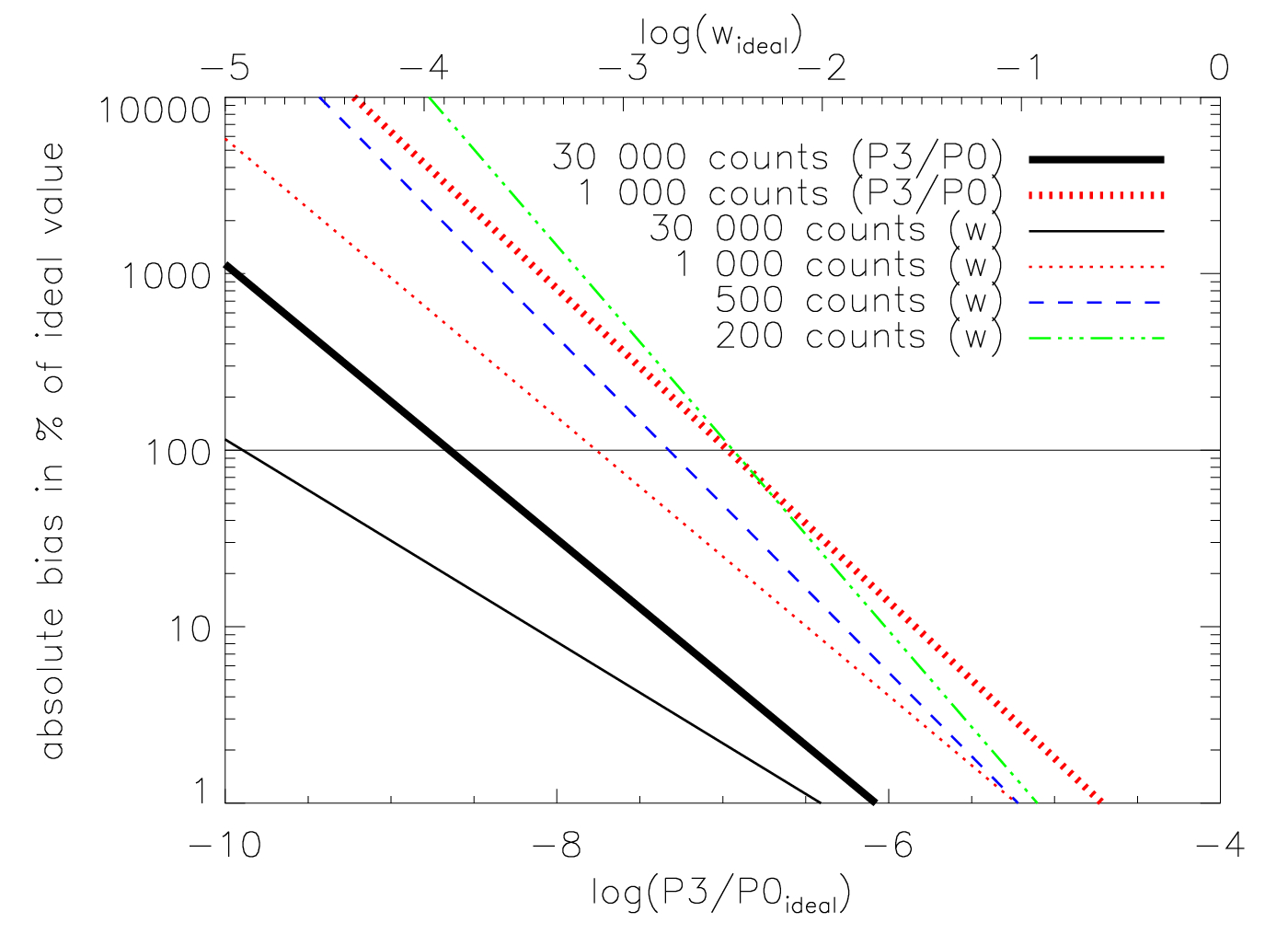}
\end{minipage}
\begin{minipage}[h]{\columnwidth}
 \includegraphics[width=\columnwidth]{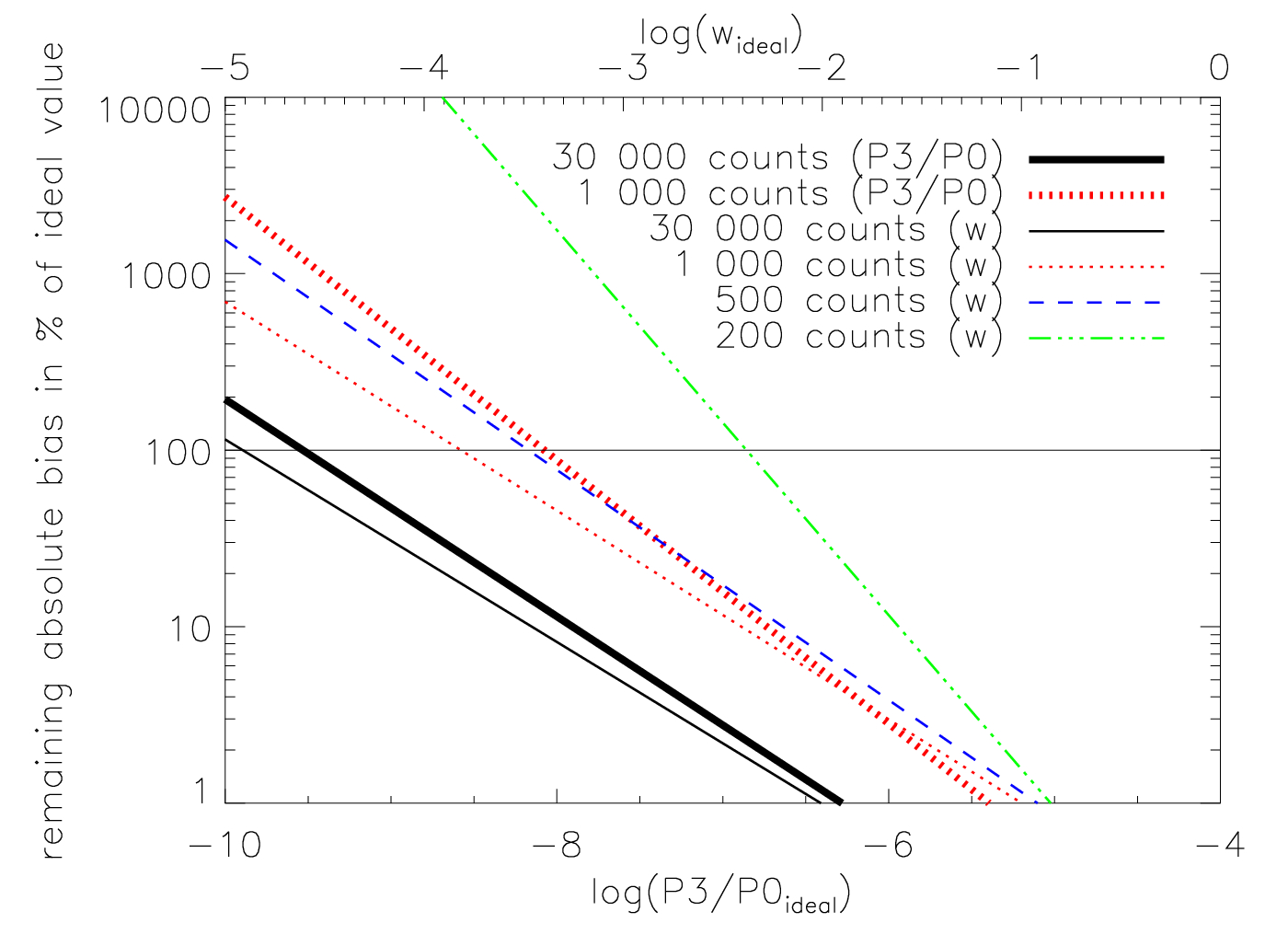}
\end{minipage}
\caption{Dependence of the bias as a function of \id_P3 (lower axis) and \i_w (upper axis). Left: Absolute value of the bias before correcting. Right: Absolute value of the remaining bias after applying the bias correction $\mathrm{B_{P3}^*}$ and $\mathrm{B_{w}^*}$. The different counts are color-coded: 30\,000 black, 1\,000 red, 500 blue, 200 green. \B_P3 is shown using thick lines, \b_w is represented by different thin lines. The dependencies are fits to all 121 simulated cluster images using the BCES linear regression method \citep{Akritas1996}.}
\label{Figbias}
\end{center}
\end{figure*}

For the center shift parameter $w$ the behavior is similar, but less pronounced. \b_w is more robust and in general significantly smaller than \B_P3. In addition, the distributions are narrower, which shows that $w$ is less sensitive to photon noise than \P3P0. This allows an accurate calculation of the center shift parameter down to $\sim$200 counts. An overview of the absolute value of the bias as a function of counts (different colored lines) and the ideal value is given in Fig. \ref{Figbias}, left. We combined the bias of the substructure parameters (\B_P3 thick black solid and red dotted line, \b_w different thin lines) as a function of \id_P3 (lower x-axis) and \i_w (upper x-axis) for a direct comparison. However, while the simulated clusters occupy the full \P3P0 range, they only have $w$ parameters between 3.1$\times10^{-4}$ and 2.4$\times10^{-1}$. This and all other fits which will be displayed later are obtained using the orthogonal BCES linear regression method \citep{Akritas1996}. \newline

The dependence of the bias on photon statistics and the substructure measurement encourages a bias correction as a function of these parameters. However, the bias depends also on the morphology of the cluster itself. We thus performed the following test: if we consider two clusters with the same \P3P0 or center shift value, they have nominally the same amount of structure. If this cluster pair also has the same amount of counts, only the intrinsic shape of the cluster remains variable. We chose six pairs of clusters with the same \i_w and \id_P3 value and four different counts: 1\,000, 2\,000, 30\,000 and 170\,000 counts. For a dependence of the bias on the amount of structure and counts only, one would expect very similar distributions and mean values. However, this is not the case. Especially for the unstructured cluster pair and low counts the offset and the behavior of \B_P3 is significant. For high counts or structured clusters, this offset decreases. We thus cannot give a general correction factor as a function of counts and \P3P0 or $w$ but have to treat the estimate of the bias correction for each cluster individually.

\subsection{Significance threshold}

As we have shown, shot noise can introduce spurious structure. While our bias correction alleviates this to some extent it is useful to relate the measured (and corrected) signal to its error. In order to do so, we define a significance $S$ as

\begin{equation}
 S=\frac{bias\ corrected\  signal}{error},
\end{equation}
and call values with $S\ge3$ significant signals. This value however strongly depends on the photon statistics. \newline

\begin{figure}[h]
\centering
\begin{minipage}[h]{\columnwidth}
    \includegraphics[width=\columnwidth]{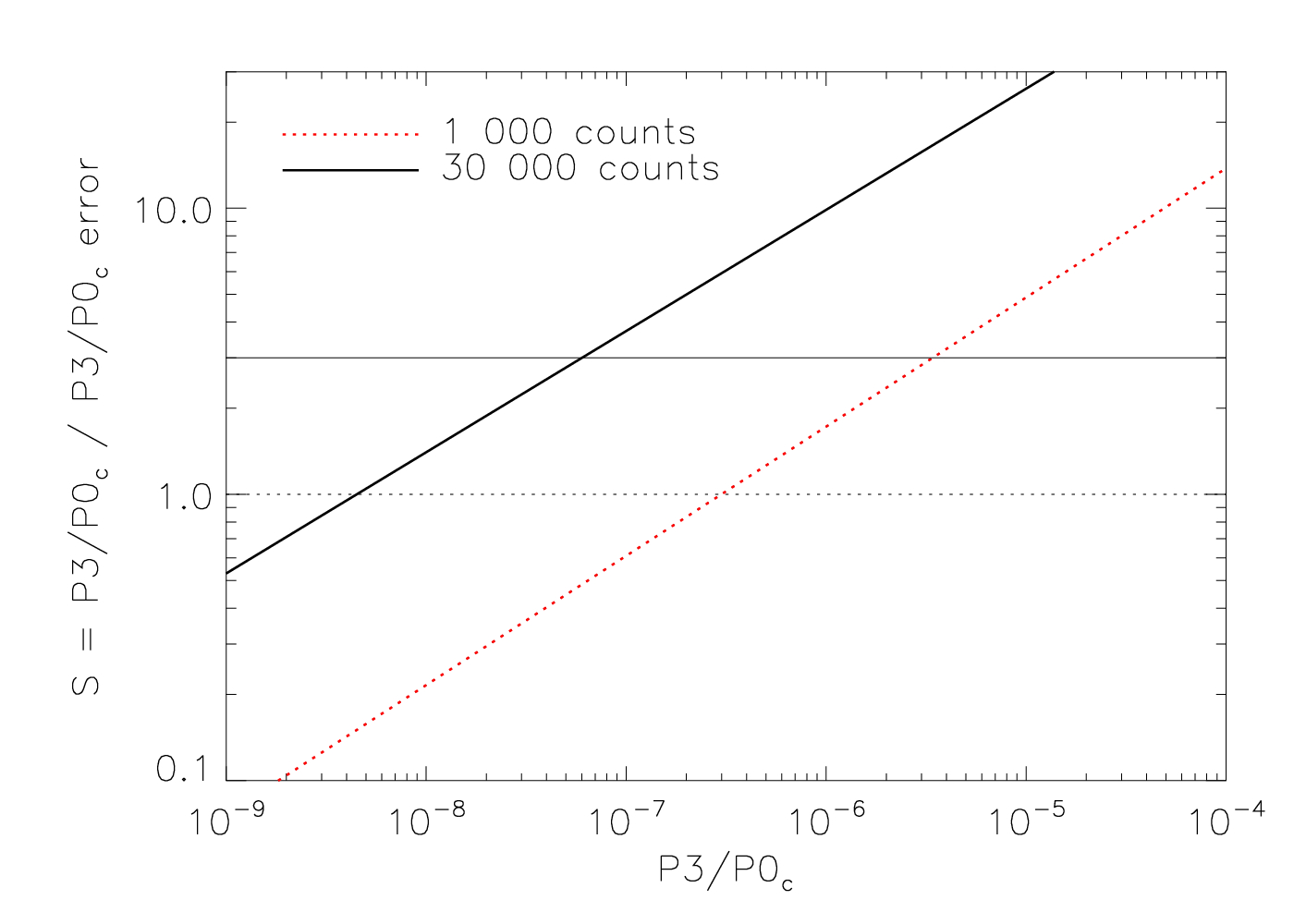}
\end{minipage}
\begin{minipage}[h]{\columnwidth}
 \includegraphics[width=\columnwidth]{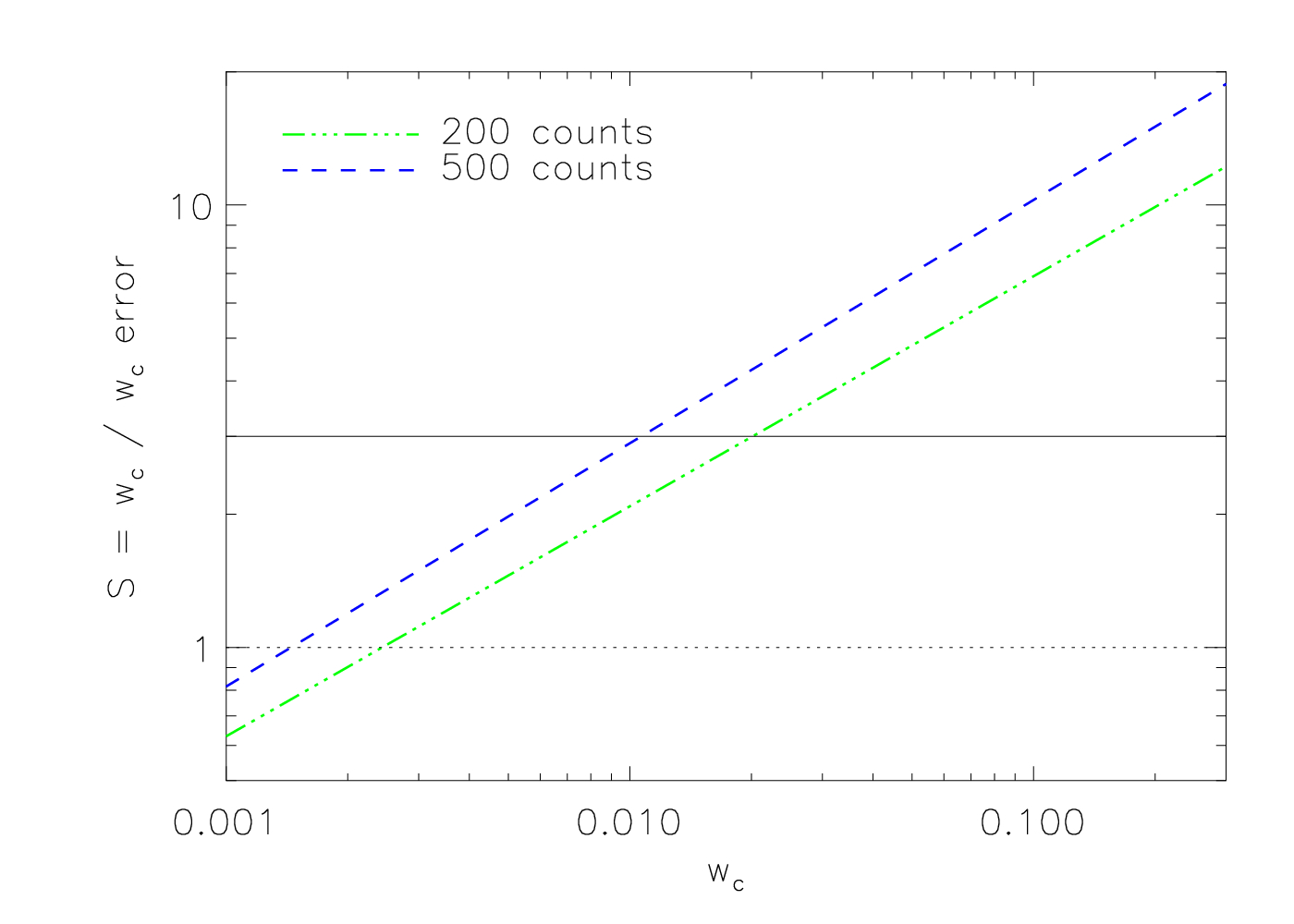}
\end{minipage}
\caption{Significance $S$ of the \P3P0 (top) and $w$ (bottom) measurements for different counts. Top: 1\,000 red dotted line and 30\,000 black solid line, bottom: 200 green dot-dashed, 500 blue dashed line. The different thresholds are marked ($S=3$: solid line, $S=1$ dotted line).}
\label{figthreshold}
\end{figure}

We studied the significance $S$ as a function of the bias corrected substructure parameters for different total count numbers and show some results in Fig. \ref{figthreshold}. The bias correction was done using the method described in Sect. \ref{method}. Different total count numbers are color-coded and displayed using different linestyles (Top: 1\,000 red dotted and 30\,000 black solid line; bottom: 200 green dot-dashed and 500 blue dashed line) for \P3P0 (top) and  $w$ (bottom). The lines represent a BCES fit to all 121 simulated clusters. The significance thresholds ($S=3$) for both structure parameters and several total count numbers are given in Table \ref{sig} and displayed as horizontal lines in Fig. \ref{figthreshold}. \newline

We will take a closer look at \P3P0 first. For a typical observation of 30\,000 counts we are able to detect intrinsic structures corresponding to $\mathrm{\P3P0}=6\times10^{-8}$ at $S=3$ confidence level ($\mathrm{\P3P0} \ll 10^{-8}$ at $S=1$ level). This shows that the errors are small enough to ensure significant results even for clusters with little intrinsic structure. In the case of a low counts observation with only 1\,000 counts, the $S=3$ confidence level is located around $3.4\times10^{-6}$, which means that we can only obtain significant results for very structured clusters. In such cases, we use a less conservative and lower value like $S=1$. However, when dealing with such low counts observations special care has to be taken. The well-defined behavior of the center shift parameter is confirmed by the significance of the measurements. We find the $S=3$ values to be in the lower center shift range and can thus obtain significant results even for relaxed clusters. This result holds well below 1\,000 counts. For 200 and 500 counts we find $S=3$ to coincide with the median of the sample. A discussion of the implications of these results for a morphological analysis will be provided in a later section. 

\begin{table}
\begin{center}
  \caption{Dependence of the significance of the signal ($S$=signal/error) on total number counts (net counts within \r500) for \p3_c and \w_c. We call values with $S>3$ significant signals, however for low counts observations a less conservative value like $S=1$ has to be used.}
 \begin{tabular}{lccc}
\hline\hline\\[-1.2ex]
\multirow{1}{*}{\p3_c} &\multicolumn{1}{c}{Total count number} & \multicolumn{1}{c}{$S=1$}  & \multicolumn{1}{c}{$S=3$} \\[0.5ex]
\hline\\[-1.2ex]
&1\,000&  $3.0\times10^{-7}$ & $3.4\times10^{-6}$ \\
&2\,000&  $1.4\times10^{-7}$ & $1.6\times10^{-6}$ \\
&30\,000& $4.5\times10^{-9}$ & $6.0\times10^{-8}$ \\[0.5 ex]
\hline\\[-1.2ex]
\multirow{1}{*}{\w_c} &\multicolumn{1}{c}{Total count number} & \multicolumn{1}{c}{$S=1$}  & \multicolumn{1}{c}{$S=3$} \\[0.5ex]
\hline\\[-1.2ex]
&200    & $2.4\times10^{-3}$ & $2.0\times10^{-2}$ \\ 
&500    & $1.4\times10^{-3}$ & $1.0\times10^{-2}$ \\ 
&1\,000 & $9.0\times10^{-4}$ & $6.0\times10^{-4}$ \\
&30\,000& $1.6\times10^{-4}$ & $8.0\times10^{-4}$ \\[0.5ex]
\hline
\label{sig}
\end{tabular}
\end{center}
\end{table}

\subsection{Bias correction method}
\label{method}

After characterizing the bias and its dependence on the photon statistics, we propose a statistical method to estimate and correct for the true bias B$_\mathrm{P3}$ and B$_\mathrm{w}$ (for P3/P0 and w). In Sect. 4.1 we defined the true bias as the difference between the true signal P3/P0$_\mathrm{ideal}$ or $w_\mathrm{ideal}$ and P3/P0$_\mathrm{raw}$ or $w_\mathrm{raw}$, the signal obtained after the first poissonization or the signal of the observation. Simulated images do not contain noise and give the true structure parameters. Observations however are poissonized, where this first poissonization is due to photon shot noise. They allow us to measure only P3/P0$_\mathrm{raw}$ or $w_\mathrm{raw}$ but not the true signal P3/P0$_\mathrm{ideal}$ and $w_\mathrm{ideal}$. We therefore cannot obtain the true bias directly but need to estimate it.\newline
We assume that a second poissonization step returns roughly the same bias and error as the first poissonization. Analogously to the true bias we therefore define the "estimated bias" as the difference between the signal after the first (P3/P0$_\mathrm{raw}$ or $w_\mathrm{raw}$) and second poissonization (P3/P0$_\mathrm{realization}$ or $w_\mathrm{realization}$). For simulated images, we mimiced the effect of the first poissonization by adding artifical Poisson noise creating observation-like images. The second poissonization is performed on the observation/observation-like image to create a repoissonized image (realization of the observation). Using the mean P3/P0 or $w$ value of 100 realizations of the observation in combination with P3/P0$_\mathrm{raw}$ or $w_\mathrm{raw}$ to calculate the estimated bias $\mathrm{B_{P3}^*}$ or $\mathrm{B_{w}^*}$ yields a good approximation for the true bias. Subtracting $\mathrm{B_{P3}^*}$ or $\mathrm{B_{w}^*}$ from the substructure parameters of the cluster image returns the corrected substructure parameters $\mathrm{P3/P0_{c}}$ and $w_\mathrm{c}$. The remaining bias after this correction approaches zero for high-quality observations and is defined as \C_B_P3 and \c_b_w, respectively.\newline

Considering this, we present a refined version of the B10 method including the following steps:
\begin{enumerate}
\item Calculate the substructure parameters (\P3P0 and $w$) of the cluster image: P3/P0$_\mathrm{raw}$ and $w_\mathrm{raw}$.
\item Create 100 poissonized realizations of the cluster image.
\item Calculate the substructure parameters (\P3P0 and $w$) of all 100 realizations and their mean: $\langle \mathrm{P3/P0}_\mathrm{realizations} \rangle$ and $\langle w_\mathrm{realizations} \rangle$.
\item Obtain the estimated bias $\mathrm{B_{P3}^*}$ and $\mathrm{B_{w}^*}$ as the difference of the mean parameters of these 100 realizations and P3/P0$_\mathrm{raw}$ and $w_\mathrm{raw}$:\\
$\mathrm{B_{P3}^*}=\langle \mathrm{P3/P0_{realizations}} \rangle - \mathrm{{P3/P0}_{raw}}$ and \\
$\mathrm{B_{w}^*}=\langle w_\mathrm{realizations} \rangle - w_\mathrm{raw}$
\item Subtracting the estimated bias from the substructure parameters of the cluster image yields the corrected parameters: \\
$\mathrm{P3/P0_{c}}= \mathrm{P3/P0_{raw}} - \mathrm{{B}_{P3}^*}$ and \\
$w_\mathrm{c}= w_\mathrm{raw} - \mathrm{B}_{w}^*$
\item Obtain the uncertainty as the standard deviation $\sigma$ of the structure parameters of the 100 realizations of the cluster image.
\end{enumerate}

In case of a real observation, also the background needs to be considered (see Sect. \ref{bkg}). After testing several methods, the generally best performing method for power ratios is to subtract the moments $a_{m}$ and $b_{m}$ (where $m$=1,2,3,4; see Eq. 2 and 3) of the background image from the measured moments of the cluster image and its realizations before calculating the powers \citep{Jeltema2005}. Since $w$ is not additive as $a_{m}$ and $b_{m}$, we have taken a different approach in the case of the center shift method and subtract the background prior to the calculation of $w_\mathrm{raw}$ and $w_\mathrm{realizations}$. This rather simple method works very well in a statistical way, as is shown below.\newline

In some cases, we do not gain any information about the cluster because the estimated bias is larger than the true bias. We then obtain a negative \p3_c with a large uncertainty which indicates that the signal is consistent with zero. For a few \% of realizations we found that the repoissonization leads to a change of the brightest pixel and thus the zero-point of the center shift calculation. This can change the values significantly, however does not influence the mean $w$ value of all 100 realizations. Also the centroid, which is calculated using the surface brightness distribution, can change for different realizations, especially when dealing with low photon statistics. This shift is included in the error estimation when recalculating the centroid for each realization. In our analysis however we found that the remaining bias of the corrected \P3P0 values (\C_B_P3) vary only slightly. The mean change in the absolute \P3P0 value for 1\,000 counts is a factor of 2, however in 19 cases the increase is larger (max. 17). All clusters with such a considerable change in the centroid are very obvious merging systems with two distinct surface brightness peaks. The error increases especially for large \P3P0 values, but still remains small compared to the \P3P0 value itself.

\subsection{Testing of the method}
\label{testp3} 
We tested and refined this method using simulated images as described in Sect. \ref{Simulations}. As with the characterization of the bias, we used different counts to simulate different depths of observations. We poissonized each simulated image 100 times and treated each of those 100 images as an "observation" which are subject to a second poissonization step. After the bias correction of all 100 "observations" using the estimated bias from the second poissonization, we obtain a mean value of the corrected parameter to show the statistical strength of this method. \newline

The results of the bias correction method are shown in Fig. \ref{Figbias} for \P3P0 and $w$. The figure on the left shows the absolute value of the bias before noise correction (discussed in Sect. \ref{characterizationofbias}), while the right side displays the remaining bias after applying the correction method. In both panels, we simultaneously show the absolute value of the bias as defined in Sect. \ref{characterizationofbias} for \P3P0 (thick lines and lower x-axis) and $w$ (thin lines and upper x-axis) for different counts. The decrease of \C_B_P3 is apparent for cases with $1\,000$ counts, where the correction method is successful down to the detection limit ($S=1$ at $3\times10^{-7}$). In the insignificant range ($S<3$) \C_B_P3 lies below 10\% after noise-correction. The solid black line shows the case for high-counts observations, where a drop below 10\% can be seen around the $S=3$ cut at $6\times10^{-8}$. \newline

The center shift parameter is more robust, even at 200 counts, where \c_b_w$\ \sim 10\%$ for $S=3$. Center shifts are less sensitive to shot noise and their bias is smaller. This is especially interesting when looking at relaxed clusters ($w<0.01$), where \c_b_w is significantly smaller than \C_B_P3. Motivated by these results at low counts, we decided to test even lower photon statistics - 500 and 200 counts. With such observations the power ratios are not reliable anymore, but the center shifts show remarkably good results. \newline

\begin{figure}[!h]
\begin{center}
 \includegraphics[width=\columnwidth]{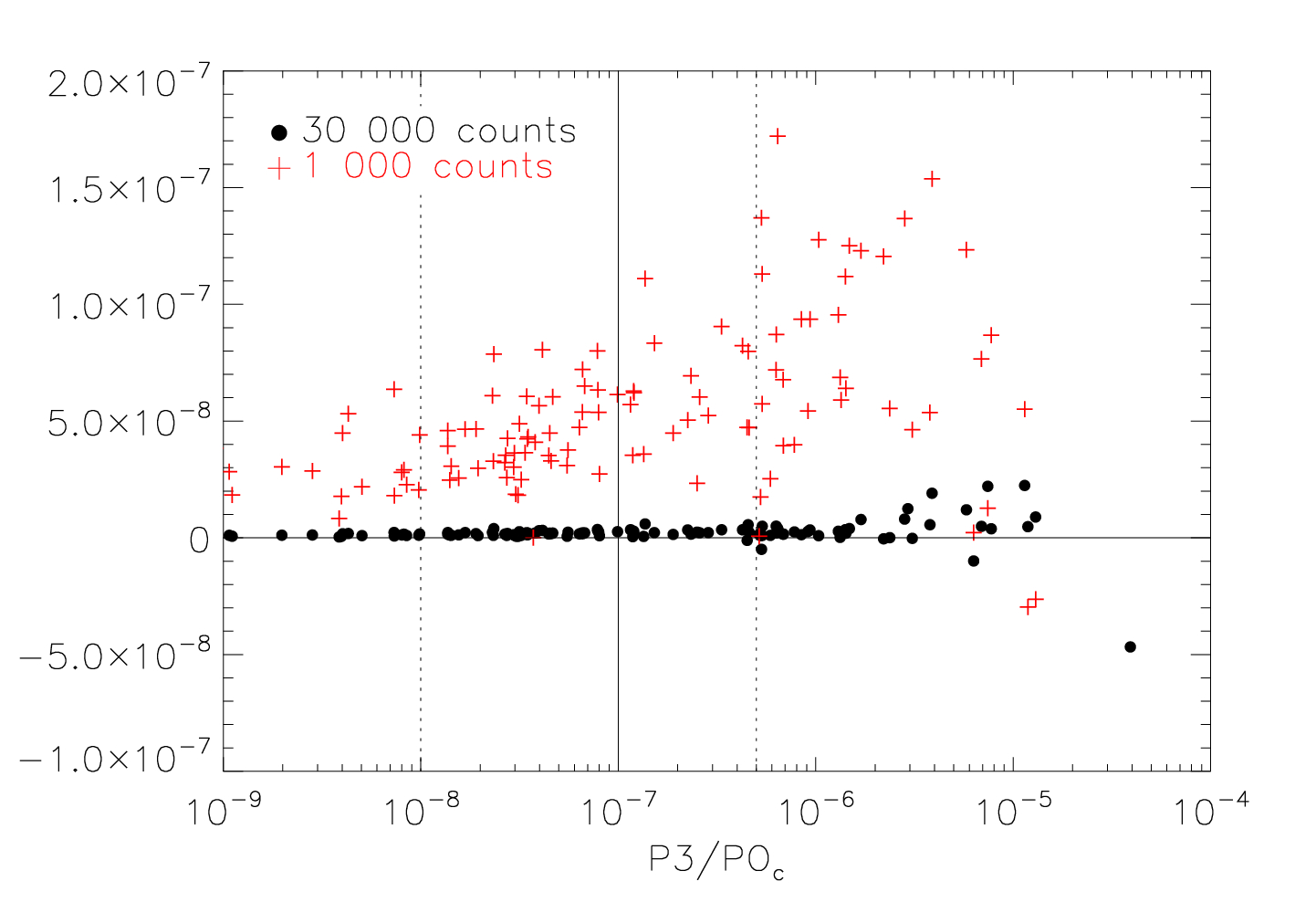}
 \caption{Illustration of the probability of a negative bias. We show the applied bias correction $\mathrm{B_{P3}^*}$ (mean of 100 realizations) as a function of \p3_c. The colors indicate the different counts within \r500: 1\,000 (red crosses) and 30\,000 (black circles). A negative bias correction is only needed for very structured clusters and even then it is only of the order of 1\%. The solid and dotted lines show the different morphological ranges as discussed below in Sect. \ref{threshold}.}
\label{negbias}
\end{center}
\end{figure}

In some of the 100 realizations of the poissonized images we find that a negative bias correction is needed, where the structure in the poissonized images has a too small value. However, the mean of the bias correction of all poissonizations is always positive, except for a few cases with very high structure parameters. For these clusters the bias correction is only around 1\% as is shown in Fig. \ref{negbias}, where we plot the applied bias correction $\mathrm{B_{P3}^*}$ (mean of 100 realizations) as a function of \p3_c.
 
\subsection{Effect of the X-ray background}
\label{bkg}

The quality of X-ray observations suffers from several components - including photon noise which was discussed in Sect. \ref{method} and the X-ray background, which was not taken into account yet. We thus investigated how the background and different cluster-to-background count (S/B) ratios influence the measurements. Motivated by the work of \citet{Jeltema2005}, in which the authors use an analytic approach to assess and correct for the background contribution for power ratios by subtracting moments due to noise, we inspected the behavior of power ratios and moments when adding or subtracting them. Power ratios are not additive, moments ($a_0$ to $b_4$) however are and thus can be used for background noise subtraction. \newline

When correcting for the background two issues have to be addressed: the increase of total counts (normalization) and the noise component of the background image. Depending on S/B, the noise in the background image can influence the power ratio and center shift calculation. In order to account for the noise in both the cluster signal and in the background, we add a poissonized cluster and a poissonized flat background image to obtain an "observation". As during the bias study, we create 100 "observations" per simulated cluster image and show mean values. The correction of the bias is done using a two-step process. In step one, the background is treated by subtracting the moments ($a_0$ to $b_4$) of the background image from the moments of the observation before calculating the power ratios (in Sect. \ref{method}, Step 1). For a flat background image without noise, only $a_0$ should be non-zero. However, vignetting and other instrumental artifacts cause also higher moments to be non-zero. The background-subtracted moments should thus (statistically) only contain the cluster emission and the signal noise component. The background moments have to be subtracted also from the 100 realizations of the observation. As a second step, the power ratios and the bias are calculated using the background-subtracted moments. We therefore recommend the following power ratio treatment of the observation: 

\begin{enumerate}
 \item Calculate moments of the observation (incl. background) and of the background model/image.
 \item Create 100 poissonizations of the observation and obtain their moments.
 \item Subtract the background moments from the moments of the observation and the 100 poissonizations.
 \item Calculate power ratios of the observation and the 100 poissonizations
 \item Correct the bias and obtain the $\sigma$ as described in Sect. \ref{method}, Step 4-6. 
\end{enumerate}

\begin{figure*}[!h]
\begin{center}
\begin{minipage}[h]{\columnwidth}
 \includegraphics[width=\columnwidth]{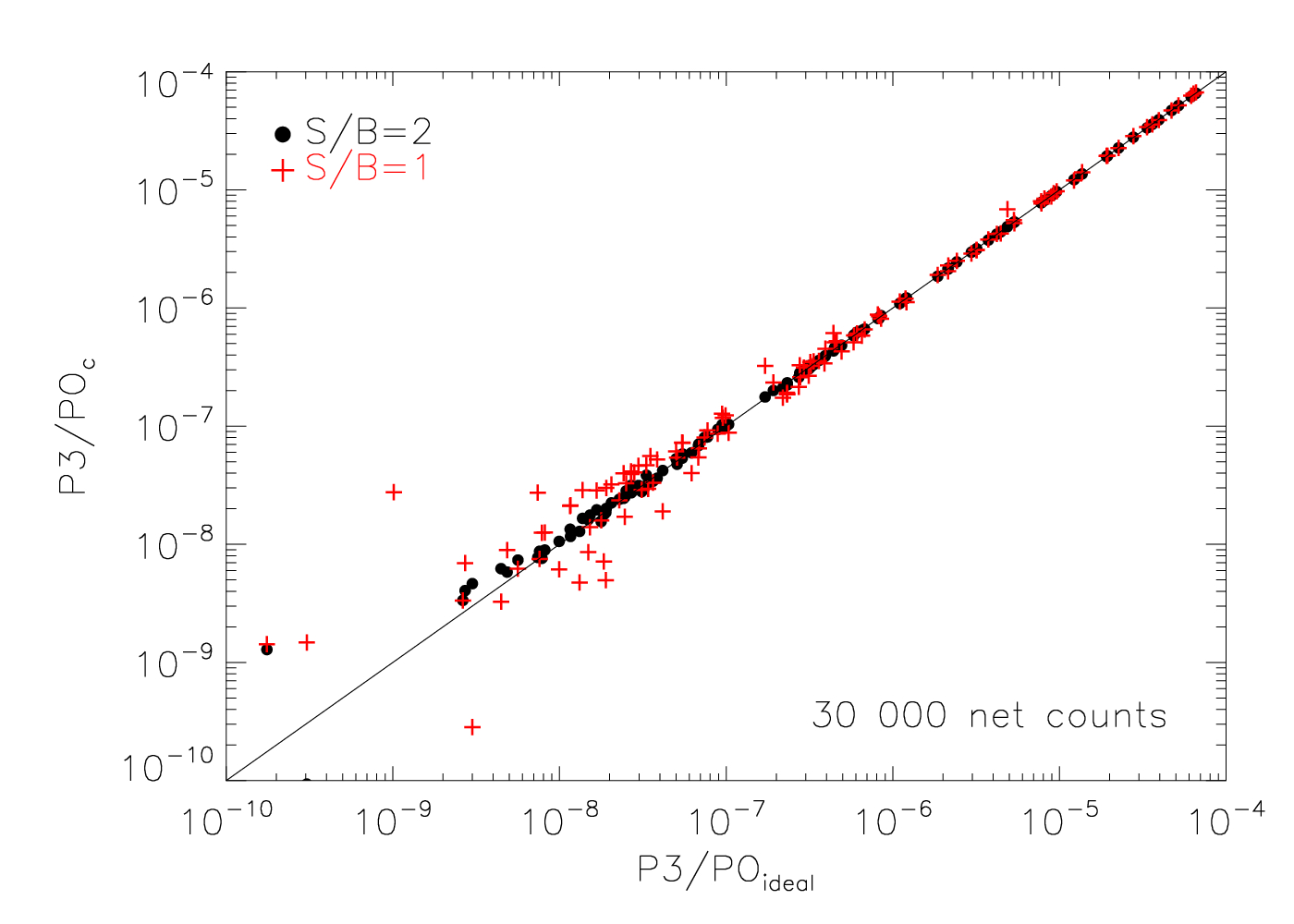}
\end{minipage}
\begin{minipage}[h]{\columnwidth}
 \includegraphics[width=\columnwidth]{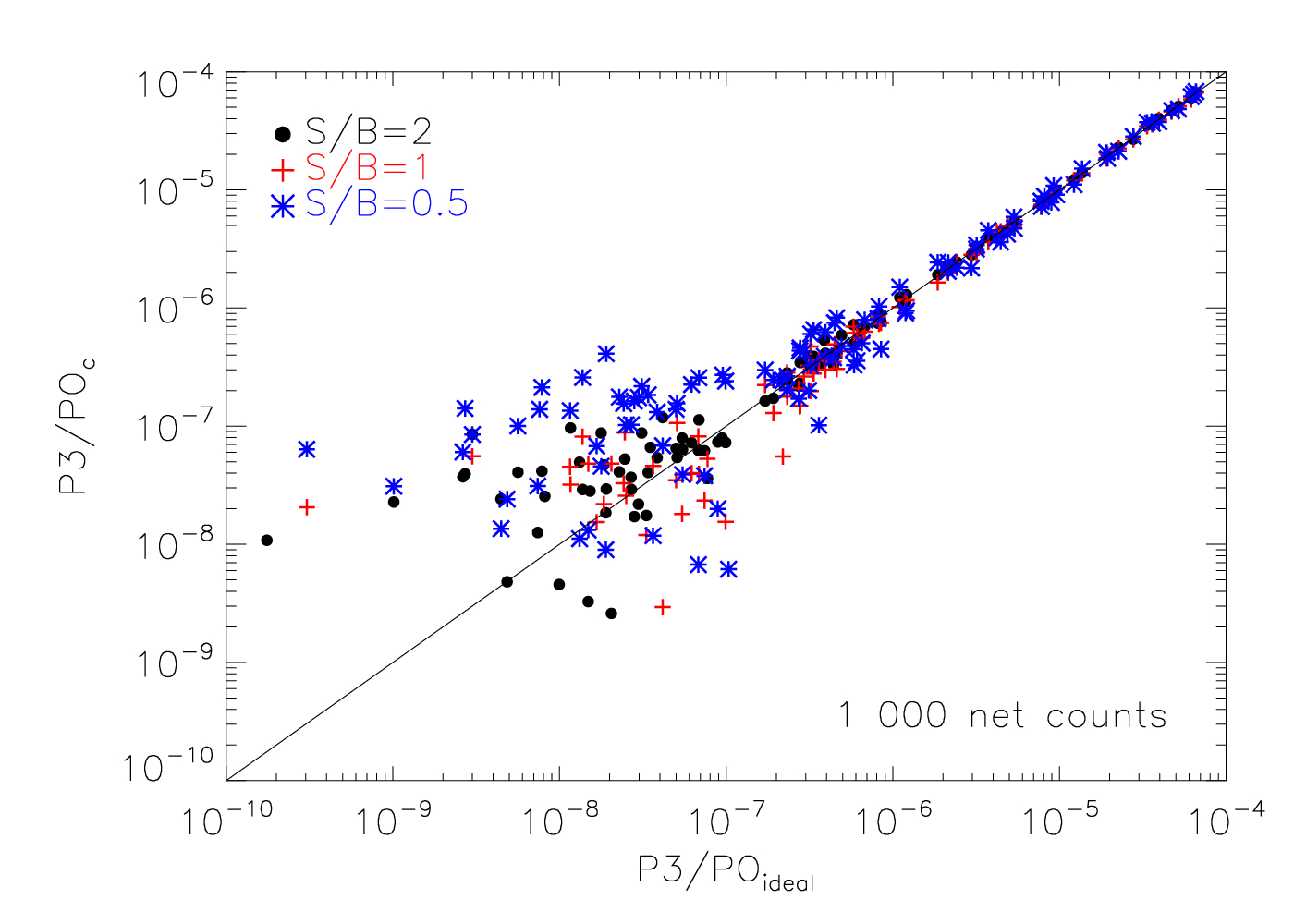}
\end{minipage}
\caption{Background and bias corrected \P3P0 as a function of \id_P3. Left: Testing the bias correcting method simulating an observation with 30\,000 net counts but poor S/B ratios (S/B=2 black circles, S/B=1 red crosses). Right: Simulating a high-z observation with 1\,000 net counts and S/B=2 (black circles), 1 (red crosses), 0.5 (blue asterisks). The increasing influence of the background for decreasing net counts and S/B is shown.} 
\label{bkg_bias}
\end{center}
\end{figure*}

In the next step, we studied the influence of the background noise component as a function of net/background counts using typical XMM-\emph{Newton} values. We first discuss the power ratios and show the results for 30\,000 net counts and a S/B (net/background counts within \r500) of 2:1 and 1:1. We chose these values to test the method simulating an observation with a large number of net counts but poor S/B ratios. Figure \ref{bkg_bias} (left) compares the background and bias corrected power ratio \p3_c with \id_P3 for these cases and shows that we can very accurately determine \P3P0 well below $10^{-8}$ for an observation with 30\,000 net counts and a S/B=2 (black circles). For a higher background (red crosses) the method still works well, however below $10^{-7}$ the scatter increases. Our sample of 80 observed low-z clusters includes only 6 clusters with a S/B$<$2 of which RXCJ0225.1-2928 shows the lowest with S/B=1.2. For observations with more than 30\,000 net counts, we find a mean S/B of 6.7 and a median S/B of 5.6. In such cases, the background noise component is not significant. \newline
This situation changes when analyzing high-z observations which typically have low-photon statistics ($<$ 1\,000 net counts) and where the S/B can become $<1$. We therefore show on the right side of Fig. \ref{bkg_bias} the results of the background and bias correction for 1\,000 net counts and a S/B of 0.5 (blue asterisks), 1 (red crosses) and 2 (black circles). Although the relation shows more scatter than for the high-counts case, the method works well down to $10^{-7}$ for S/B=1. For observations with higher background the scatter increases, however even under such conditions we can distinguish well between high power ratios values ($S>1$) and values below $10^{-7}$, with typically $S<1$. \newline   

\begin{figure*}[!h]
\begin{center}
\begin{minipage}[h]{\columnwidth}
 \includegraphics[width=\columnwidth]{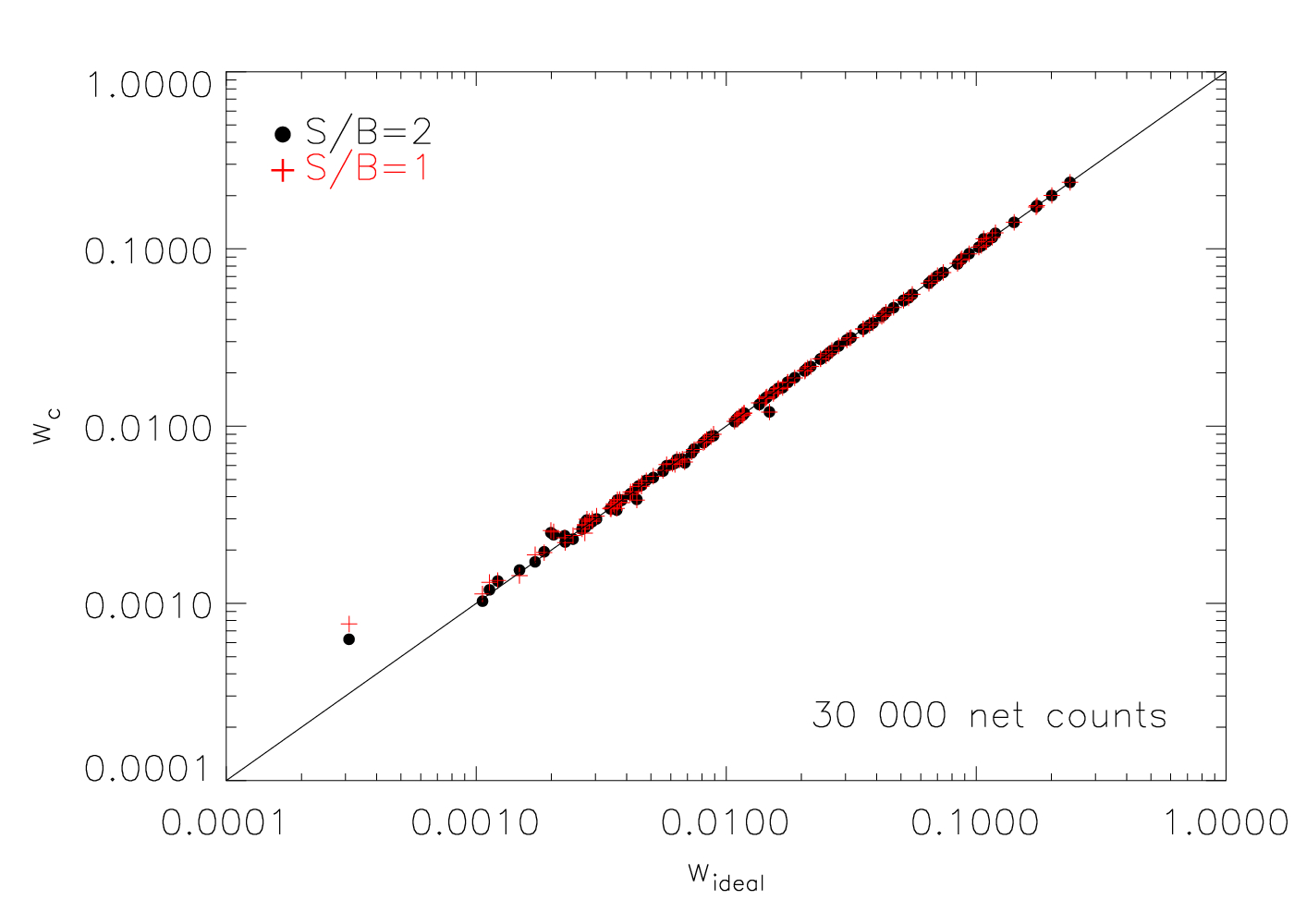}
\end{minipage}
\begin{minipage}[h]{\columnwidth}
 \includegraphics[width=\columnwidth]{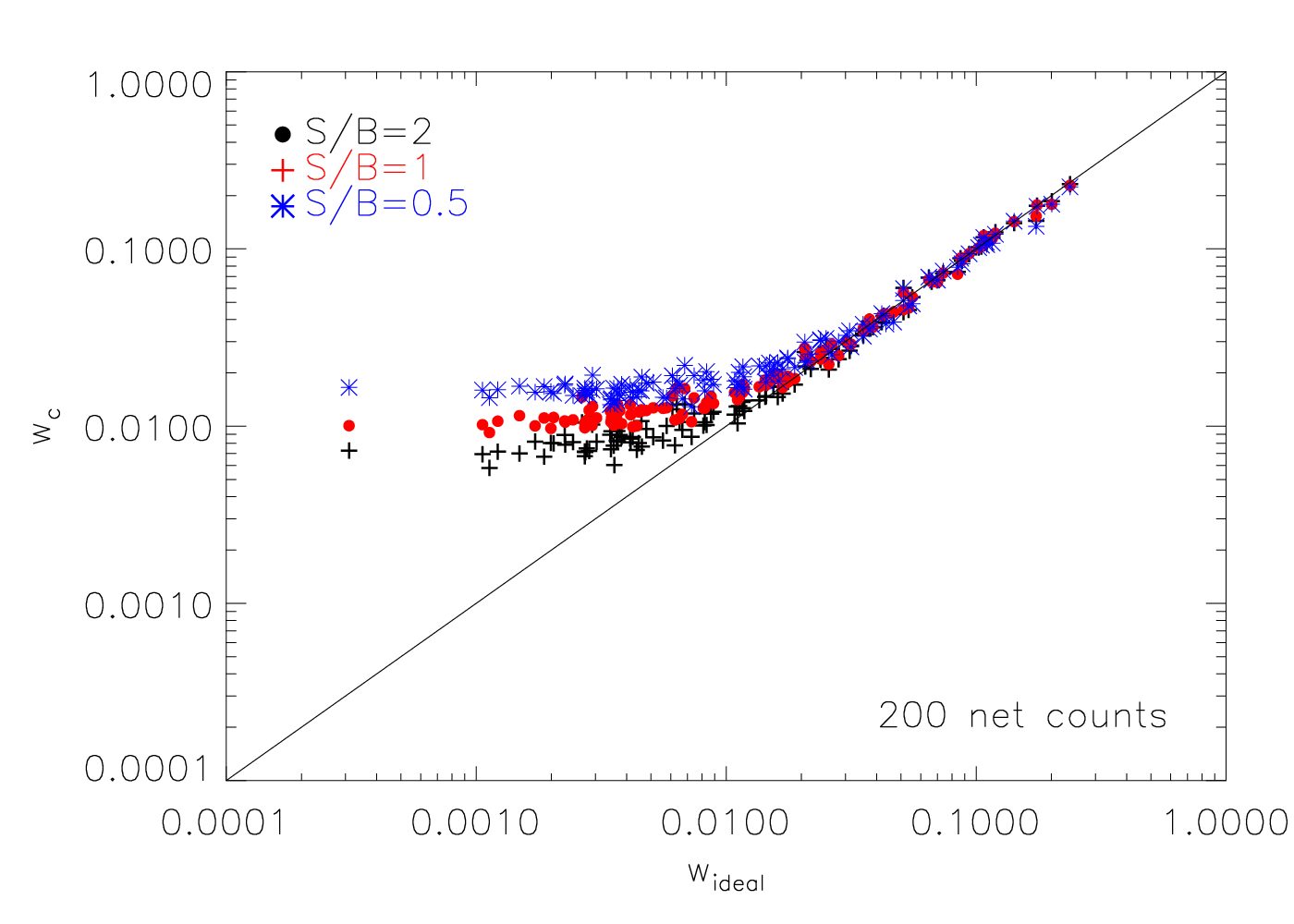}
\end{minipage}
\caption{Background and noise corrected center shifts as a function of \i_w for good photon statistics (left, same S/B ratio as in Fig. \ref{bkg_bias} on the left side) and low counts observations (right).} 
\label{FigidBKGw}
\end{center}
\end{figure*}

In the case of center shifts, the background noise influences mainly the position of the centroid. This effect is more pronounced for smaller center shifts and higher backgrounds. Analogous to the power ratios, we correct the bias using poissonizations of the observation (incl. background). However, we subtract the background counts for each pixel (instead of the moments) from the observation and its 100 poissonizations before calculating the X-ray peak and the centroid. The bias is then obtained as described in Sect. \ref{method}. We again tested the method for the above mentioned cases and found that the correction works very well down to $10^{-3}$ for the 30\,000 net counts case, even for S/B=1. This behavior is due to the lower sensitivity to noise and shown in Fig. \ref{FigidBKGw} on the left. In addition, it enables us to probe even lower photon statistics, going down to 200 net counts. Even in such an extreme case, the method works well down to about $w=10^{-2}$. A plateau forms which characterizes the remaining noise level (Fig. \ref{FigidBKGw}, right). As expected, the plateau level moves to lower values for larger S/B, representing the decreasing influence of the background with larger S/B.

\section{Morphology}
\label{threshold}

After establishing in which parameter range we can obtain significant results, we want to discuss the strength of power ratios and center shifts in distinguishing different cluster morphologies. One aim of this analysis is to find a substructure value below which a cluster can be considered essentially relaxed. An overview of the results is given in Table \ref{morph_thres}. \newline

\begin{figure}[!h]
\begin{center}
\begin{minipage}[h]{\columnwidth}
 \includegraphics[width=\columnwidth]{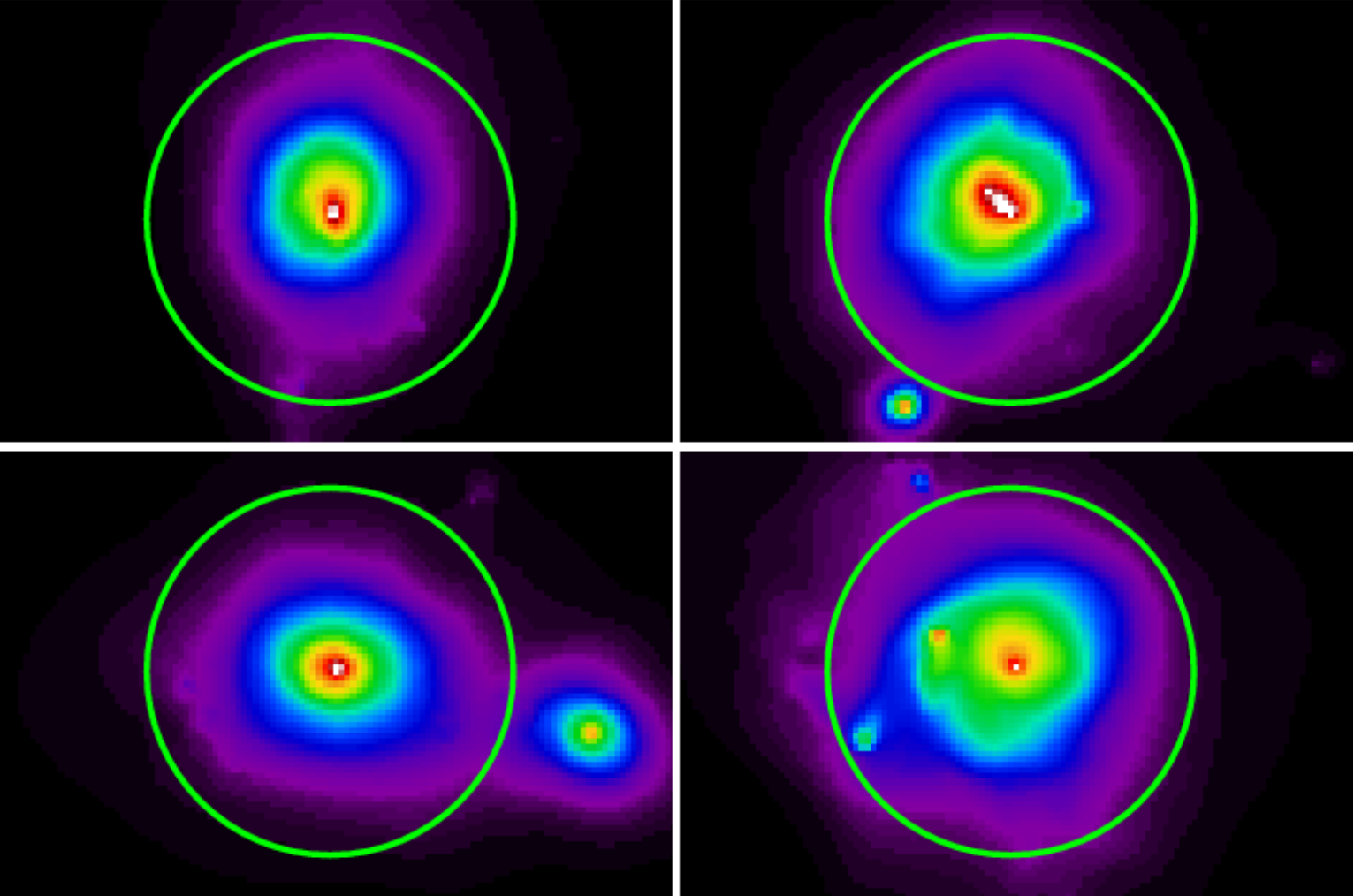}
\end{minipage}
\begin{minipage}[h]{\columnwidth}
 \includegraphics[width=\columnwidth]{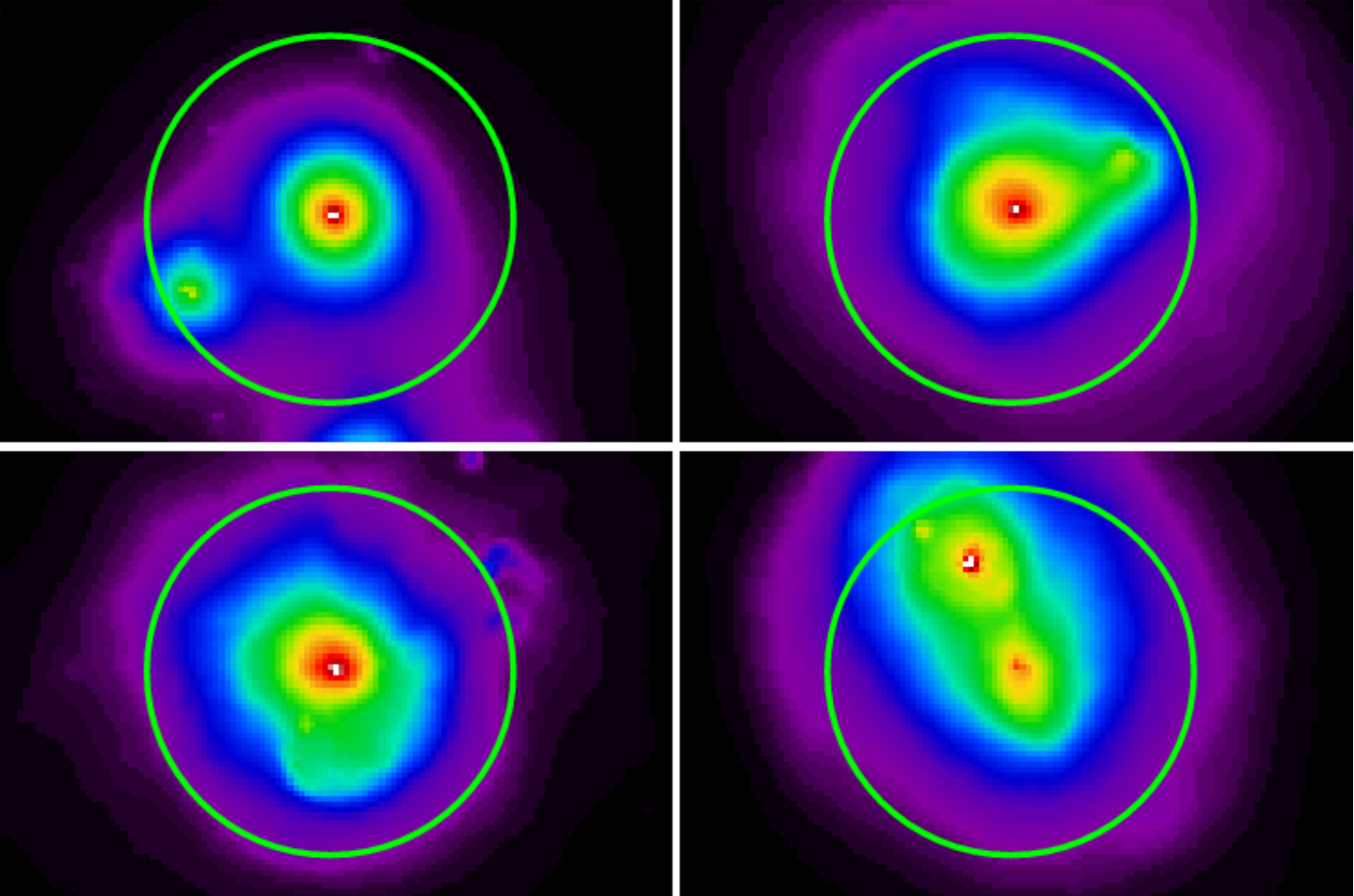}
\end{minipage}
\caption{Example gallery of clusters visually classified as essentially relaxed (top four panels) and disturbed (bottom four panels). The classification is not unambiguous in all cases, however the overall visual appearance within \r500 (green circle) was more important than small-scale disturbances.}
\label{rel_distmorph}
\end{center}
\end{figure}

We first consider \P3P0. As a result of the visual screening of the ideal simulated cluster images (no noise or background contribution), we classified all clusters as essentially relaxed (relaxed hereafter) or disturbed, depending on whether they show some signs of substructure (asymmetries, second component of comparable size, general disturbed appearance) within \r500 or not. A few examples are given in Fig. \ref{rel_distmorph}, which also illustrates that this division is not always unambiguous, however the overall visual appearance within \r500 (green circle) was more important than small-scale disturbances. \newline

\begin{figure}[!h]
\begin{center}
 \includegraphics[width=\columnwidth]{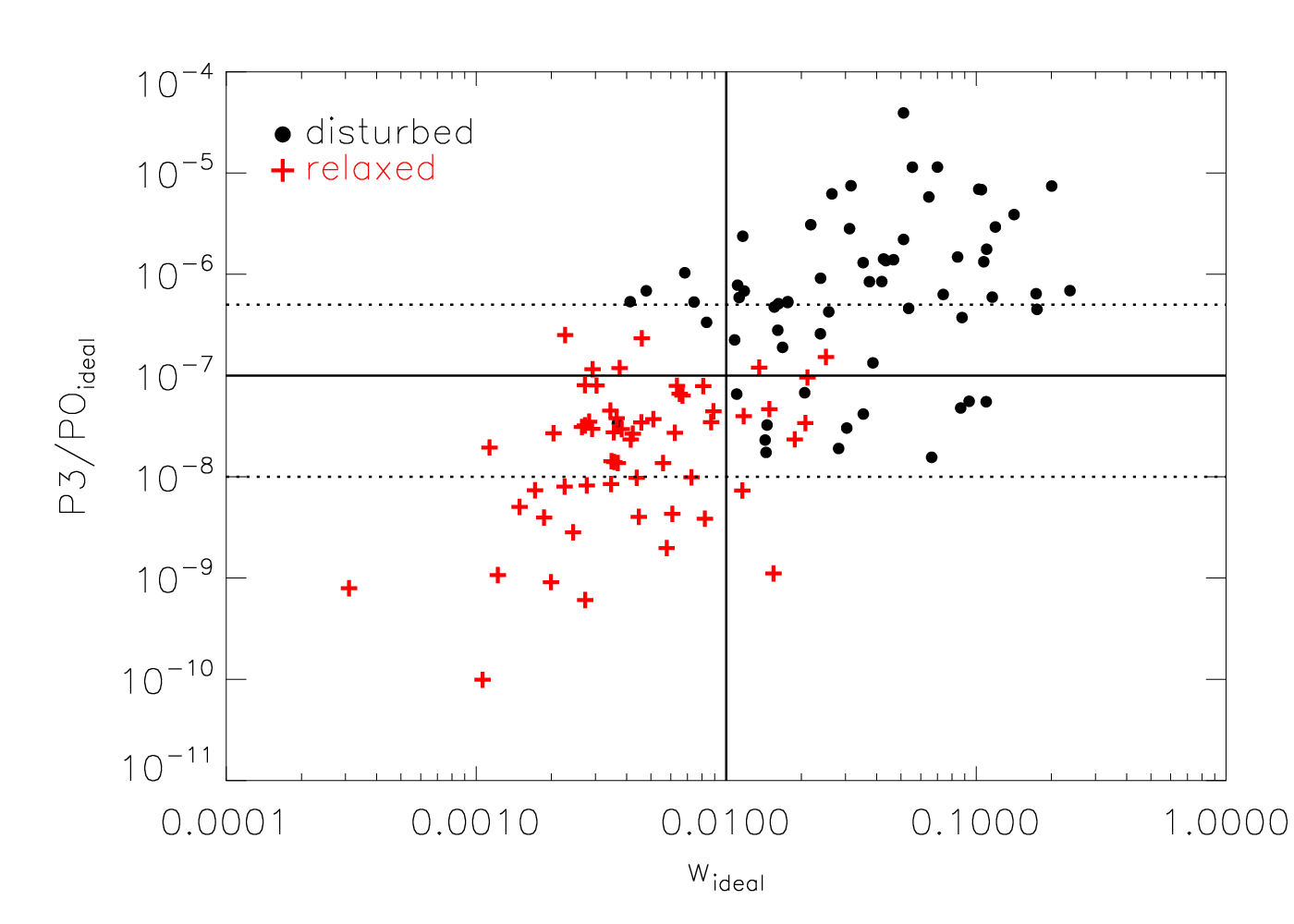}
\caption{Motivation for the {\it simple} and {\it morphological boundaries} for \P3P0 and $w$. We show the \P3P0-$w$ plane for ideal simulated cluster images. The classification into relaxed (red crosses) and disturbed (black circles) was done visually. The boundaries are displayed by horizontal and vertical lines and fit the data well.} 
\label{P3morph}
\end{center}
\end{figure}

Taking all this into account, we found \P3P0 ranges for relaxed and disturbed morphologies with a boundary value of about $10^{-7}$, which we call {\it simple \P3P0 boundary}. The motivation for this condition is shown in Fig. \ref{P3morph}, where we give the substructure parameters for all 121 simulated ideal cluster images including their visual classification as relaxed or disturbed. The horizontal line at $\mathrm{\P3P0}=10^{-7}$ divides the sample into the two populations. Out of 121 we find 6/58 ($\sim$10\%) relaxed and 13/63 ($\sim$20\%) disturbed clusters to be differently classified. For two of these 6 relaxed clusters however a merging subcluster is just entering \r500 and thus boosting the \P3P0 signal while the main cluster still seems relaxed. The remaining 4 show a slight elongation but no clear sign of structure or disturbance. For the 13 disturbed clusters we found that they have structure mostly in the inner region of the aperture radius which is not picked up by the power ratio method.\newline

\begin{table}
\begin{center}
 \caption{Overview of the boundaries for \P3P0 and $w$ including statistics when applying them to the simulated cluster sample.}
\scalebox{0.69}{
 \begin{tabular}{lccccc}
\hline\hline\\[-1.2ex]
 & Boundary & Relaxed & Disturbed & &\\[0.5 ex]
\hline \\[-1.2 ex]
\id_P3 & simple &$<10^{-7} $ &  $>10^{-7}$ & \\
1\,000 counts & & $<3\times10^{-7}$ & $>3\times10^{-7}$ &\\
Number of clusters && 58 (48\%) & 63 (52\%) &&\\
Classified differently && 10\% & 20\% &&\\
\hline \\[-1.2 ex]
\i_w & $w$ &$<10^{-2}$ & $>10^{-2}$ &\\
Number of clusters && 55 (45\%) & 66 (55\%) &&\\
Classified differently && 7\% & 5\% &&\\
\hline \\[-1.2 ex]
 &Boundary& Relaxed & Mildly disturbed & Disturbed & \\[0.5 ex]
\hline \\[-1.2 ex]
\id_P3 & morphological & $<10^{-8} $ & $10^{-8}-5\times10^{-7} $ & $>5\times10^{-7}$ \\
Number of clusters && 20 (17\%) & 62 (51\%) & 39 (32\%)\\
\hline
\label{morph_thres}
\end{tabular}
}
\end{center}
\end{table}

For high-quality observations a more detailed morphological analysis is possible because power ratios can be obtained more precisely. Taking a closer look again at Fig. \ref{P3morph}, three distinct regions present themselves: $\mathrm{\P3P0}<10^{-8}$, $10^{-8}<\mathrm{\P3P0}<5\times10^{-7}$ and $\mathrm{\P3P0}>5\times10^{-7}$. These three regions are occupied by only relaxed, a mix of relaxed and disturbed and only disturbed clusters and are indicated by the dotted lines in the figure. The borders between these regions at $\mathrm{\P3P0}=10^{-8}$ and $\mathrm{\P3P0}=5\times10^{-7}$ are named {\it morphological boundaries}. At the lower boundary of $10^{-8}$ we reach $S=2$ for 30\,000 counts images. With lower photon statistics such a classification is not possible. Making use of the {\it morphological boundaries}, we find 32\% of our simulated clusters to be significantly disturbed, while only 17\% show no signs of structure (see Table \ref{morph_thres}). The majority however (51\%) is found somewhere in the middle and called mildly disturbed objects.\newline

\begin{figure}[h]
\begin{center}
    \includegraphics[width=\columnwidth]{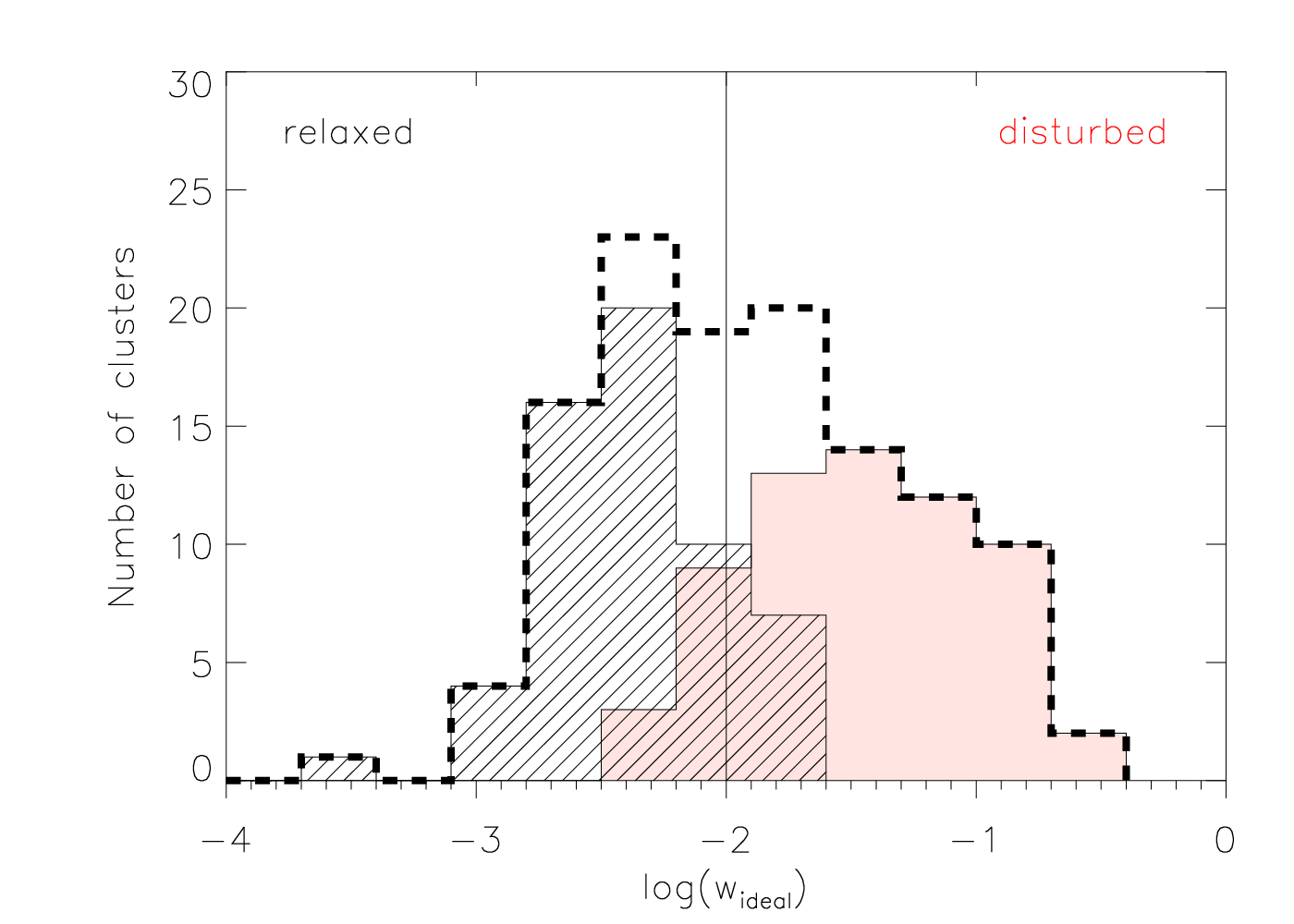}
\caption{Center shift histogram of all simulated clusters (thick black dashed line) defining the {\it $w$ boundary}. Relaxed clusters are represented by the filled black (left) and disturbed ones by the red filled histogram (right). The vertical line marks the {\it $w$ boundary} at log(\i_w)=-2.}
\label{wthres}
\end{center}
\end{figure}

\begin{figure*}[!h]
\begin{center}
\begin{minipage}[h]{\columnwidth}
 \includegraphics[width=\columnwidth]{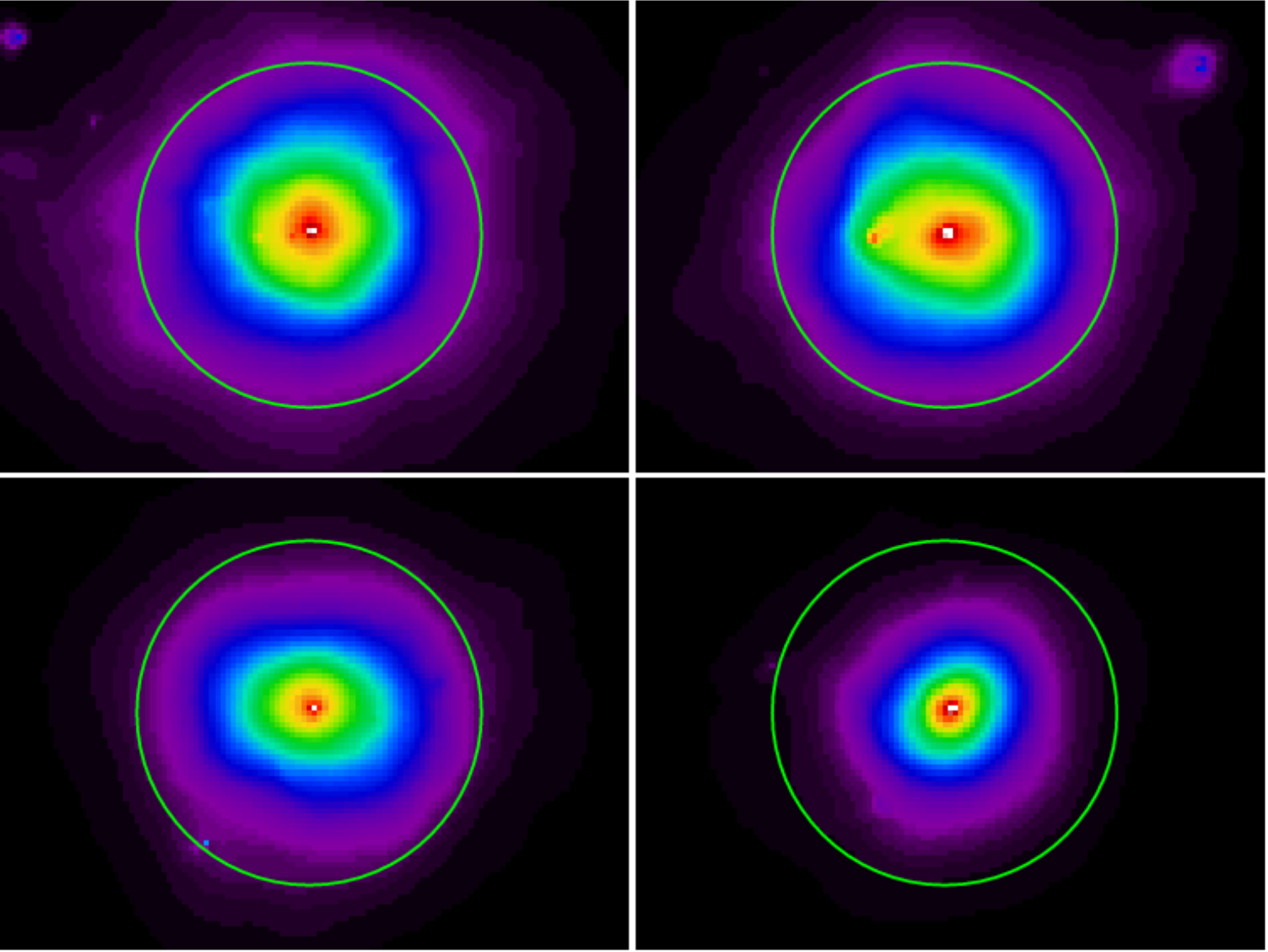}
\end{minipage}
\begin{minipage}[h]{\columnwidth}
 \includegraphics[width=\columnwidth]{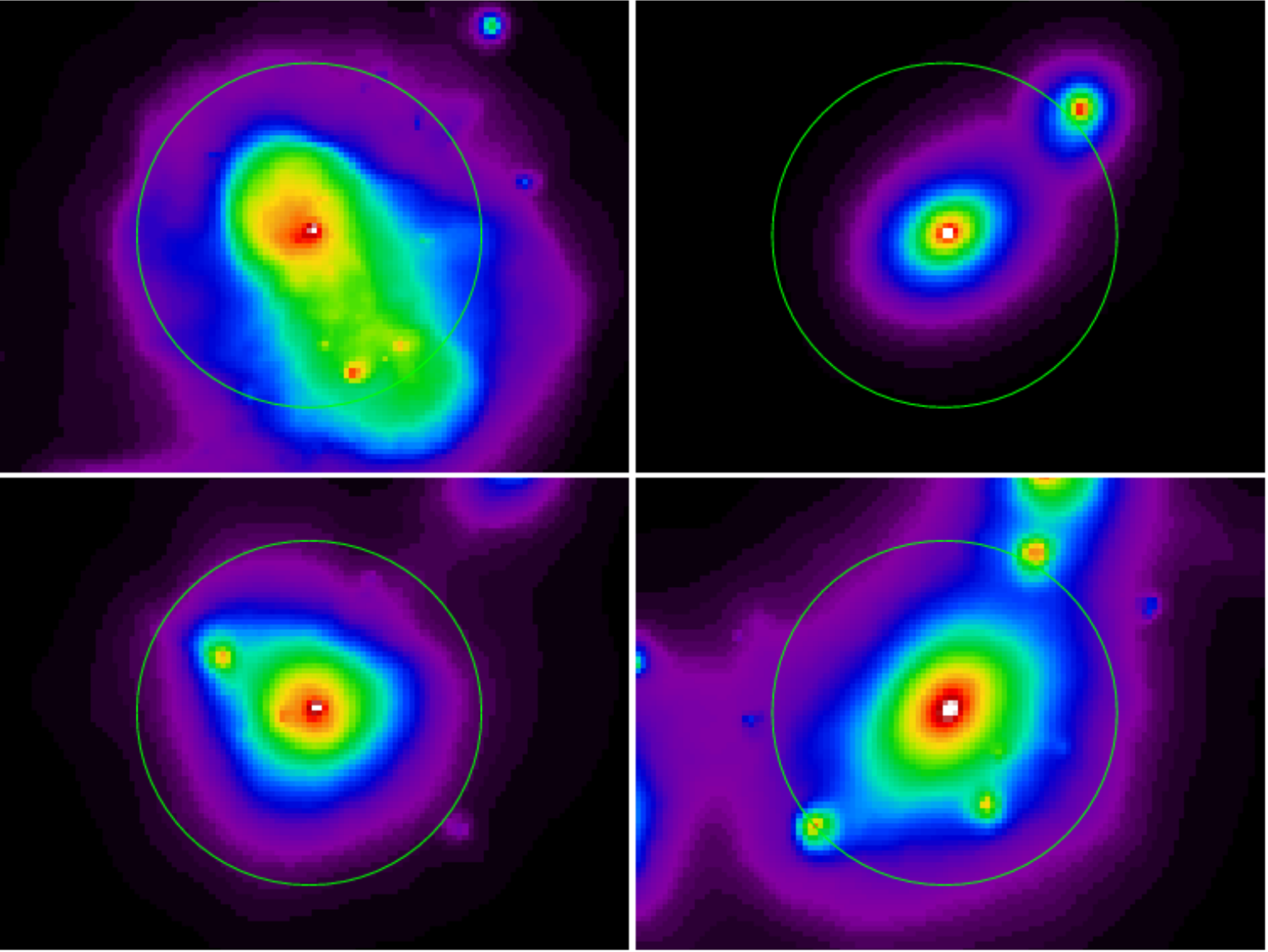}
\end{minipage} 
\caption{Example of cluster images classified using the {\it $w$ boundary}. Left 4 panels: $w < 0.01$, right 4 panels: $w > 0.01$.}
\label{wthres_pics}
\end{center}
\end{figure*}

For the center shift parameter we define a boundary at $w=0.01$. This value also agrees with our visual classification and analysis (see Fig. \ref{wthres_pics}). Figure \ref{wthres} shows \i_w histograms for relaxed (filled black histogram) and disturbed (filled red histogram) clusters, including the distribution of all clusters (thick black line). This {\it $w$ boundary} at log(\i_w)$=-2$ is apparent and the misclassification lies below 10\%. The {\it $w$ boundary} is significant with $S>2$ down to lowest counts (e.g. 200).

\section{Cluster sample}
\label{Section5}

Our sample comprises 80 galaxy clusters which are part of different larger samples observed with XMM-\emph{Newton}. An overview of the samples from which the clusters were taken and their redshift are given in Table \ref{SampleInfo}. For this study we use 31 targets from the Representative X-ray Cluster Substructure Survey \citep[REXCESS,][]{Boehringer2007}, which was created as a morphologically and dynamically unbiased sample, selected mainly by X-ray luminosity and restricted to redshifts $z<0.2$. Except for RXCJ2157.4-0747 (OBSID: 0404910701) and RXCJ2234.5-3744 (OBSID: 0404910801), where we were able to obtain longer exposures, we used the observation IDs as described in \citet[][Table 5]{Boehringer2007}.  \newline
From the Local Cluster Substructure Survey (LoCuSS, Smith et al.), we use a small subsample of 30 clusters, which was published by \citet{Zhang2008}. Except for A2204 (OBSID: 0306490401), we use the same observations as stated in \citet[][Table A.1.]{Zhang2008}. \newline
34 targets were taken from the Snowden Catalog \citep{Snowden2008}, while 10 clusters are part of the REFLEX-DXL sample \citep{Zhang2006}. In addition, we use 9 clusters discussed in \citet{Buote1996}, from which only A1651 (properties taken from \citet{Arnaud2005}) is not part of the Snowden sample. In total, 28 clusters are found in at least two samples. In such cases, the cluster properties are taken from the larger sample - as indicated in Table \ref{SampleInfo}. The clusters were chosen to be well-studied, nearby ($0.05<z<0.45$) and publicly available (in 2009) in the XMM-Archive\footnote{http://xmm.esac.esa.int/xsa/}. In addition, we required \r500 to fit on the detector. Our full sample populates the whole observed substructure range, as is shown in Fig. \ref{SIMREXP3w}. In addition, except for 13 cases, all clusters are high-quality observations with $>30\,000$ net counts. Of those 13 observations, only RXCJ2308.3-0211 has less than 9\,000 net counts ($\sim$2130 net counts with a S/B$\sim$4.6). This merged sample has no unique selection function, but a wide spread in luminosity, temperature and mass. A large fraction of the clusters comes from representative samples like REXCESS and LoCuSS and we therefore expect the sample to have a very roughly representative character. In addition, the aim to test the presented structure estimators does not necessarily need a representative sample but a large number of clusters with different morphologies which is fulfilled with this sample.


\section{Data analysis}
\label{Section6}
\subsection{XMM-\emph{Newton} data reduction}

The XMM-\emph{Newton} observations were analyzed with the XMM-\emph{Newton} SAS\footnote{Science Analysis Software: http://xmm.esa.int/sas/} v. 9.0.0. The data reduction is described in detail in B10 and \citet[][]{Boehringer2007}. We followed their recipe except for the point source removal and background subtraction. Our method of detecting point sources is consistent with B10 and \citet[][]{Boehringer2007}, where the SAS task \textit{ewavelet} is run on the combined image from all 3 detectors in order to increase the sensitivity of the point source detection. However, we removed the point sources from each detector image in the 0.5-2 keV band individually and refilled the gaps using the CIAO\footnote{CHANDRA Interactive Analysis of Observations software package: http://cxc.harvard.edu/ciao/} task \textit{dmfilth}. In the next step we subtracted the background from the point source corrected images and combined them. This method yields point source corrected images without visible artifacts of the cutting regions.

\subsection{Structure parameters}

Power ratios and center shifts were calculated according to the repoissonization method described in Sect. \ref{method}, subtracting the background moments from the full (background included) image to obtain power ratios and correcting the bias due to shot noise. For center shifts we subtract the background pixel values before calculating the positions of the X-ray peak and centroid. Errors were taken as the $\sigma$ of 100 poissonized realizations. Unless stated otherwise, all displayed \P3P0 and $w$ values are background and bias corrected and calculated in the full \r500 aperture. 


\section{Morphological analysis of 80 observed clusters}
\label{secMorph}

In this section, we will apply the substructure estimation method to our sample of 80 observed clusters and show that power ratios can give more than just a global picture of the cluster. We will briefly recapitulate the dependence of the power ratio signal on the aperture size and discuss improved morphology estimators based on these findings. In order to do so, we visually classify and divide the sample into 4 categories: a) DOUBLE - clusters with two distinct maxima, b) COMPLEX - clusters without two distinct maxima but global complex structure, c) INTERMEDIATE - overall regular clusters which show some kind of locally restricted structure or slight asymmetry, d) REGULAR - regular clusters without structure. The classification was done visually using two smoothed images (smoothed with a Gaussian with $\sigma$=4 and 8 arcseconds). This classification can then be compared to the boundaries defined in the morphological analysis of simulated cluster images. All 80 clusters are sorted according to their morphology and displayed in Figs. \ref{Morphregular}-\ref{Morphdouble}. We give the three different structure parameters (\p3_c, \w_c and \P3P0$_{\mathrm{max}}$) and the morphology for each cluster in Table \ref{SampleInfo_structure}, while an overview of the dynamical state of the sample using these three morphology estimators is detailed in Table \ref{sample_morph}. 

\begin{table}
\begin{center}
 \caption{Sample statistics. Clusters defined as relaxed, disturbed and mildly disturbed objects using different boundary conditions and three different substructure estimators \p3_c, \P3P0$_{\mathrm{max}}$ and \w_c.}
\scalebox{0.89}{
 \begin{tabular}{lccc}
\hline\hline\\[-1.2ex]
\multicolumn{1}{c}{Boundary} &\multicolumn{1}{c}{Relaxed} & \multicolumn{1}{c}{Disturbed} &\multicolumn{1}{c}{Mildly disturbed} \\[0.5ex]
\hline\\[-1.2ex]
simple \P3P0  &  59\% & 41\% &  \\
$w $	      & 53\% & 47\% &  \\
morphological \P3P0& 25\% & 10\% & 65\% \\[0.5ex]
\hline\\[-1.2ex]
simple \P3P0$_{\mathrm{max}}$ & 33\% & 67\% & \\
morphological \P3P0$_{\mathrm{max}}$ & 5\% & 24\% & 71\% \\[0.5ex]
\hline
\label{sample_morph}
\end{tabular}
}
\end{center}
\end{table}

\subsection{Improved structure estimator}
\label{Sect81}
A simple application of \P3P0 and $w$ using the repoissonization method to estimate the bias yields good results. As expected, we find very structured and in particular double clusters at high \P3P0 and $w$, while regular clusters are found to have low power ratios, but have a large spread in the $w$ range. This was already shown by \citet{Buote1996} for power ratios and several authors afterwards for both substructure measures. The center shift parameter was already discussed in detail \citep[e.g.][B10]{Mohr1995,Ohara2006,Poole2006} and shows a wide spread for disturbed and regular clusters. We therefore focus on the \P3P0 parameter and discuss it in more detail.\newline

\begin{figure}[!h]
\begin{center}
 \includegraphics[width=\columnwidth]{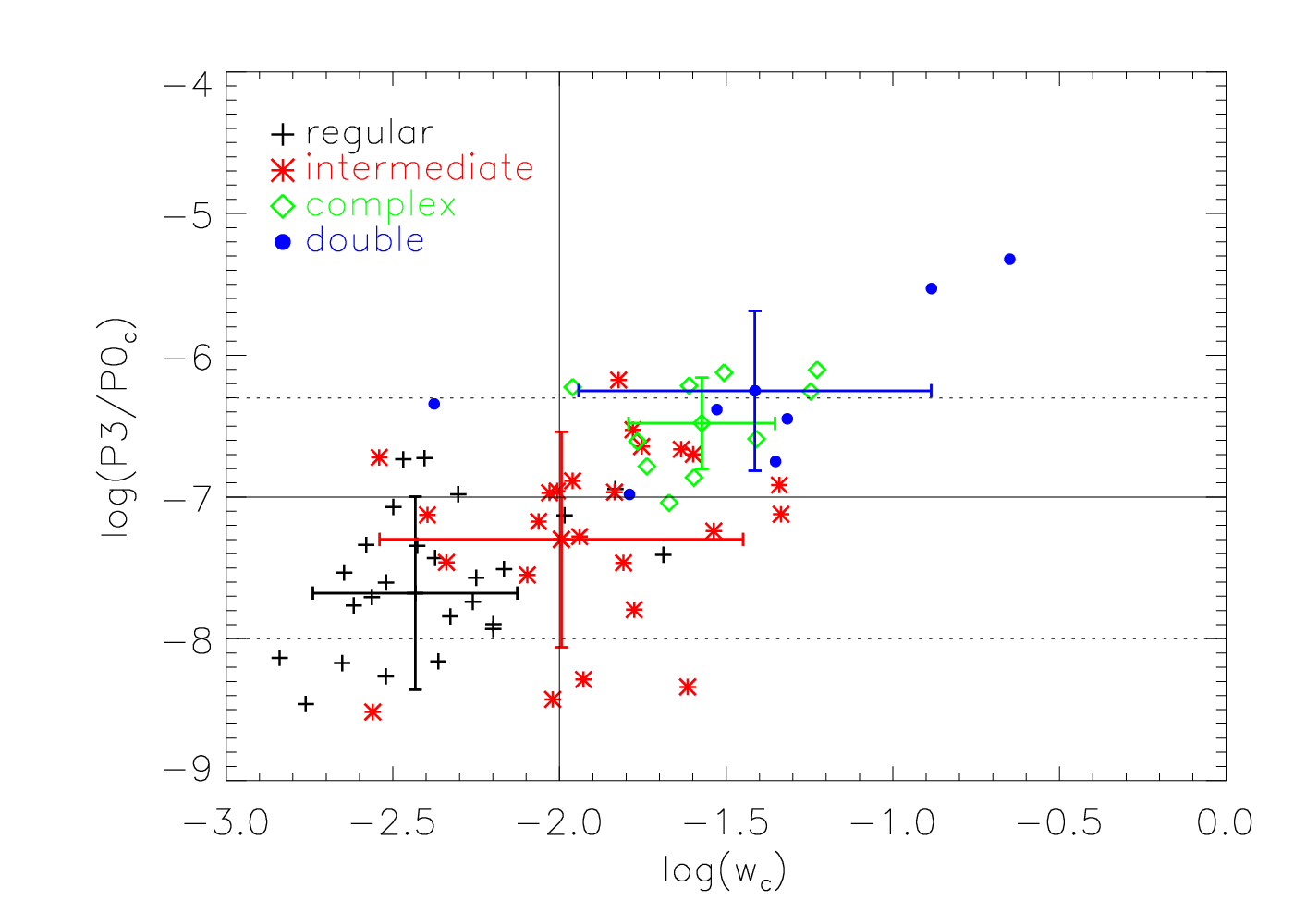}
\caption{\P3P0 - $w$ plane for all 80 observed clusters including the  {\it $w$ boundary} at $10^{-2}$ and both the {\it simple} ($10^{-7}$, black solid line) and the {\it morphological \P3P0 boundaries} ($10^{-8}$, $5\times10^{-7}$, dotted lines) in an \r500 aperture. The substructure parameters are background and bias-corrected. The outlier at $w <10^{-4}$ can be considered as $w=0$ and is excluded from the analysis. The different morphological types show a rough segregation with double (blue circles) and complex (green diamonds) with high structure parameters, while intermediate (red asterisks) and regular (black crosses) clusters are found to have very low structure values. In addition, we show the mean of the 4 populations and their spread (standard deviation).}
\label{Sample_Amorph}
\end{center}
\end{figure}

\begin{figure}[!h]
\begin{center}
 \includegraphics[width=\columnwidth]{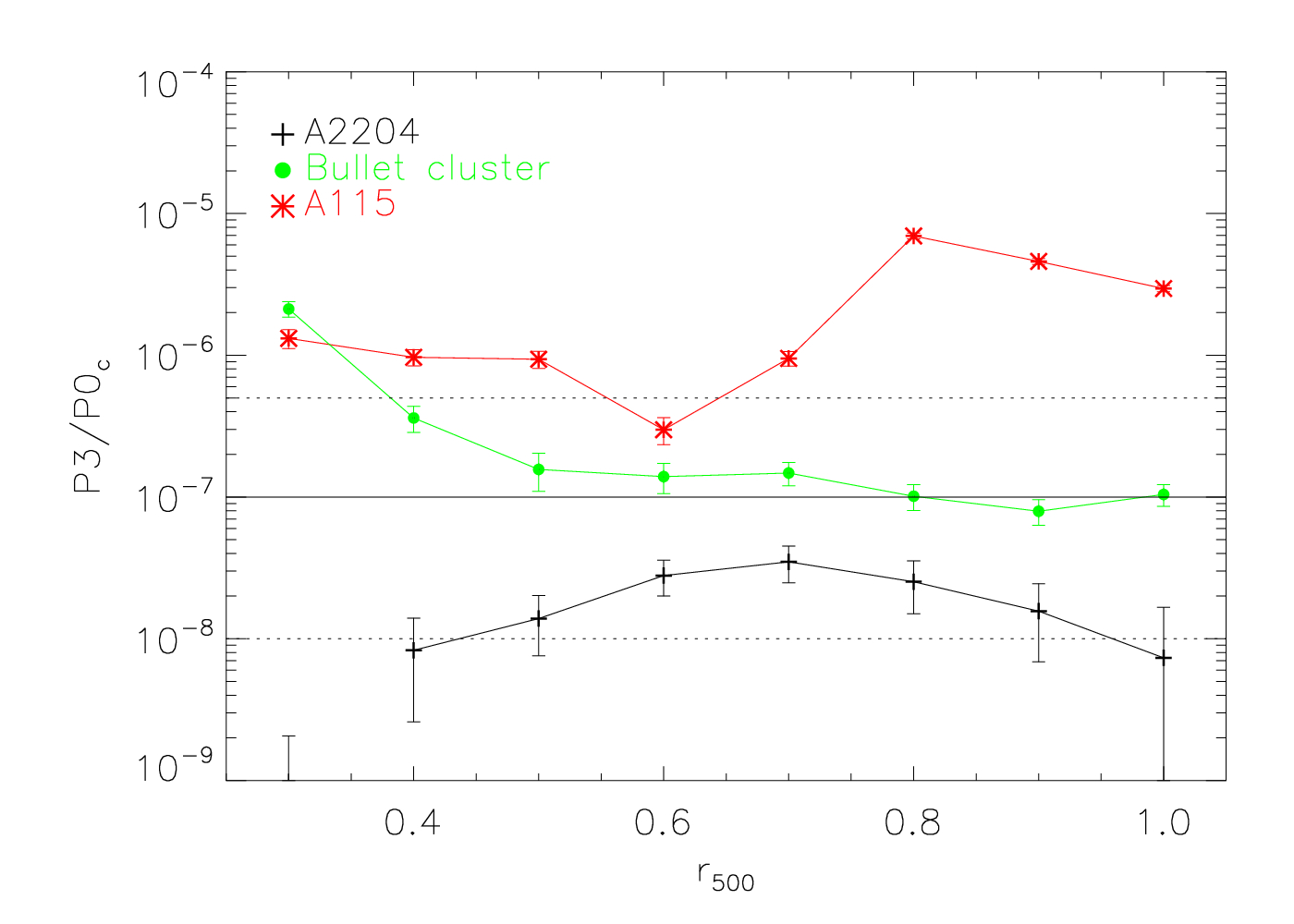}
\caption{\P3P0 profile. \P3P0 calculated in 8 apertures (0.3-1 \r500) is shown for 3 different clusters. The horizontal lines show the {\it simple} (solid line) and the {\it morphological \P3P0 boundaries} (dotted lines). A115 (red asterisks) shows a clear second component, which is located around 0.8 \r500. In the \r500 aperture it is thus classified as highly disturbed. The Bullet cluster (green circles) also clearly shows a second component, however this component lies at 0.3 \r500 and thus \P3P0 becomes less important for larger apertures. A2204 (black crosses) on the other hand is a regular cluster, which does not reach a large \P3P0 value in any aperture.}
\label{P3profile}
\end{center}
\end{figure}

For a sizeable cluster sample with the \P3P0 parameter calculated in the \r500 aperture we are able to distinguish between very structured clusters ($\mathrm{\P3P0} > 5\times10^{-7}$ - double in our classification), clusters which show some kind of structure ($5\times10^{-7} < \mathrm{\P3P0}<10^{-8}$ - complex and intermediate) and regular clusters (\P3P0 $<10^{-8}$ - regular). However, as is shown in Fig. \ref{Sample_Amorph}, there is an overlap of all three classifications in the $\mathrm{\P3P0}=5\times10^{-7}-10^{-8}$ range. This is due to the definition of the powers (see Sect. \ref{powerratios}) and the stronger weighting of structures closer to the aperture radius. In a large aperture like \r500 structures in the cluster center are less important than e.g. a merging subcluster at \r500. In order to illustrate this and motivate the next step, we show a \P3P0-profile (\P3P0 calculated in different aperture sizes) in Fig. \ref{P3profile} for three different clusters. In addition to the profiles we show the {\it simple} (solid line) and {\it morphological \P3P0 boundaries} (dotted lines). The different behavior of the three clusters is clearly visible. While both the Bullet cluster (green circles, RXCJ0658.5-5556 in Fig. \ref{Morphdouble}) and A115 (red asterisks) show prominent substructure in the visual inspection, only A115 is classified as such in the \r500 aperture. This is due to the fact that the "bullet" in the Bullet cluster lies at 0.3 \r500 and is less prominent in the full \r500 aperture. However, in the smaller aperture it would be detected as prominent substructure. As a reference cluster, we use the regular object A2204 (black crosses), which shows low substructure values in all apertures. \newline

\begin{figure}[!h]
\begin{center}
 \includegraphics[width=\columnwidth]{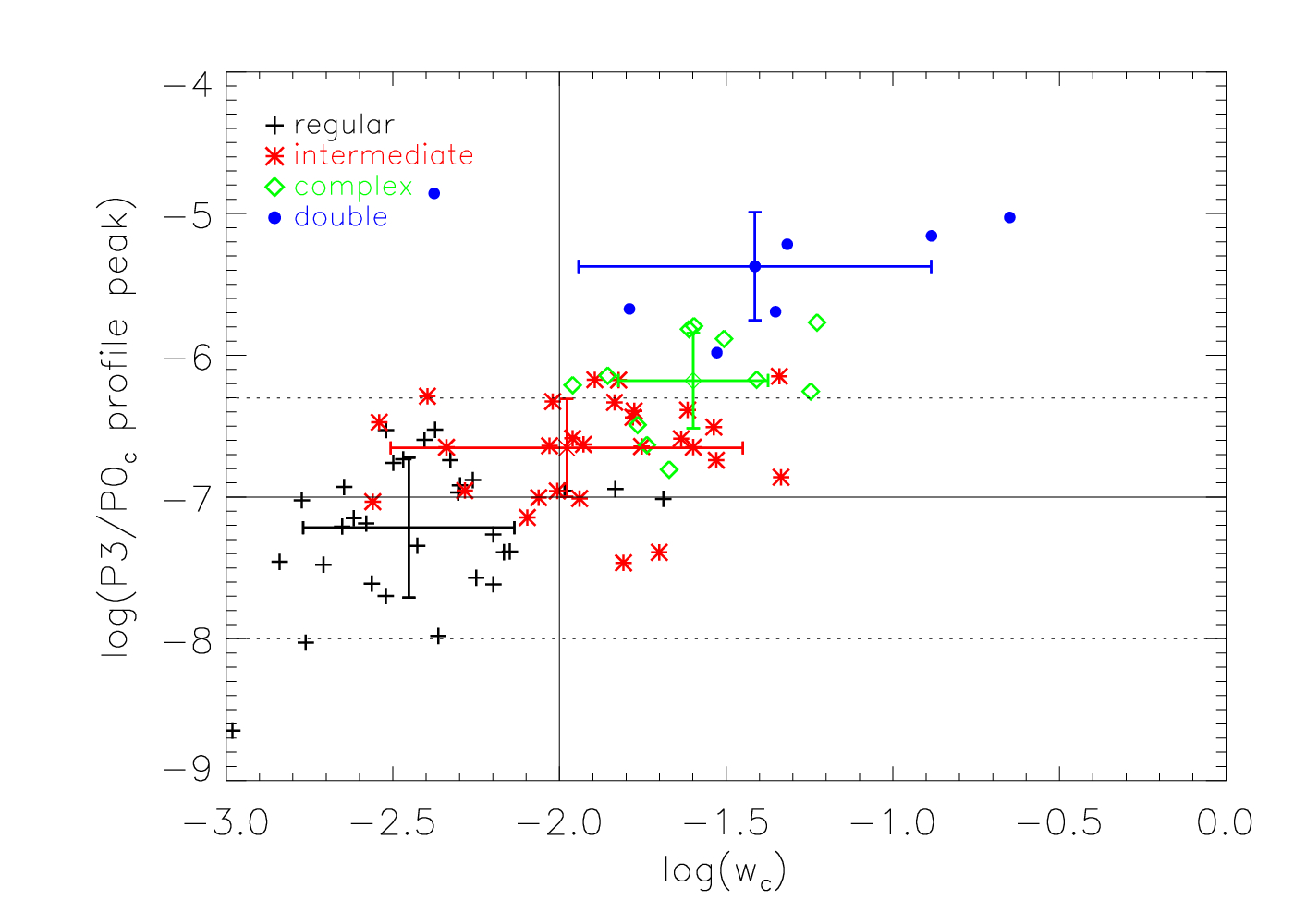}
\caption{Relation between the significant peak ($S>0$) of the \P3P0 profile and the center shift parameter for different morphologies. A tighter correlation than in the \P3P0 - $w$ plane (Fig. \ref{Sample_Amorph}) can be seen. In addition, a clearer separation between the different morphological categories is apparent. The horizontal lines mark the \P3P0 boundaries (solid: {\it simple} at $10^{-7}$, dotted: {\it morphological} at $5\times10^{-7}$ and $10^{-8}$), the vertical line displays the {\it $w$ boundary} at $10^{-2}$. The colors are as described in Fig. \ref{Sample_Amorph}.}
\label{P3maxsignw}
\end{center}
\end{figure}

We use this characteristic to introduce an improved substructure estimator, which will be detailed in the next section: the peak of the \P3P0 profile (0.3-1 \r500, in 0.1 \r500 steps), thereafter called \P3P0$_{\mathrm{max}}$. If the peak is not significant ($S<1$ or \P3P0 $<0$), we take the next highest significant value. \P3P0$_{\mathrm{max}}$ correlates well with \P3P0 in all apertures (Spearman $\rho$ between 0.5 and 0.75, prob $< 10^{-7}$). The relation between \P3P0$_{\mathrm{max}}$ and $w$ is stronger than that of \P3P0 and $w$, no matter in which aperture. Figure \ref{P3maxsignw} shows the relation between \P3P0$_{\mathrm{max}}$ and $w$, details are given in Table \ref{Sample_corr}. In addition, one can see the separation of double (blue), complex and intermediate (green and red) and regular (black) much clearer than in the \P3P0-$w$ plane in Fig. \ref{Sample_Amorph}.

\begin{table*}
\begin{center}
  \caption{Correlations between structure estimators. For correlations with \P3P0 we only show the strongest and most interesting apertures.}
 \begin{tabular}{lccccc}
\hline\hline\\[-1.2ex]
Relation & \P3P0 radius & Spearman $\rho$ & prob. & Kendall $\tau$ & prob. \\ [0.5 ex]
\hline\\[-1.2ex]
\P3P0 - $w$ & \r500	&0.55 & $6.4\times10^{-7}$ & 0.40 & $7.7\times10^{-7}$ \\
\P3P0 - $w$ & 0.9 \r500	&0.62 & $2.1\times10^{-8}$ & 0.46 & $6.0\times10^{-8}$\\
\P3P0 - $w$ & 0.3 \r500	&0.47 & $3.0\times10^{-5}$ & 0.34 & $2.7\times10^{-5}$\\
\P3P0$_{\mathrm{max}}$\tablefootmark{a} - $w$ &&0.58 & $1.8\times10^{-8}$ & 0.42 & $6.0\times10^{-8}$\\	  	
\P3P0$_{\mathrm{max}}$ - \P3P0 & \r500 &0.66 & $4.3\times10^{-10}$ & 0.51 &     0.0\\	
\P3P0$_{\mathrm{max}}$ - \P3P0 & 0.3 \r500 &0.75 & $8.4\times10^{-15}$ & 0.62 &     0.0\\[0.5 ex]
\hline\hline
\label{Sample_corr}
\end{tabular}
\tablefoot{
\tablefoottext{a}{peak of the 0.3-1 \r500 \P3P0 profile}
}
\end{center}
\end{table*}

\section{Discussion}
\label{Discussion}
\subsection{Substructure estimation and bias correction}

The reliability of these substructure estimators suffers from shot noise, especially when dealing with observations with low photon statistics. We therefore performed a detailed analysis of power ratios and center shifts using 121 simulated cluster images to study the influence of shot noise for different observational set-ups (net counts and background). \newline

We find that the center shift parameter is only affected by shot noise at very low photon statistics. This is due to the definition of this parameter, which uses the distance between the X-ray peak and the centroid in several apertures. The position of the X-ray peak is determined from a smoothed image and robust to noise (shift of the position of the brightest pixel in $<5$\% of the realizations of ~5\% of the most disturbed cluster cores). The centroid is slightly more influenced by low photon statistics than the X-ray peak. However, in units of \r500, this shift of the centroid due to shot noise is rather small. This assures a reliable center shift measurement down to low net counts ($\sim$200).\newline

The possible effect of noise on power ratios can be severe, because they are calculated in an aperture, where each pixel can be influenced by shot noise. We find a clear dependence of the bias (spuriously detected structure due to noise) on the photon statistics and the amount of intrinsic structure. Very structured clusters can be identified even in shallow observations (e.g. 1\,000 net counts in \r500). Clusters without prominent substructure (e.g. without a visible second component) might be misclassified in some cases. We therefore present an improved method to estimate the shot noise and correct for background contributions which suffer from additional noise. We use 100 poissonized realizations of the X-ray image (background included) and calculate moments ($a_0$ to $b_4$) for the image, each realization and the background image. We subtract the background moments from the image moments and those of the 100 realizations before calculating power ratios. The mean power ratio of the poissonized versions of the image gives the bias, which is subtracted from the signal of the original image. \newline

This method was influenced by several previous studies. \citet{Hart2008} estimates the bias in a similar way using a smoothed (Gaussian with 1-pixel width) image and 20 poissonized realizations of the cluster. \citet{Jeltema2005} use an analytic approach to correct the noise in the cluster but also in the background image. \newline
B10 introduced two methods to estimate the bias. The approach of poissionizing observed cluster images is the basis of our refined method presented above (see Sect. \ref{method}). The second method they proposed estimates the bias by azimuthally redistributing the counts in all pixels at a certain radial distance with random angles. Thus only the radial information is stored, but all azimuthal structure is now randomly distributed. The final bias is the mean of 100 such randomizations. Ideally, this mean gives the power ratio of a regular cluster with the same amount of shot noise as the real observation. We performed a direct comparison with this method (thereafter called azimuthal redistribution) using all 121 simulated cluster images and found that both methods yield very similar results for 1\,000 counts. Our method yields slightly better results at high counts (e.g. 30\,000) because it determines the bias more accurately than the azimuthal redistribution. In addition, our method gives better results at low \P3P0 values, partly already above the lower {\it morphological boundary} of 10$^{-8}$. However, for the high-quality observations like our sample of XMM-\emph{Newton} observations, the differences are small. \newline

Our method to correct the bias for the center shift parameter is analogous to the one for power ratios, however with the subtraction of background pixel values instead of moments before the calculation. \citet{Mohr1995} already investigated the influence of photon noise on $w$, however they define their center shift parameter in a different way and thus a direct comparison is not possible. \newline

Having established a method to correct the bias in the power ratio and center shift calculation to obtain meaningful results, we defined parameter ranges for different morphologies. Due to the variety and complexity of the morphologies of the simulated (and observed) cluster sample, a direct link between a certain substructure value and a distinct morphology could not be found. However, we showed that different types of morphologies occupy on average different regions of the substructure parameter space. Our aim to characterize a large sample can be reached using two types of boundaries for \P3P0 ({\it simple boundary} at $10^{-7}$ or {\it morphological boundaries} at $5\times10^{-7}$ and $10^{-8}$) and a center shift value of $10^{-2}$ to divide the sample into relaxed, mildly disturbed and disturbed objects. In previous studies, similar values for significant substructure were found. B10 used 1.5$\times10^{-7}$ and 2-4$\times10^{-8}$ for significant and insignificant structure, while \citet{Jeltema2008} define all clusters with $\mathrm{\P3P0}>4.5\times10^{-7}$ as disturbed and $< 10^{-8}$ as relaxed. This agrees well with our findings.\newline

The definition of the boundaries shows the large range of cluster morphologies. Merging clusters with two clear components or very irregular structure can be identified under almost all conditions because of their strong signal. Clusters which appear relaxed (spherical or elongated) yield very low substructure values, however noise might increase their signal and some relaxed clusters might have $\mathrm{\P3P0}>10^{-8}$. Applying the {\it morphological boundaries} to our sample of simulated clusters, we identify 32\% as significantly disturbed. On the other hand, only 17\% of our simulated sample show no signs of structure. This leaves the majority of clusters (51\%) to be mildly disturbed objects. They show a slightly disturbed surface brightness distribution but no clear sign of a second component, the beginning of a merger, where the merging body lies outside of the aperture radius but already influences the ICM or a post-merger. This agrees well with observed values in X-rays which range between 40-70\% of disturbed clusters \citep[e.g.][]{Mohr1995, Jones1999,Kolokotronis2001,Schuecker2001}. Using the same visual analysis as for the power ratios, a useful boundary for the center shift parameter was found to be $w=0.01$. This value agrees well with the values of B10 and \citet{Cassano2010} who also give $w=0.01$, and of \citet{Maughan2008} and \citet{Ohara2006} with $w=0.02$.\newline

In general, using this method, we can significantly lower the influence of noise, especially for power ratios. For a shallow observation (1\,000 counts), we find significant results ($S>1$) for $\mathrm{\cp3} > 3\times10^{-7}$ and are able to reduce the mean bias for this subsample of disturbed clusters from 13\% to 5\%. At 30\,000 counts, even relaxed clusters yield significant results ($S=1$ at $\mathrm{\P3P0}=4.5\times10^{-9}$) and reach a mean bias of 7\% of the ideal value after applying the correction. Using the {\it morphological boundaries} at $5\times10^{-7}$ and $10^{-8}$ to divide the sample into relaxed, mildly disturbed and disturbed objects, we see that the high bias is mainly due to truly relaxed objects with $\mathrm{\P3P0}<10^{-8}$. 

\subsection{Morphological analysis of cluster sample}

We investigated the morphologies of a sample of 80 galaxy clusters observed with XMM-\emph{Newton} in detail to give a profound and detailed illustration of these two structure estimators. In addition, we want to demonstrate the statistical strength of power ratios and center shifts and test the above defined boundaries.  \newline

While power ratios are mainly used in a large aperture of \r500 and are more sensitive to structures close to the aperture (e.g. merging component just inside \r500), center shifts are sensitive to the change of the centroid in different apertures and should thus be more sensitive to central gas properties. The center shift parameter indeed shows a tighter correlation with e.g. the central cooling time than \P3P0 (aperture of \r500), but also the power ratios are not insensitive to central gas properties \citep[e.g.][]{Croston2008}. In agreement with B10, we find the best correlation between $w$ and power ratios for a large aperture of 0.9 \r500 of \P3P0 (B10: 0.7 \r500). This indicates that while power ratios are most sensitive to substructures close to the aperture radius, they are also sensitive to large central disturbances and strong cool core activity. Merging clusters like the "Bullet cluster" however are not identified as very disturbed in large apertures when the second component is well within the aperture. Although we can see clear signs of merging, the disturbance in the outer region of \r500 is not severe enough to be identified as such. Simulations show that a powerful event like a merger influences the global cluster properties and boosts the luminosity and temperature of the cluster for a few hundred Myrs \citep{Poole2006}. In such a case, a misclassification might lead to a false interpretation. \newline

The dependence of the power ratios on the aperture size was already discussed in detail by e.g. \citet{Buote1996, Buote2001,Jeltema2005}. Looking at \P3P0 profiles, peaks due to substructure are visible in dynamically unrelaxed clusters. Not taking only one aperture size but the whole profile into account increases the probability of finding clusters with prominent structure - also in the central parts of the cluster. We thus introduced a new substructure estimator: \P3P0$_{\mathrm{max}}$, the peak of the \P3P0 profile. Comparing the detection of a merging cluster (\P3P0 $> 5\times10^{-7}$; morphological type double - as defined in Sect. \ref{secMorph}) using \P3P0 in \r500 and \P3P0$_{\mathrm{max}}$, we see that the probability of detecting substructure increases from 33\% to 100\% (compare Figs. \ref{Sample_Amorph} and \ref{P3maxsignw}). Also complex clusters are more likely to be identified as disturbed using the \P3P0 profile (45\% for \P3P0 and 73\% for \P3P0$_{\mathrm{max}}$). This is due to a shift towards larger power ratio values when using the maximum of the profile. In the lower power ratio range this increase leads to a jump of all relaxed clusters (regular and some intermediate) to power ratio values higher than $10^{-8}$. This shows that for this new parameter, the upper {\it morphological boundary} at $5\times10^{-7}$ yields best results in dividing the sample into relaxed and disturbed clusters. A few intermediate clusters cross this value, however the statistical strength remains. In order to demonstrate this again, we show the mean of each subsample and the width of the distribution in Figs. \ref{Sample_Amorph} and \ref{P3maxsignw}. It is visible that disturbed clusters (double (circles) and complex (diamonds)) are in a more defined region and better separated from the relaxed clusters (crosses). \newline

\begin{figure}[!h]
\begin{center}
\begin{minipage}{\columnwidth}
 \includegraphics[width=\columnwidth]{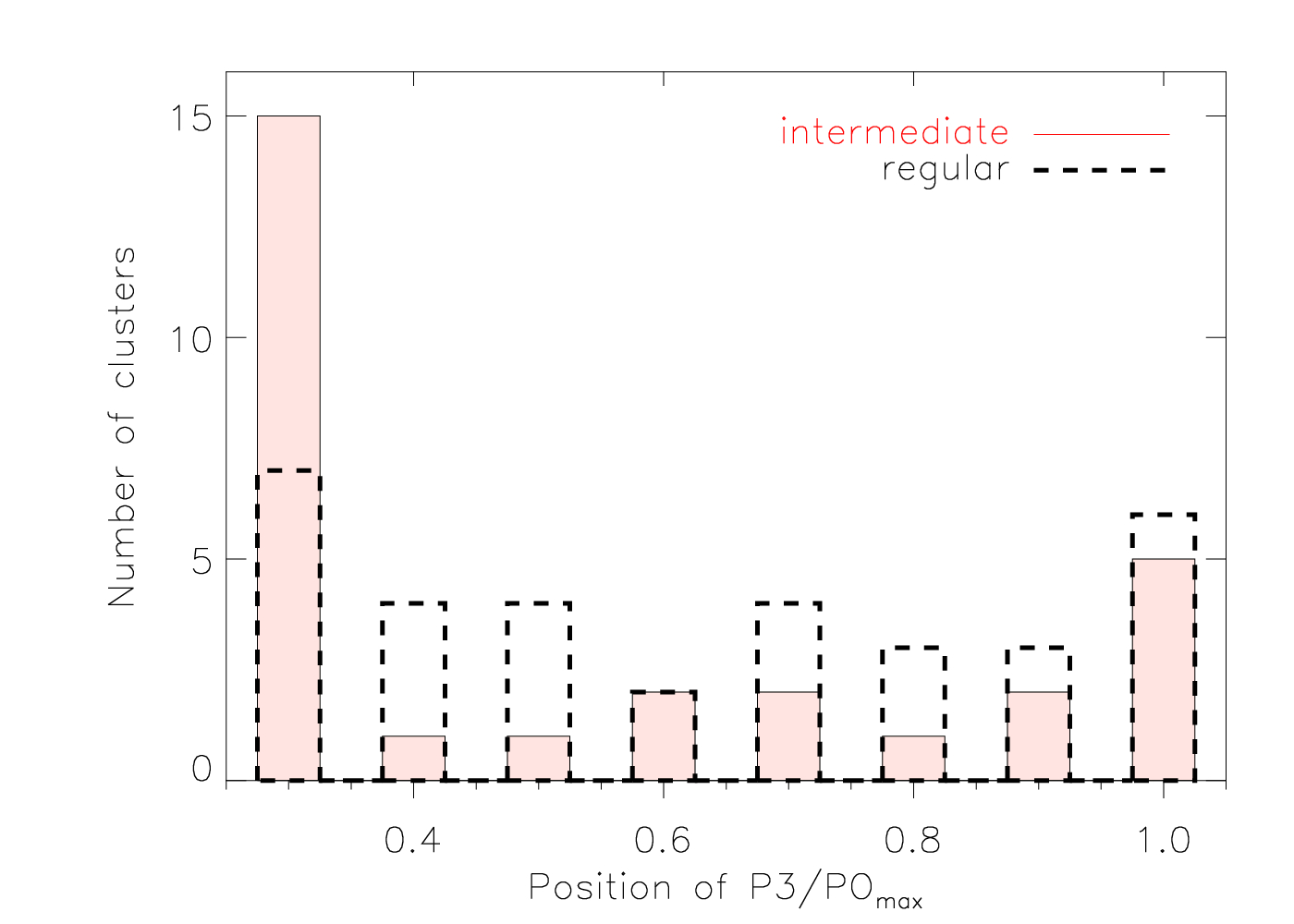}
\end{minipage}
\begin{minipage}{\columnwidth}
 \includegraphics[width=\columnwidth]{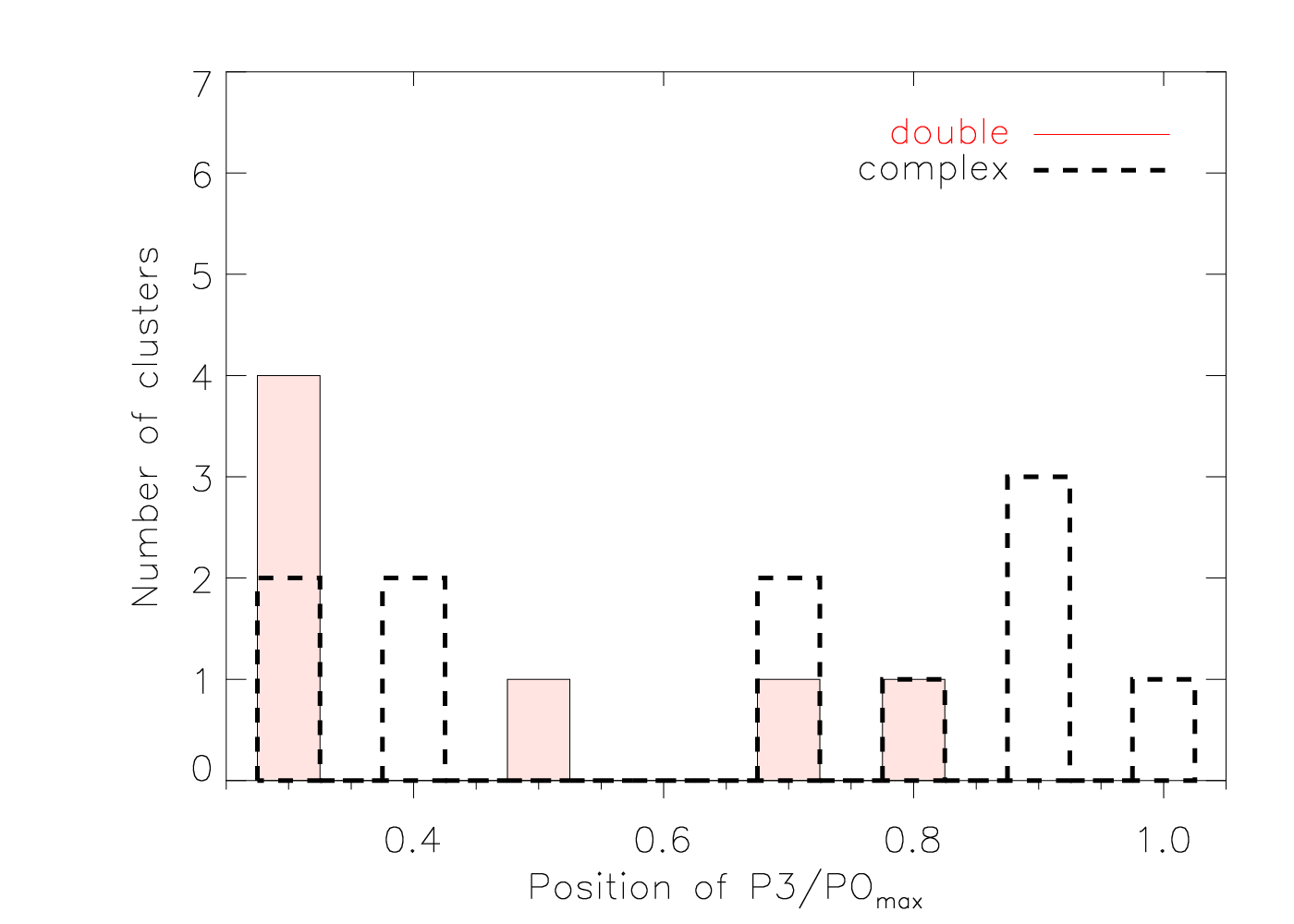}
\end{minipage}
 \caption{Histogram for all four morphological types showing the position of \P3P0$_{\mathrm{max}}$. Top: Regular (black dashed line) and intermediate (red filled histogram) clusters are shown. There is a clear excess in the 0.3 \r500 aperture. Bottom: Complex (black dashed line) and double (red filled histogram) clusters are displayed. The distribution is homogeneous since the position of the peak depends on the location of the second component or structure.}
\label{maxPeakhisto}
\end{center}
\end{figure}

In addition to the improved classification when using \P3P0$_{\mathrm{max}}$, it is interesting to see in which aperture this peak resides. A histogram of the position of the \P3P0 peak is shown in Fig. \ref{maxPeakhisto} for regular and intermediate (top) and complex and double (bottom) clusters. For complex and double clusters the distribution is as expected with no favored position. While regular clusters show a very homogeneous distribution, intermediate clusters mostly peak in small apertures. This is partly due to noise, which is larger in smaller apertures. However, these values are significant ($S>1$). This suggests that the distribution reflects the visual classification. While double and complex clusters are characterized as having two maxima in the surface brightness distribution or a complex global appearance, intermediate clusters show no global structure but slight inhomogeneities or asymmetry in the central region. \newline

Comparing our morphological classification to other works with clusters used for our analysis, we find a good agreement. \citet{Okabe2010} use asymmetry (A) and fluctuations of the X-ray surface brightness distribution in the 0.2-7 keV band (F) to divide their sample of 12 LoCuSS clusters into relaxed (low A and F) and disturbed (high A or F or both) clusters. For 9 overlapping clusters, we both find A115 to be very disturbed and agree on 2 relaxed clusters. The remaining 6 clusters are found in the \P3P0 range of mildly disturbed clusters and with not too high $w$ values. They show a low A but a spread in the F range, which fits to our definition of intermediate, showing only slight asymmetries and/or some kind of locally restricted structure. We find a large overlap of 59 clusters with \citet{Andersson2009} who used power ratios to study the evolution of structure with redshift. However, they use a fixed aperture of 500 kpc for all redshifts (0.069 to 0.89), which relates to very different apertures sizes in our analysis. \citet{Bauer2005} also use a radius of 500 kpc to obtain power ratios, however their sample is more restricted in redshift (0.15-0.37). In addition they give a visual classification and divide their sample into relaxed, disturbed and double clusters. We have 11 common clusters and our morphology classification agrees well. Other studies having an overlapping sample but using a different aperture radius are e.g. \citet{Jeltema2005} or \citet{Cassano2010}. \newline
For their comparison of X-ray and lensing scaling relations, \citet{Zhang2008} visually classified a subsample of the LoCuSS clusters according to \citet{Jones1992} as single, primary with small secondary, elliptical, off-center and complex. The last 4 classes characterize disturbed clusters. Comparing the overlapping 30 clusters to our visual classification, we find all 14 "single" clusters to be either regular or intermediate, which agrees well with our definition. Three "primary with a small secondary" are found to be complex (A1763, A13) or intermediate (RXCJ2234.5-3744). For the elliptical class, we find 4 intermediate, 1 complex and 3 regular clusters, showing that the definition of "elliptical" seems not very precise to asses the dynamical state of a cluster. The same holds for the definition of "off-center" for which we find 1 complex, 1 double and 2 intermediate clusters. The last morphological type, "complex", does not agree with our definition of complex. The only cluster defined as such, A115, is a clear double cluster. Placing these 30 clusters in the \P3P0-$w$ and \P3P0$_{\mathrm{max}}$-$w$ plane, we find the "single" clusters at low \P3P0 and $w$ value, agreeing with our definition of regular and intermediate. For 4 cases, we find either \P3P0 slightly $>10^{-7}$ (RXCJ2308.3-0211 and RXCJ0547.6-3152) or $w$ slightly $>0.01$ (A209, A2218 and RXCJ0547.6-3152). \newline
The result of a direct comparison between power ratios and the "primary" class depends on the position and size of the second component. The same holds for center shifts. A small second component close to the center will lead to a much smaller shift than one further outside. Clusters of this class can thus be found almost in the whole \P3P0 range and spread around the $w=0.01$ boundary. The same is expected for "elliptical" clusters. They will not reach a center shift value as high as for merging clusters due to the lack of a strong second component. On the other hand one would not expect extremely high or low power ratio values. The lower limit is set by the fact that the cluster is elliptical and not completely symmetric and will thus show $\mathrm{\P3P0}>10^{-8}$. Due to an asymmetric elliptical structure however the centroid on which the aperture is centered shifts when going to larger radii, therefore setting an upper limit of a few times $10^{-7}$ to the expected value. The "off-center" class showing no clear sign of substructure has similar characteristics as the elliptical one and is thus found in the same \P3P0 and $w$ range. Therefore only the morphological type "complex" remains to be discussed, which characterizes clusters with complex, multiple structures. This fits to our definition of double clusters with two distinct maxima in the surface brightness distribution. Overall one can conclude that clearly relaxed clusters and apparent mergers are very well described using both morphology schemes. The intermediate range, however, is defined ambiguously. We discussed these two classification schemes in detail because we have a large overlap of clusters and therefore can derive statistics from it.\newline

It is important to point out again that morphological classifications are very often done using visual impressions and are dependent on the observer. Power ratios and center shifts on the other hand give numbers, which - using the results of our analysis - can be related to different, simple morphologies. Recalling the {\it morphological boundaries} defined for \P3P0 at $10^{-8}$ and $5\times10^{-7}$, we find a clear overlap between our mildly disturbed class and the three intermediate classes of \citet{Jones1992}, "primary with small secondary", "elliptical" and "off-center". Using \P3P0$_{\mathrm{max}}$ would help to better filter out clusters of the "primary" class, due to the sensitivity in all aperture sizes. \newline

It is clearly shown that each of the three discussed parameters (\P3P0 in one aperture, \P3P0 profile and $w$) is sensitive on different scales. We therefore propose to use all three substructure estimators to characterize the dynamical state of large cluster samples. This can be done without a large computational effort for a large number of objects and help in identifying the potentially most interesting clusters for further analysis. 


\section{Conclusions}
\label{Conclusions}

In this paper we provide a well tested method to obtain bias and background corrected substructure measures (power ratio \P3P0 and center shift $w$). We studied the influence of shot noise in detail and are able to correct for it sufficiently. We demonstrate that a simple parametrized bias correction is not possible and thus we propose a non-parametric bias correction method applicable to each cluster individually. We tested the method for different observational set-ups (net counts and background) using typical XMM-\emph{Newton} values. We conclude that for low counts observations the influence of the background and bias can be severe. In general, the center shift parameter $w$ is less sensitive to noise and more reliable than power ratios, especially for low photon statistics. However, one should be reminded that this method is statistically strong but might not be completely accurate for each individual cluster. We thus looked in more detail into the power ratio method and how certain parameter ranges can be related to different morphologies. 

\begin{itemize}
 \item Using a sample of 121 simulated X-ray cluster images, we visually inspected each cluster and established two kinds of substructure boundaries for \P3P0 (simple and morphological) and similarly one boundary for the center shift parameter.
 \item The {\it simple \P3P0 boundary} at \P3P0 = $10^{-7}$ or the {\it $w$ boundary} at $w=0.01$ divide a large sample into relaxed and disturbed clusters. For a more detailed morphological analysis, we introduce the {\it morphological \P3P0 boundaries} at $10^{-8}$ and $5\times10^{-7}$, which divide the sample into relaxed, mildly disturbed and disturbed objects. The two classification schemes can be used for low ({\it simple \P3P0 boundary}) and high photon statistics ({\it simple} and {\it morphological \P3P0 boundaries}).
 \item We applied the bias correction method and the defined boundaries to a sample of 80 galaxy clusters observed with XMM-\emph{Newton}. We give structure parameters (\P3P0 in \r500, $w$ and \P3P0$_{\mathrm{max}}$) for all clusters which are mostly part of well-known samples like LoCuSS or REXCESS. 
 \item Applying the simple \P3P0 ($w$) substructure boundary, we find 41\% (47\%) of our observed clusters to be disturbed. The {\it morphological boundaries} yield 10\% disturbed, 65\% mildly disturbed and 25\% relaxed objects. This large difference in the number of disturbed objects using the different conditions shows that most objects are not significantly but only mildly disturbed and do not show a clear second component.
 \item We visually classified all clusters into 4 groups (regular, intermediate, complex, double) to further test the strength of the structure estimators and find 8.75\% double, 13.75\% complex, 36.25\% intermediate and 41.25\% regular objects.
 \item We introduce the use of the \P3P0 profile, which picks up structures at all distances from the cluster center and in all aperture sizes. 
 \item At last, we propose to use the maximum of the \P3P0 profile because it is not sensitive to the aperture size but finds clusters with structure on all scales. This parameter is more correlated with $w$ than \P3P0 at any fixed aperture.  
\end{itemize}

Using the proposed methods is especially interesting when dealing with a large sample, where visual classification of each individual cluster is not required but the global dynamical state of the whole sample is of interest. Applying the modified structure estimators like \P3P0$_{\mathrm{max}}$ gives additional constraints and helps to single out very structured or very relaxed clusters. Finding cluster mergers to study structure evolution or strong cool core clusters (very relaxed clusters) requires only a small computational effort, but gives a first indication about the dynamical state and properties of the cluster and whether a detailed analysis is desired.


\begin{acknowledgements}
We would like to thank the anonymous referee for constructive comments and suggestions. AW wants to thank Heike Modest for helpful discussions. This work is based on observations obtained with XMM-\emph{Newton}, an ESA science mission with instruments and contributions directly funded by ESA Member States and NASA. The XMM-\emph{Newton} project is supported by the Bundesministerium f\"{u}r Wirtschaft und Technologie/Deutsches Zentrum f\"{u}r Luft- und Raumfahrt (BMWI/DLR, FKZ 50 OX 0001), the Max-Planck Society and the Heidenhain-Stiftung. AW acknowledges the support from and participation in the International Max-Planck Research School on Astrophysics at the Ludwig-Maximilians University. An early part of this work was financed by the Bundesministerium f\"{u}r Wissenschaft und Forschung, Austria, through a "F\"{o}rderungsstipendium nach dem Studienf\"{o}rderungsgesetz" granted by the University of Vienna. 
\end{acknowledgements}

\bibliographystyle{aa}
\bibliography{library_paper1}

\FloatBarrier

\begin{appendix}
\section{Tables}

\begin{longtable}{lccc|clcc}
\caption{Details about the cluster sample.}\\
\hline\hline\\[-1.2ex]
\multicolumn{1}{c}{Cluster} & \multicolumn{1}{c}{$z$} & \multicolumn{1}{c}{Source}& & &\multicolumn{1}{c}{Cluster} & \multicolumn{1}{c}{$z$} & \multicolumn{1}{c}{Source}\\ [0.5ex]
\hline\\[-1.2ex]
\endfirsthead
\caption{continued.}\\
\hline\hline\\[-1.2ex]
\multicolumn{1}{c}{Cluster} & \multicolumn{1}{c}{$z$} & \multicolumn{1}{c}{Source}& & &\multicolumn{1}{c}{Cluster} & \multicolumn{1}{c}{$z$} & \multicolumn{1}{c}{Source}\\ [0.5ex]
\hline\\[-1.2ex]
\endhead
\hline\\[-1.2ex]
\endfoot
 \object{RXCJ0307.0-2840}	& 0.2580 & 1,2	& &  & \object{A2597}		& 0.0804 & 3,4	\\
 \object{RXCJ0516.7-5430}	& 0.2940 & 1,2	& &  & \object{A1775}		& 0.0754 & 3	\\
 \object{RXCJ0528.9-3927}	& 0.2840 & 1,2	& & &  \object{A1837}		& 0.0663 & 3,5	\\
 \object{RXCJ0532.9-3701}	& 0.2750 & 1,2	& &  & \object{RXCJ0014.3-3022} & 0.3066 & 2	\\
 \object{RXCJ0658.5-5556} 	& 0.2960 & 1,2,3,5& &  & \object{RXCJ1131.9-1955}	& 0.3075 & 2	\\
 \object{RXCJ0945.4-0839}	& 0.1530 & 1	& &  & \object{A1651}		& 0.0845 & 5	\\
 \object{RXCJ2129.6+0005}	& 0.2350 & 1	& &  & \object{A133}		& 0.0575 & 3	\\
 \object{RXCJ2308.3-0211}	& 0.2970 & 1,2	& &  & \object{A2626}		& 0.0549 & 3	\\
 \object{RXCJ2337.6+0016}	& 0.2750 & 1,2	& &  & \object{A2065}		& 0.0728 & 3	\\
 \object{A68}		& 0.2550 & 1,3	& &  & \object{RXCJ0003.8+0203} & 0.0924 & 6	\\ 
 \object{A115}		& 0.1970 & 1	& &  & \object{RXCJ0006.0-3443} & 0.1147 & 6	\\
 \object{A209}		& 0.2090 & 1,3	& &  & \object{RXCJ0020.7-2542} & 0.1410 & 6	\\
 \object{A267}		& 0.2300 & 1	& &  & \object{RXCJ0049.4-2931} & 0.1084 & 6	\\
 \object{A383}		& 0.1870 & 1,3	& &  & \object{RXCJ0145.0-5300} & 0.1168 & 6	\\
 \object{A773}		& 0.2170 & 1,3	& &  & \object{RXCJ0211.4-4017} & 0.1008 & 6	\\
 \object{A963}		& 0.2060 & 1	& &  & \object{RXCJ0225.1-2928} & 0.0604 & 6	\\
 \object{A1413}		& 0.1430 & 1,3,4,5& &  & \object{RXCJ0345.7-4112} & 0.0603 & 6	\\
 \object{A1763}		& 0.2280 & 1	& &  & \object{RXCJ0547.6-3152} & 0.1483 & 1,6	\\
 \object{A1914}		& 0.1710 & 1,3,5& &  & \object{RXCJ0605.8-3518} & 0.1392 & 6	\\
 \object{A2390}		& 0.2330 & 1	& &  & \object{RXCJ0616.8-4748} & 0.1164 & 6	\\
 \object{A2667}		& 0.2300 & 1,3	& & &  \object{RXCJ0645.4-5413} & 0.1644 & 1,6	\\
 \object{A2204}		& 0.1520 & 1,3,4,5& &  & \object{RXCJ0821.8+0112} & 0.0822 & 6	\\
 \object{A2218}		& 0.1760 & 1,3,5& &  & \object{RXCJ0958.3-1103} & 0.1669 & 1,6	\\
 \object{RXCJ0232.2-4420}	& 0.2840 & 1,2	& &  & \object{RXCJ1044.5-0704} & 0.1342 & 6	\\
 \object{A13}		& 0.1035 & 3	& &  & \object{RXCJ1141.4-1216} & 0.1195 & 6	\\
 \object{A520}		& 0.1946 & 3	& &  & \object{RXCJ1236.7-3354} & 0.0796 & 6	\\
 \object{A665}		& 0.1788 & 3,5	& &  & \object{RXCJ1302.8-0230} & 0.0847 & 6	\\
 \object{A1068}		& 0.1471 & 3,4,5& &  & \object{RXCJ1311.4-0120}	& 0.1832 & 1,3,6\\
 \object{A1589}		& 0.0722 & 3	& &  & \object{RXCJ1516.3+0005} & 0.1181 & 6	\\
 \object{A2163}		& 0.2021 & 3	& &  & \object{RXCJ1516.5-0056} & 0.1198 & 6	\\
 \object{A2717}		& 0.0510 & 3,4,5& &  & \object{RXCJ2014.8-2430} & 0.1538 & 6	\\
 \object{A3112}		& 0.0723 & 3	& &  & \object{RXCJ2023.0-2056} & 0.0564 & 6	\\
 \object{A3827}		& 0.0959 & 3	& &  & \object{RXCJ2048.1-1750} & 0.1475 & 6	\\
 \object{A3911}		& 0.0958 & 3	& &  & \object{RXCJ2129.8-5048} & 0.0796 & 6	\\
 \object{A3921}		& 0.0919 & 3	& &  & \object{RXCJ2149.1-3041} & 0.1184 & 6	\\
 \object{1E1455.0+2232}	& 0.2583 & 3	& &  & \object{RXCJ2217.7-3543} & 0.1486 & 6	\\
 \object{PKS0745-19}	& 0.0986 & 3,4	& &  & \object{RXCJ2218.6-3853}	& 0.1411 & 1,6	\\ 
 \object{RXJ1347.5-1145}	& 0.4477 & 3	& &  & \object{RXCJ2234.5-3744} & 0.1510 & 1,6	\\
 \object{Sersic159-3}	& 0.0563 & 3	& &  & \object{RXCJ2319.6-7313} & 0.0984 & 6	\\
 \object{ZwCl3146}	& 0.2817 & 3	& &  & \object{RXCJ2157.4-0747} & 0.0579 & 6	
 \label{SampleInfo}
 \end{longtable} 
 \tablebib{(1) LoCuSS: \citet{Zhang2008}; (2) REFLEX-DXL: \citet{Zhang2006}; (3) \citet{Snowden2008}; (4) \citet{Arnaud2005};\\ (5) \citet{Buote1996}; (6) REXCESS: \citet{Boehringer2009}.}
\FloatBarrier

\begin{longtable}{lcccc}
\caption{Structure parameters of the cluster sample. We show the bias and background corrected parameters \p3_c in \r500, \w_c and the new morphology estimator \P3P0$_{\mathrm{max}}$, the peak of the 0.3-1 \r500 P3/P0 profile. Details can be found in Sect. \ref{method} for \p3_c and \w_c and Sect. \ref{Sect81} for \P3P0$_{\mathrm{max}}$. In addition, the morphology as defined in Sect. \ref{secMorph} is given.}\\
\hline\hline\\[-1.2ex]
\multicolumn{1}{c}{Cluster} & \multicolumn{1}{c}{\p3_c} & \multicolumn{1}{c}{\w_c} & \multicolumn{1}{c}{\P3P0$_{\mathrm{max}}$} & \multicolumn{1}{c}{Morphology}\\ [0.5ex]
\hline\\[-1.2ex]
\endfirsthead
\caption{continued.}\\
\hline\hline\\[-1.2ex]
\multicolumn{1}{c}{Cluster} & \multicolumn{1}{c}{\p3_c} & \multicolumn{1}{c}{\w_c} & \multicolumn{1}{c}{\P3P0$_{\mathrm{max}}$}& \multicolumn{1}{c}{Morphology}\\ [0.5ex]
\hline\\[-1.2ex]
\endhead
\hline\\[-1.2ex]
\endfoot

 RXCJ0307.0-2840	&$  -3.3\times10^{-9} \pm 1.4\times10^{-8}   $&$  1.7\times10^{-3} \pm 4.5\times10^{-4} $&$ 9.5\times10^{-8} \pm 4.3\times10^{-8}  $& regular \\
 RXCJ0516.7-5430	&$   7.9\times10^{-7} \pm 3.3\times10^{-7}   $&$  6.0\times10^{-2} \pm 4.3\times10^{-3} $&$ 1.7\times10^{-6} \pm 6.0\times10^{-7}  $& complex\\
 RXCJ0528.9-3927	&$   9.1\times10^{-8} \pm 6.3\times10^{-8}   $&$  2.1\times10^{-2} \pm 1.3\times10^{-3} $&$ 1.6\times10^{-7} \pm 8.6\times10^{-8}  $& complex\\
 RXCJ0532.9-3701	&$   7.5\times10^{-8} \pm 5.7\times10^{-8}   $&$  4.0\times10^{-3} \pm 6.6\times10^{-4} $&$ 5.1\times10^{-7} \pm 2.6\times10^{-7}  $&intermediate\\
 RXCJ0658.5-5556 	&$   1.0\times10^{-7} \pm 1.8\times10^{-8}   $&$  1.6\times10^{-2} \pm 5.6\times10^{-4} $&$ 2.1\times10^{-6} \pm 2.6\times10^{-7}  $& double\\
 RXCJ0945.4-0839	&$   2.3\times10^{-7} \pm 1.4\times10^{-7}   $&$  1.8\times10^{-2} \pm 1.7\times10^{-3} $&$ 2.3\times10^{-7} \pm 1.4\times10^{-7}  $&intermediate\\
 RXCJ2129.6+0005	&$   1.3\times10^{-8} \pm 6.9\times10^{-9}   $&$  6.3\times10^{-3} \pm 3.5\times10^{-4} $&$ 2.4\times10^{-8} \pm 1.9\times10^{-8}  $& regular\\
 RXCJ2308.3-0211\tablefootmark{*}	&$   1.9\times10^{-7} \pm 4.0\times10^{-7}   $&$  2.9\times10^{-3} \pm 1.7\times10^{-3} $&$ 3.4\times10^{-7} \pm 4.5\times10^{-7}$ &intermediate\\   
 RXCJ2337.6+0016	&$  -1.4\times10^{-8} \pm 3.1\times10^{-8}   $&$  3.0\times10^{-2} \pm 3.4\times10^{-3} $&$ 1.8\times10^{-7} \pm 9.6\times10^{-8}  $&intermediate\\
 A68			&$   1.3\times10^{-7} \pm 4.1\times10^{-8}   $&$  1.1\times10^{-2} \pm 6.3\times10^{-4} $&$ 2.6\times10^{-7} \pm 1.5\times10^{-7} $&intermediate\\
 A115			&$   3.0\times10^{-6} \pm 1.5\times10^{-7}   $&$  1.3\times10^{-1} \pm 6.9\times10^{-4} $&$ 6.9\times10^{-6} \pm 3.1\times10^{-7} $& double\\
 A209			&$   5.3\times10^{-8} \pm 3.5\times10^{-8}   $&$  1.1\times10^{-2} \pm 9.8\times10^{-4} $&$ 9.8\times10^{-8} \pm 4.1\times10^{-8} $&intermediate\\
 A267			&$   1.1\times10^{-7} \pm 4.8\times10^{-8}   $&$  9.8\times10^{-3} \pm 1.1\times10^{-3} $&$ 1.1\times10^{-7} \pm 4.8\times10^{-8} $&intermediate\\
 A383			&$   1.7\times10^{-8} \pm 1.0\times10^{-8}   $&$  2.4\times10^{-3} \pm 3.0\times10^{-4} $&$ 7.1\times10^{-8} \pm 2.5\times10^{-8} $& regular\\
 A773			&$  -2.0\times10^{-8} \pm 2.2\times10^{-8}   $&$  5.2\times10^{-3} \pm 6.8\times10^{-4} $&$ 1.1\times10^{-7} \pm 7.7\times10^{-8} $&intermediate\\
 A963			&$   1.4\times10^{-8} \pm 1.3\times10^{-8}   $&$  4.7\times10^{-3} \pm 4.0\times10^{-4} $&$ 1.8\times10^{-7} \pm 7.1\times10^{-8} $& regular\\
 A1413			&$   1.9\times10^{-7} \pm 2.9\times10^{-7}   $&$  3.9\times10^{-3} \pm 1.5\times10^{-3} $&$ 2.5\times10^{-7} \pm 2.1\times10^{-7} $& regular\\
 A1763			&$   6.0\times10^{-7} \pm 1.1\times10^{-7}   $&$  1.1\times10^{-2} \pm 1.1\times10^{-3} $&$ 6.2\times10^{-7} \pm 1.1\times10^{-7} $& complex\\
 A1914			&$   3.5\times10^{-8} \pm 8.8\times10^{-9}   $&$  4.6\times10^{-3} \pm 1.9\times10^{-4} $&$ 2.2\times10^{-7} \pm 2.9\times10^{-8} $&intermediate\\
 A2390			&$   6.7\times10^{-8} \pm 2.0\times10^{-8}   $&$  8.7\times10^{-3} \pm 5.4\times10^{-4} $&$ 9.9\times10^{-8} \pm 6.4\times10^{-8} $&intermediate\\
 A2667			&$   5.2\times10^{-9} \pm 7.0\times10^{-9}   $&$  1.2\times10^{-2} \pm 3.4\times10^{-4} $&$ 2.4\times10^{-7} \pm 5.7\times10^{-8} $&intermediate\\
 A2204			&$   7.3\times10^{-9} \pm 9.4\times10^{-9}   $&$  1.4\times10^{-3} \pm 1.4\times10^{-4} $&$ 3.5\times10^{-8} \pm 1.0\times10^{-8} $& regular\\
 A2218			&$   1.6\times10^{-8} \pm 1.4\times10^{-8}   $&$  1.7\times10^{-2} \pm 1.2\times10^{-3} $&$ 4.1\times10^{-7} \pm 1.8\times10^{-7} $&intermediate\\   
 RXCJ0232.2-4420	&$   1.6\times10^{-7} \pm 6.3\times10^{-8}   $&$  1.8\times10^{-2} \pm 5.6\times10^{-4} $&$ 2.3\times10^{-7} \pm 1.1\times10^{-7} $& complex\\
 A13			&$   3.0\times10^{-7} \pm 6.3\times10^{-8}   $&$  1.7\times10^{-2} \pm 6.2\times10^{-4} $&$ 3.6\times10^{-7} \pm 1.1\times10^{-7} $& intermediate\\
 A520			&$   1.4\times10^{-7} \pm 3.4\times10^{-8}   $&$  2.5\times10^{-2} \pm 3.0\times10^{-3} $&$ 1.6\times10^{-6} \pm 2.4\times10^{-7} $& complex\\
 A665			&$   1.2\times10^{-7} \pm 6.1\times10^{-8}   $&$  4.6\times10^{-2} \pm 8.0\times10^{-4} $&$ 7.1\times10^{-7} \pm 1.4\times10^{-7} $& intermediate\\
 A1068			&$  -4.1\times10^{-9} \pm 7.4\times10^{-9}   $&$  7.1\times10^{-3} \pm 3.3\times10^{-4} $&$ 4.1\times10^{-8} \pm 1.4\times10^{-8} $& regular\\
 A1589			&$   1.1\times10^{-7} \pm 4.2\times10^{-8}   $&$  1.5\times10^{-2} \pm 1.2\times10^{-3} $&$ 4.6\times10^{-7} \pm 9.7\times10^{-8} $& intermediate\\
 A2163			&$   4.1\times10^{-7} \pm 5.7\times10^{-8}   $&$  3.0\times10^{-2} \pm 6.6\times10^{-4} $&$ 1.0\times10^{-6} \pm 1.7\times10^{-7} $& double\\
 A2717			&$   4.6\times10^{-8} \pm 2.1\times10^{-8}   $&$  2.6\times10^{-3} \pm 5.3\times10^{-4} $&$ 6.5\times10^{-8} \pm 2.3\times10^{-8} $& regular\\
 A3112			&$   1.8\times10^{-7} \pm 1.7\times10^{-8}   $&$  3.4\times10^{-3} \pm 1.5\times10^{-4} $&$ 1.8\times10^{-7} \pm 1.7\times10^{-8} $& regular\\
 A3827			&$   7.4\times10^{-8} \pm 1.8\times10^{-8}   $&$  1.0\times10^{-2} \pm 3.2\times10^{-4} $&$ 1.1\times10^{-7} \pm 1.7\times10^{-8} $& regular\\
 A3911			&$   4.7\times10^{-9} \pm 8.6\times10^{-9}   $&$  2.4\times10^{-2} \pm 1.1\times10^{-3} $&$ 4.1\times10^{-7} \pm 9.4\times10^{-8} $& intermediate\\
 A3921			&$   7.5\times10^{-7} \pm 1.1\times10^{-7}   $&$  3.1\times10^{-2} \pm 8.7\times10^{-4} $&$ 1.3\times10^{-6} \pm 1.2\times10^{-7} $& complex\\
 1E1455.0+2232		&$   4.5\times10^{-8} \pm 1.2\times10^{-8}   $&$  3.7\times10^{-3} \pm 1.7\times10^{-4} $&$ 4.5\times10^{-8} \pm 1.2\times10^{-8} $& regular\\
 PKS0745-19		&$  -1.1\times10^{-8} \pm 7.6\times10^{-9}   $&$  1.0\times10^{-3} \pm 2.4\times10^{-4} $&$ 2.3\times10^{-9} \pm 1.2\times10^{-9} $& regular\\   
 RXJ1347.5-1145		&$   1.8\times10^{-8} \pm 6.1\times10^{-9}   $&$  5.5\times10^{-3} \pm 2.7\times10^{-4} $&$ 1.3\times10^{-7} \pm 4.3\times10^{-8} $& regular\\
 Sersic159-3		&$   3.5\times10^{-9} \pm 5.6\times10^{-10}  $&$  1.7\times10^{-3} \pm 5.2\times10^{-5} $&$ 9.4\times10^{-9} \pm 2.0\times10^{-9} $& regular\\
 ZwCl3146		&$   6.7\times10^{-9} \pm 2.0\times10^{-9}   $&$  2.2\times10^{-3} \pm 1.4\times10^{-4} $&$ 6.2\times10^{-8} \pm 1.3\times10^{-8} $& regular\\
 A2597			&$   1.2\times10^{-8} \pm 1.1\times10^{-8}   $&$  9.4\times10^{-4} \pm 1.6\times10^{-4} $&$ 1.2\times10^{-8} \pm 1.1\times10^{-8} $& regular\\
 A1775			&$   2.5\times10^{-7} \pm 5.0\times10^{-8}   $&$  1.7\times10^{-2} \pm 3.0\times10^{-4} $&$ 3.2\times10^{-7} \pm 5.2\times10^{-8} $& complex\\
 A1837			&$   1.1\times10^{-7} \pm 3.0\times10^{-8}   $&$  9.3\times10^{-3} \pm 3.1\times10^{-4} $&$ 2.3\times10^{-7} \pm 3.4\times10^{-8} $& intermediate\\
 RXCJ0014.3-3022 	&$   3.6\times10^{-7} \pm 7.3\times10^{-8}   $&$  4.8\times10^{-2} \pm 1.7\times10^{-3} $&$ 6.1\times10^{-6} \pm 5.7\times10^{-7} $& double\\
 RXCJ1131.9-1955	&$   2.6\times10^{-7} \pm 1.0\times10^{-7}   $&$  3.9\times10^{-2} \pm 1.5\times10^{-3} $&$ 6.8\times10^{-7} \pm 2.0\times10^{-7} $& complex\\
 A1651			&$   5.0\times10^{-10}\pm 8.3\times10^{-9}   $&$  2.0\times10^{-3} \pm 6.6\times10^{-4} $&$ 3.3\times10^{-8} \pm 1.1\times10^{-8} $& regular\\
 A133			&$   3.1\times10^{-8} \pm 1.8\times10^{-8}   $&$  6.8\times10^{-3} \pm 4.0\times10^{-4} $&$ 4.1\times10^{-8} \pm 1.7\times10^{-8} $& regular\\
 A2626			&$   6.9\times10^{-9} \pm 4.1\times10^{-9}   $&$  4.3\times10^{-3} \pm 2.9\times10^{-4} $&$ 1.0\times10^{-8} \pm 4.5\times10^{-9} $& regular\\
 A2065			&$   3.4\times10^{-8} \pm 2.3\times10^{-8}   $&$  1.6\times10^{-2} \pm 3.3\times10^{-4} $&$ 3.4\times10^{-8} \pm 2.3\times10^{-8} $& intermediate\\
 RXCJ0003.8+0203 	&$  -3.3\times10^{-9} \pm 1.3\times10^{-8}   $&$ -1.8\times10^{-4} \pm 8.9\times10^{-4} $&$  2.7\times10^{-7} \pm 6.7\times10^{-8}$& regular\\   
 RXCJ0006.0-3443 	&$   2.2\times10^{-7} \pm 1.0\times10^{-7}   $&$  2.3\times10^{-2} \pm 1.5\times10^{-3} $&$  2.6\times10^{-7} \pm 9.5\times10^{-8}$& intermediate\\
 RXCJ0020.7-2542 	&$  -7.9\times10^{-9} \pm 1.3\times10^{-8}   $&$  1.4\times10^{-2} \pm 9.3\times10^{-4} $&$  7.2\times10^{-7} \pm 1.8\times10^{-7}$& complex\\
 RXCJ0049.4-2931 	&$   2.9\times10^{-8} \pm 5.7\times10^{-8}   $&$  2.3\times10^{-3} \pm 1.2\times10^{-3} $&$  1.2\times10^{-7} \pm 8.1\times10^{-8}$& regular\\
 RXCJ0145.0-5300 	&$   7.6\times10^{-8} \pm 6.0\times10^{-8}   $&$  4.6\times10^{-2} \pm 1.5\times10^{-3} $&$  1.4\times10^{-7} \pm 8.5\times10^{-8}$& intermediate\\
 RXCJ0211.4-4017 	&$   3.7\times10^{-8} \pm 5.0\times10^{-8}   $&$  4.2\times10^{-3} \pm 7.1\times10^{-4} $&$  3.0\times10^{-7} \pm 1.0\times10^{-7}$& regular\\
 RXCJ0225.1-2928 	&$   4.3\times10^{-7} \pm 2.0\times10^{-7}   $&$  5.8\times10^{-5} \pm 1.5\times10^{-3} $&$  4.3\times10^{-7} \pm 2.0\times10^{-7}$& intermediate\\
 RXCJ0345.7-4112 	&$   3.4\times10^{-7} \pm 8.7\times10^{-8}   $&$  8.3\times10^{-4} \pm 1.1\times10^{-3} $&$  3.4\times10^{-7} \pm 8.7\times10^{-8}$& regular\\
 RXCJ0547.6-3152	&$   1.1\times10^{-7} \pm 4.3\times10^{-8}   $&$  1.5\times10^{-2} \pm 6.5\times10^{-4} $&$  1.1\times10^{-7} \pm 4.3\times10^{-8}$& regular\\
 RXCJ0605.8-3518 	&$   1.2\times10^{-8} \pm 4.1\times10^{-9}   $&$  6.3\times10^{-3} \pm 2.6\times10^{-4} $&$  5.4\times10^{-8} \pm 2.7\times10^{-8}$& regular\\
 RXCJ0616.8-4748 	&$   6.7\times10^{-7} \pm 1.6\times10^{-7}   $&$  1.5\times10^{-2} \pm 1.2\times10^{-3} $&$  6.7\times10^{-7} \pm 1.6\times10^{-7}$&intermediate\\
 RXCJ0645.4-5413 	&$   3.1\times10^{-10}\pm 2.1\times10^{-8}   $&$  1.3\times10^{-2} \pm 4.9\times10^{-4} $&$  6.7\times10^{-7} \pm 1.5\times10^{-7}$&intermediate\\
 RXCJ0821.8+0112 	&$   4.5\times10^{-7} \pm 2.4\times10^{-7}   $&$  4.2\times10^{-3} \pm 1.2\times10^{-3} $&$  1.4\times10^{-5} \pm 2.7\times10^{-6}$& double\\
 RXCJ0958.3-1103	&$   2.5\times10^{-8} \pm 2.4\times10^{-8}   $&$  3.0\times10^{-3} \pm 6.7\times10^{-4} $&$  3.0\times10^{-7} \pm 1.4\times10^{-7}$& regular\\
 RXCJ1044.5-0704 	&$   2.8\times10^{-10}\pm 2.0\times10^{-9}   $&$  5.0\times10^{-3} \pm 2.2\times10^{-4} $&$  1.2\times10^{-7} \pm 3.4\times10^{-8}$& regular\\
 RXCJ1141.4-1216 	&$   2.0\times10^{-8} \pm 1.4\times10^{-8}   $&$  2.7\times10^{-3} \pm 3.8\times10^{-4} $&$  2.5\times10^{-8} \pm 1.7\times10^{-8}$& regular\\
 RXCJ1236.7-3354 	&$   3.0\times10^{-9} \pm 3.7\times10^{-8}   $&$  2.8\times10^{-3} \pm 5.3\times10^{-4} $&$  9.3\times10^{-8} \pm 8.3\times10^{-8}$&intermediate\\
 RXCJ1302.8-0230 	&$   2.0\times10^{-7} \pm 5.6\times10^{-8}   $&$  2.5\times10^{-2} \pm 6.3\times10^{-4} $&$  2.2\times10^{-7} \pm 6.5\times10^{-8}$&intermediate\\
 RXCJ1311.4-0120	&$   5.4\times10^{-9} \pm 2.2\times10^{-9}   $&$  3.0\times10^{-3} \pm 2.1\times10^{-4} $&$  2.0\times10^{-8} \pm 6.7\times10^{-9}$& regular\\
 RXCJ1516.3+0005 	&$   2.8\times10^{-8} \pm 1.8\times10^{-8}   $&$  8.0\times10^{-3} \pm 5.3\times10^{-4} $&$  7.2\times10^{-8} \pm 4.2\times10^{-8}$&intermediate\\
 RXCJ1516.5-0056 	&$   6.1\times10^{-7} \pm 1.6\times10^{-7}   $&$  2.4\times10^{-2} \pm 1.3\times10^{-3} $&$  1.5\times10^{-6} \pm 2.7\times10^{-7}$& complex\\
 RXCJ2014.8-2430	&$   2.7\times10^{-8} \pm 7.1\times10^{-9}   $&$  5.6\times10^{-3} \pm 1.9\times10^{-4} $&$  2.7\times10^{-8} \pm 7.1\times10^{-9}$& regular\\
 RXCJ2023.0-2056 	&$   5.8\times10^{-8} \pm 5.1\times10^{-8}   $&$  2.9\times10^{-2} \pm 1.1\times10^{-3} $&$  3.1\times10^{-7} \pm 1.3\times10^{-7}$&intermediate\\
 RXCJ2048.1-1750 	&$   5.6\times10^{-7} \pm 1.2\times10^{-7}   $&$  5.7\times10^{-2} \pm 6.1\times10^{-3} $&$  5.6\times10^{-7} \pm 1.1\times10^{-7}$& complex\\
 RXCJ2129.8-5048 	&$   1.8\times10^{-7} \pm 9.8\times10^{-8}   $&$  4.4\times10^{-2} \pm 1.5\times10^{-3} $&$  2.0\times10^{-6} \pm 4.1\times10^{-7}$& double\\
 RXCJ2149.1-3041 	&$   1.0\times10^{-7} \pm 3.3\times10^{-8}   $&$  5.0\times10^{-3} \pm 4.9\times10^{-4} $&$  1.1\times10^{-7} \pm 2.9\times10^{-8}$& regular\\
 RXCJ2217.7-3543 	&$   8.5\times10^{-8} \pm 2.8\times10^{-8}   $&$  3.2\times10^{-3} \pm 5.6\times10^{-4} $&$  1.7\times10^{-7} \pm 5.4\times10^{-8}$& regular\\
 RXCJ2218.6-3853	&$   3.9\times10^{-8} \pm 1.7\times10^{-8}   $&$  2.0\times10^{-2} \pm 7.4\times10^{-4} $&$  9.7\times10^{-8} \pm 4.6\times10^{-8}$& regular \\
 RXCJ2234.5-3744 	&$   3.7\times10^{-9} \pm 3.2\times10^{-9}   $&$  9.5\times10^{-3} \pm 4.1\times10^{-4} $&$  4.7\times10^{-7} \pm 8.0\times10^{-8}$&intermediate\\
 RXCJ2319.6-7313\tablefootmark{*}	&$  -3.2\times10^{-9} \pm 1.8\times10^{-8}   $&$  2.0\times10^{-2} \pm 1.1\times10^{-3} $&$  4.1\times10^{-8} \pm 6.8\times10^{-8}$ &intermediate\\
 RXCJ2157.4-0747 	&$   4.8\times10^{-6} \pm 1.2\times10^{-6}   $&$  2.2\times10^{-1} \pm 9.5\times10^{-2} $&$  9.4\times10^{-6} \pm 2.5\times10^{-6}$& double

 \label{SampleInfo_structure}			    				   
 \end{longtable}

     \tablefoottext{*}{no significant peak in any aperture, not shown in figures.}
  
\onecolumn
\section{Gallery}
Below we show images of our cluster sample. The clusters are sorted by morphological type and ordered as in Table \ref{SampleInfo}. All images are background subtracted, smoothed and normalized to the surface brightness at 0.3 \r500. A color version of all figures is available in the online journal.

\begin{figure}[!h]
\begin{center}
 \includegraphics[width=\columnwidth]{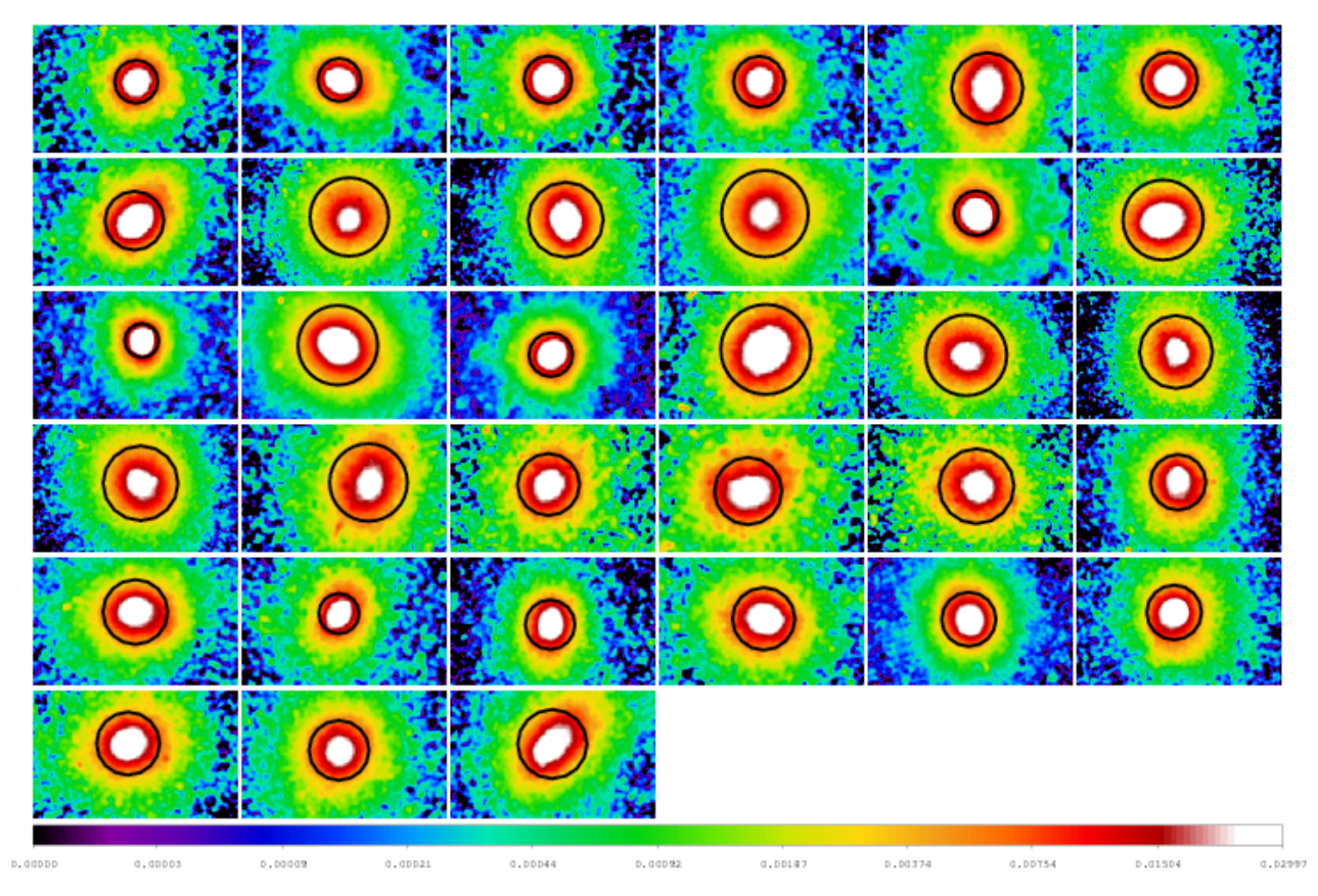}
\caption{Clusters classified as regular - regular clusters without structure. From top left to bottom right: RXCJ0307.0-2840, RXCJ2129.6+0005, A383, A963, A1413, A2204, A1068, A2717, A3112, A3827, 1E1455.0+2232, PKS0745-19, RXJ1347.5-1145, Sersic159-3, ZwCl3146, A2597, A1651, A133, A2626, RXCJ0003.8+0203, RXCJ0049.4-2931, RXCJ0211.4-4017, RXCJ0345.7-4112, RXCJ0547.6-3152, RXCJ0605.8-3518, RXCJ0958.3-1103, RXCJ1044.5-0704, RXCJ1141.4-1216, RXCJ1311.4-0120, RXCJ2014.8-2430, RXCJ2149.1-3041, RXCJ2217.7-3543, RXCJ2218.6-3853.}
\label{Morphregular}
\end{center}
\end{figure}

\begin{figure}[!h]
\begin{center}
 \includegraphics[width=\columnwidth]{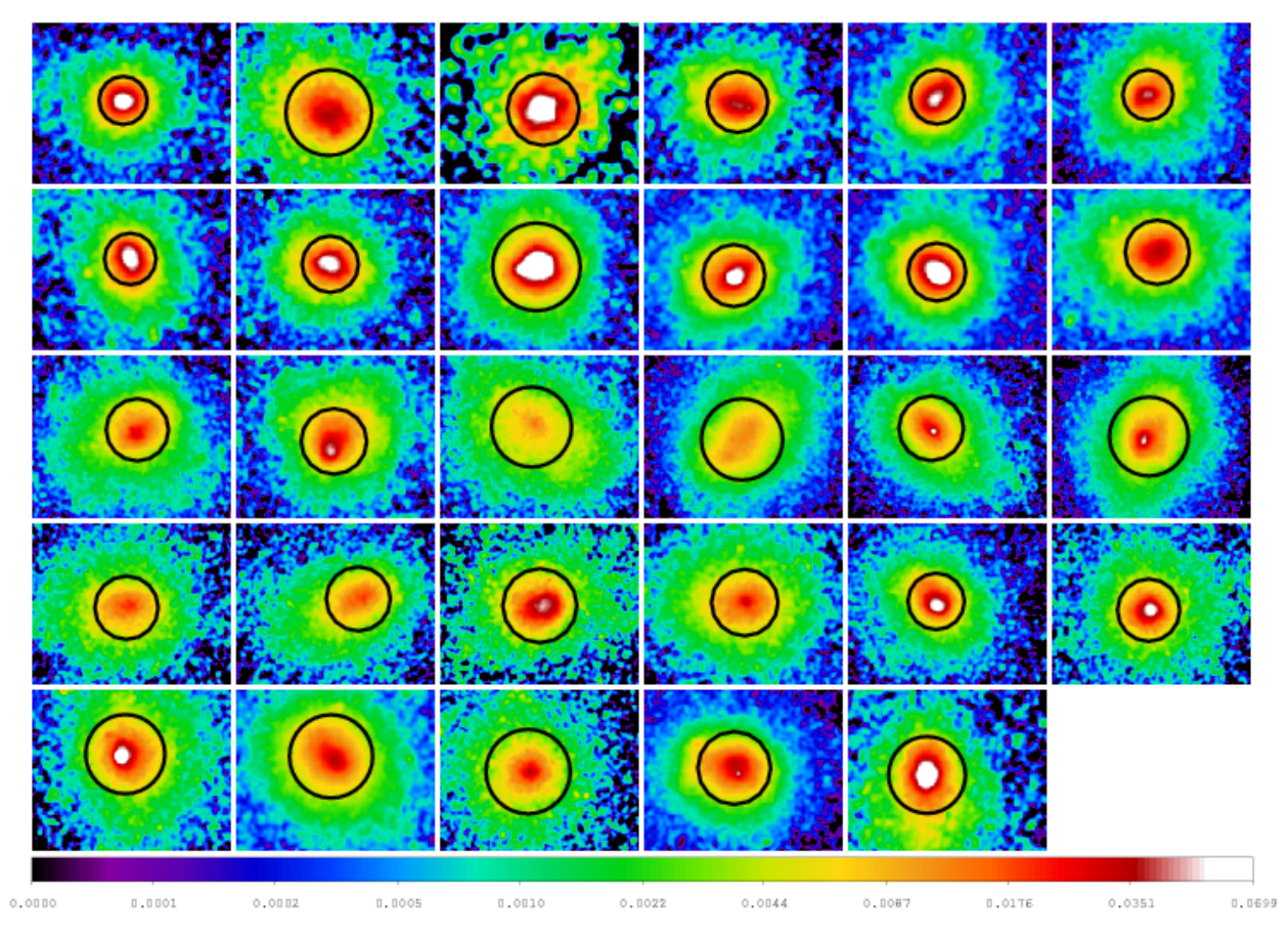}
\caption{Clusters classified as intermediate - overall regular clusters which show some kind of locally restricted structure or slight asymmetry. From top left to bottom right: RXCJ0532.9-3701, RXCJ0945.4-0839, RXCJ2308.3-0211, RXCJ2337.6+0016, A68, A209, A267, A773, A1914, A2390, A2667, A2218, A13, A665, A1589, A3911, A1837, A2065, RXCJ0006.0-3443, RXCJ0145.0-5300, RXCJ0225.1-2928, RXCJ0616.8-4748, RXCJ0645.4-5413, RXCJ1236.7-3354, RXCJ1302.8-0230, RXCJ1516.3+0005, RXCJ2023.0-2056, RXCJ2234.5-3744, RXCJ2319.6-7313}

\label{Morphintermediate}
\end{center}
\end{figure}

\begin{figure*}[!h]
\begin{center}
\begin{minipage}{0.75\columnwidth}
\includegraphics[width=\columnwidth]{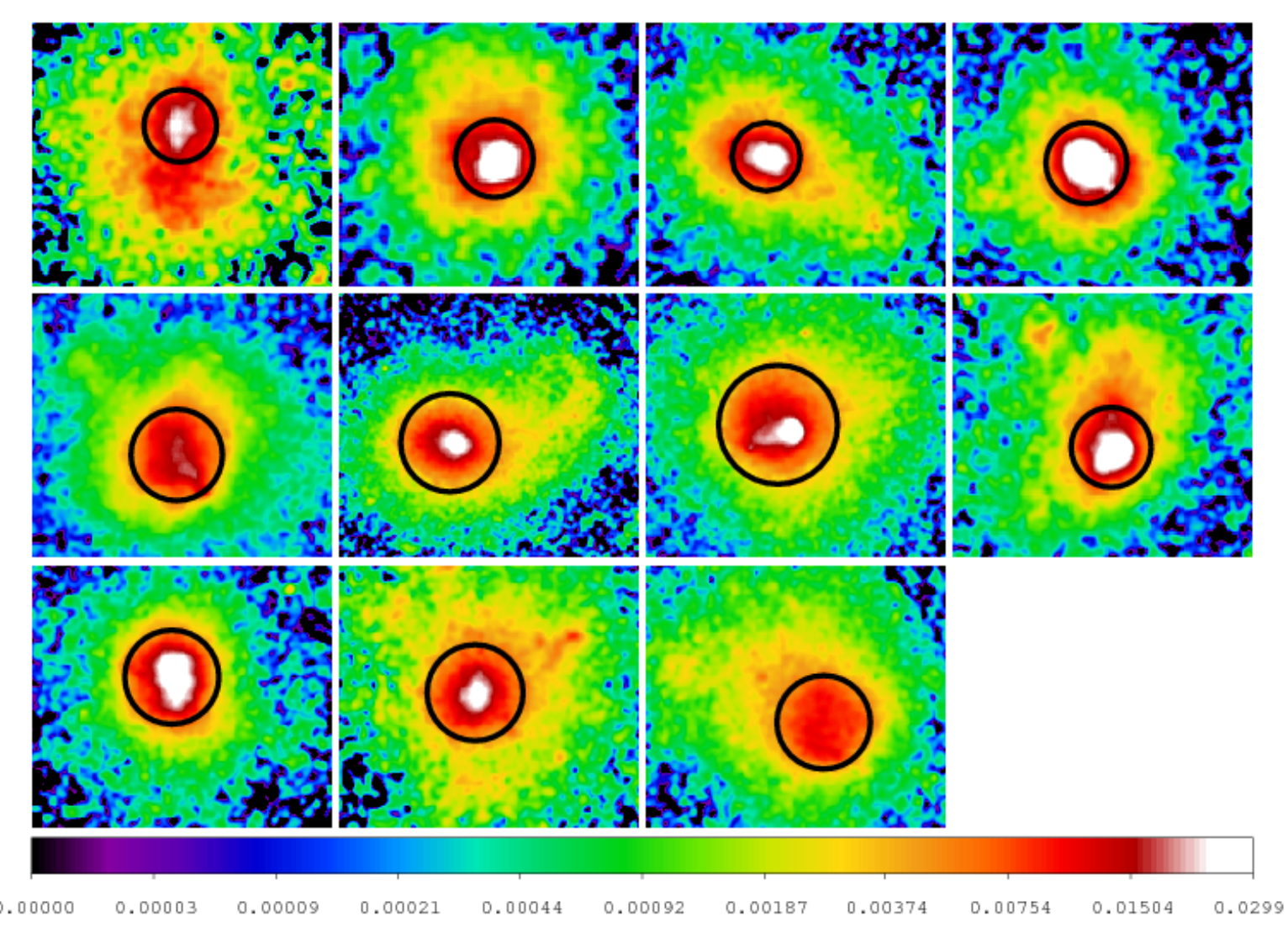}
\caption{Clusters classified as complex - clusters without two distinct maxima but global complex structure. From top left to bottom right: RXCJ0516.7-5430, RXCJ0528.9-3927, A1763, RXCJ0232.2-4420, A520, A3921, A1775, RXCJ1131.9-1955, RXCJ0020.7-2542, RXCJ1516.5-0056, RXCJ2048.1-1750.}
\label{Morphcomplex}
\end{minipage}
\begin{minipage}{0.75\columnwidth}
 \includegraphics[width=\columnwidth]{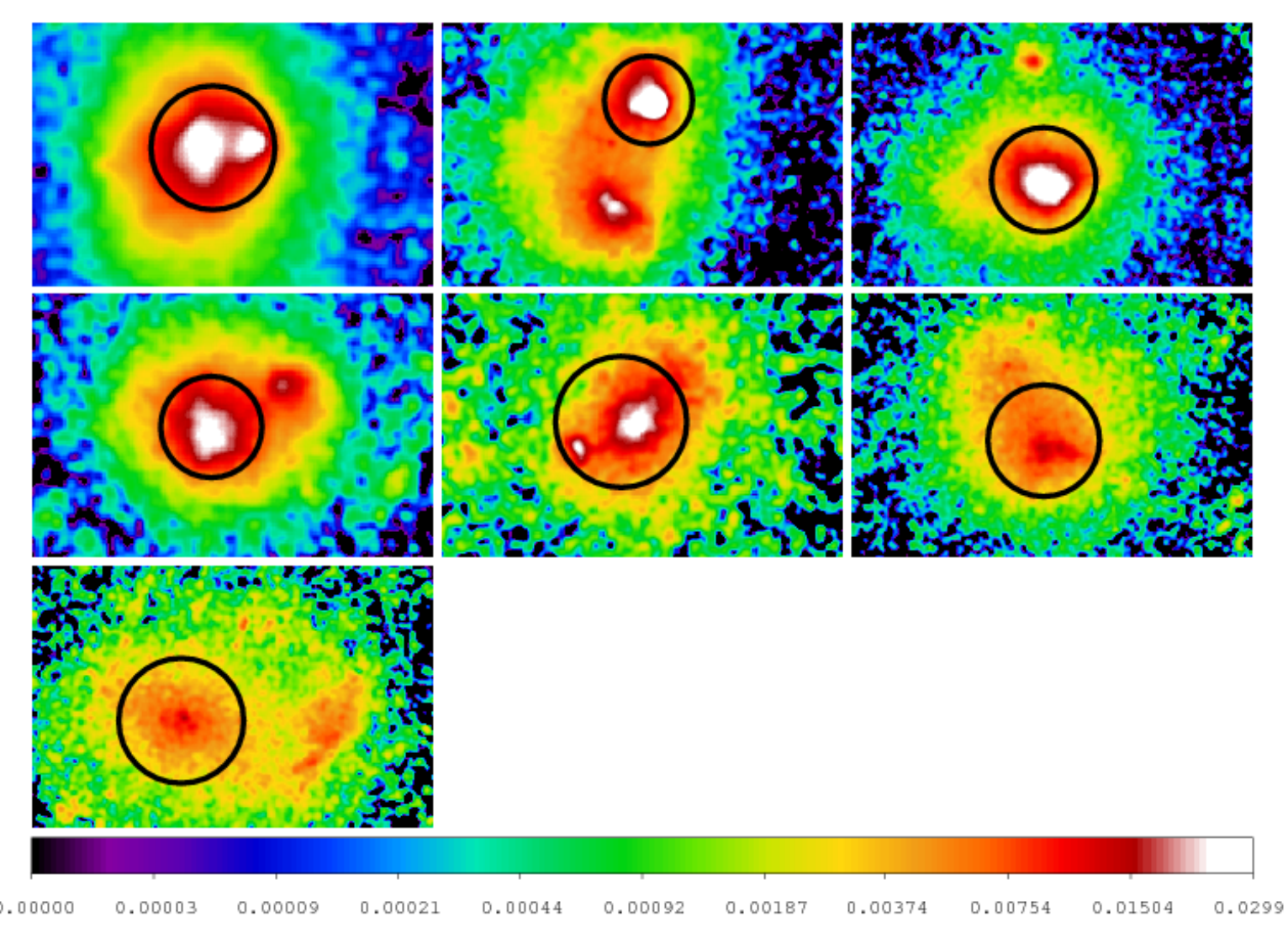}
\caption{Clusters classified as double - clusters two distinct maxima. From top left to bottom right: RXCJ0658.5-5556, A115, A2163, RXCJ0014.3-3022, RXCJ0821.8+0211, RXCJ2129.8-5048, RXCJ2157.4-0747}
\label{Morphdouble}
\end{minipage}
\end{center}
\end{figure*}

\end{appendix}

\end{document}